\documentclass[12pt]{article}
\usepackage{color}
\usepackage{latexsym}
\usepackage{epsfig,amssymb,euscript, mathrsfs}
\usepackage{amsmath}
\usepackage{caption}
\usepackage{subcaption}
\usepackage{color}
\definecolor{MyDarkBlue}{rgb}{0.15,0.15,0.45}
\usepackage[linktocpage=true]{hyperref}
\hypersetup{
colorlinks=true,
citecolor=MyDarkBlue,
linkcolor=MyDarkBlue,
urlcolor=MyDarkBlue,
pdfauthor={Daniel Farquet, Jakob Lorenzen, Dario Martelli,  James Sparks},
pdftitle={Gravity duals of supersymmetric gauge theories on three-manifolds},
pdfsubject={hep-th}
}

\usepackage[numbers,sort&compress]{natbib}

\usepackage{tikz}
\usetikzlibrary{decorations.markings}
\usetikzlibrary[arrows,snakes]

\newsavebox{\ns}
\newsavebox{\dbrane}
\newsavebox{\dbshort}

\def\be{\begin{equation}}
\def\ee{\end{equation}}
\def\bea{\begin{eqnarray}}
\def\eea{\end{eqnarray}}

\newcommand{\nn}{\nonumber}

\newcommand\R{\mathbb{R}}

\newcommand\C{\mathbb{C}}

\newcommand\diff{\mathrm{d}}
\newcommand{\de}{\partial}

\newcommand{\dd}{\mathrm{d}}

\newcommand{\ii}{\mathrm{i}}

\newcommand{\ex}{\mathrm{e}}
\newcommand{\vol}{\mathrm{vol}}
\newcommand{\Vol}{\mathrm{Vol}}
\newcommand{\Imag}{\mathrm{Im}\, }
\newcommand{\Real}{\mathrm{Re}\, }

\newcommand{\Tr}{\mathrm{Tr}}
\newcommand{\ycan}{y^{\mathrm{can}}}

\newlength{\sswidth}

\newcommand{\sign}{\mathrm{sign}}

\newcommand{\ttau}{\tau}
\newcommand{\ppsi}{\vartheta}
\newcommand{\func}{\mathcal{F}}
\newcommand{\spsi}{\nu}
\newcommand{\aspsi}{b_\spsi}
\newcommand{\aphi}{b_\varphi}
\newcommand{\btau}{b_\tau}
\newcommand{\bsigma}{b_\sigma}
\newcommand{\induced}{h}
\newcommand{\K}{L}
\newcommand{\apd}{\hat a}
\newcommand{\newvarphi}{\upsilon}
\newcommand{\anglepsi}{\varsigma}
\newcommand{\TNnu}{\varsigma}

\numberwithin{equation}{section}       

\begin{document}

\begin{titlepage}

\begin{center}

\today

\vskip 2.3 cm 

\vskip 5mm

{\Large \bf Gravity duals of supersymmetric gauge \vskip 5mm
 theories on three-manifolds}

\vskip 20mm

{Daniel Farquet${}^{\,a}$, Jakob Lorenzen${}^{\,b}$, Dario Martelli${}^{\,b}$, and James Sparks${}^{\,a}$}

\vspace{10mm}
\centerline{${}^a${\it Mathematical Institute, University of Oxford,}}
\centerline{{\it Radcliffe Observatory Quarter, Woodstock Road, Oxford OX2 6GG, UK}}
\vspace{.5cm}
\centerline{${}^b${\it Department of Mathematics, King's College London,}}
\centerline{{\it The Strand, London WC2R 2LS, UK}}

\end{center}

\vskip 2 cm

\begin{abstract}
\noindent  We study gravity duals to a broad class of $\mathcal{N}=2$ supersymmetric gauge theories defined
on a general class of three-manifold geometries. The gravity backgrounds are based on Euclidean self-dual solutions to  four-dimensional  gauged supergravity. 
As well as constructing new examples, we prove in general that for solutions defined on the four-ball 
the gravitational  free energy depends only on the supersymmetric Killing vector, finding a simple closed 
formula when the solution has $U(1)\times U(1)$ symmetry. Our result agrees with the large $N$ 
limit of the free energy of the dual gauge theory, computed using localization. This constitutes an 
exact check of the gauge/gravity correspondence for a very broad class of gauge theories with a large $N$ limit, 
defined on a general  class of background three-manifold geometries.

\end{abstract}

\end{titlepage}

\pagestyle{plain}
\setcounter{page}{1}
\newcounter{bean}
\baselineskip18pt
\tableofcontents


\section{Introduction}
\label{sec:intro}

Exact results in quantum field theories are rare and for some time the gauge/gravity duality  \cite{Maldacena:1997re,Gubser:1998bc,WittenAdS}
has been a main tool for obtaining such results in a growing variety of situations. 
More recently, it has been appreciated that  exact non-perturbative computations can be performed in certain 
supersymmetric  field theories defined on curved Riemannian manifolds,  using the technique of  localization 
  \cite{Pestun:2007rz}. On the one hand, this  has motivated  the systematic study of rigid supersymmetry in curved space 
\cite{Festuccia:2011ws},  and on the other hand it has prompted the exploration of the gauge/gravity duality in situations 
when the boundary supersymmetric field  theories are defined on non-trivial curved manifolds. 
This programme has been initiated in \cite{Martelli:2011fu}, where a simple 
(Euclidean) supersymmetric solution of four-dimensional minimal gauged supergravity
 was proposed as the  dual to three-dimensional supersymmetric Chern-Simons quiver
theories defined on a squashed three-sphere (ellipsoid), for which the exact partition function had been computed previously in 
\cite{Hama:2011ea}.  Generalizations  have been discussed by some of the authors in 
\cite{Martelli:2011fw,Martelli:2012sz,Martelli:2013aqa}.  Further examples of four-dimensional gravity solutions with curved 
boundary, where in the dual field theory the path integral can be computed exactly using localization, have been discussed in 
\cite{Huang:2014gca,Nishioka:2014mwa}. In this case, the  exactly calculable quantity on both sides of the duality is the so-called 
supersymmetric R\'enyi entropy \cite{Nishioka:2013haa}, which is a simple modification 
 of the partition function  on the ellipsoid \cite{Hama:2011ea} (see also \cite{Martelli:2011fu}). 
 In five bulk dimensions a supersymmetric solution, 
 where holographic computations have been compared with exact four-dimensional results in  ${\cal N}=1$ SCFTs,  has been recently constructed in \cite{Cassani:2014zwa}, while in \cite{Alday:2014rxa}  the gravity dual to supersymmetric gauge theories on a squashed five-sphere 
has been constructed in Romans $F(4)$ gauged supergravity in six dimensions, and the holographic free energy and BPS Wilson loops 
successfully matched to localization computations in five dimensions.
Gravity solutions dual to exact localization results have also been discussed in \cite{Freedman:2013ryh}
 (for three-dimensional ${\cal N}=2$ theories on $S^3$) and in 
\cite{Russo:2012ay,Buchel:2013id,Bobev:2013cja} (for four-dimensional ${\cal N}=2^*$ theories on $S^4$). 
These, however, have conformally flat boundaries. 

Using localization, the partition function $Z$ of a large class of ${\cal  N}=2$ three-dimensional Chern-Simons theories defined on a general manifold with three-sphere topology was
computed \emph{explicitly} in \cite{Alday:2013lba}. This has provided a unified understanding of all 
previous localization computations on deformed three-spheres \cite{Hama:2011ea,Martelli:2011fu,Imamura:2011wg,Nishioka:2013haa}, 
and has  shown that the partition function on these manifolds depends only on a single parameter $b_1/b_2$, 
related to a choice of almost contact structure.\footnote{This fact  has been recovered independently in \cite{Closset:2013vra}, using different methods.}
Specifically, for a general toric metric on the three-sphere, the real numbers $b_1$, $b_2$ specify a choice of Killing vector $K$ in the torus of isometries.  
 For a broad class of Chern-Simons quiver theories, the large $N$ limit of the free energy ${\cal F}=-\log |Z|$ can be computed using 
 saddle points methods  \cite{Martelli:2011fu}, giving the general result 
 \bea 
 \lim_{N\to \infty} {\cal F}_{\tfrac{b_1}{b_2}} & = & \frac{1}{4}\left(\sqrt{\left|\frac{b_1}{b_2}\right|} + \sqrt{\left|\frac{b_2}{b_1}\right|} \right)^2 {\cal F}_1  ~,
 \label{prediction}
 \eea
 where ${\cal F}_1$ is the large $N$ limit of the free energy on the round three-sphere, scaling with $N^{3/2}$ \cite{Drukker:2010nc}.
  
On the gravity side this yields a universal prediction for the holographically renormalized on-shell action of the corresponding 
 supergravity solutions. Indeed, the on-shell action of  the solutions of  \cite{Martelli:2011fu},  \cite{Martelli:2011fw}, \cite{Martelli:2012sz}, and \cite{Martelli:2013aqa} 
 reproduced this formula, for certain choices of metrics \emph{and} background gauge fields. More precisely, these are all supersymmetric solutions of minimal 
 four-dimensional gauged supergravity in Euclidean signature, and  comprise a negatively curved  Einstein anti-self-dual metric on the four-ball\footnote{References 
 \cite{Martelli:2011fw} and  \cite{Martelli:2012sz} also  discuss several solutions with topology different from the four-ball; however,  currently the precise field 
 theory constructions dual to these remain unknown. In the present paper we will not discuss topologies different from the four-ball.}, with a specific choice of gauge field 
 with  anti-self-dual  curvature, that we refer to as an instanton. The result of \cite{Alday:2013lba} raises two questions: 1) given an arbitrary (toric) 
 metric on the three-sphere, with a background gauge field satisfying the rigid Killing spinor equations  \cite{Klare:2012gn,Closset:2012ru}, can one construct a dual 
 supegravity solution? 2) Assuming such a supergravity solution exists, can one compute the corresponding holographic free energy and show that it matches  (\ref{prediction})?
 
 The purpose of this paper is to address these two questions. Working  in the context of minimal gauged supergravity, and assuming 
 an ansatz that the solutions are \emph{anti-self-dual} and have the topology of the \emph{ball}, we will be able to provide rather general answers
 to both these questions. 
 In the concluding section we will discuss the possibility of extending our results beyond the class of solutions considered in this paper. 
  
 Regarding the first question, we will show that given a (non-singular) 
 anti-self-dual metric on the ball with $U(1)^2$ isometry, and a choice of an arbitrary 
 Killing vector therein, we can construct a (non-singular) instanton configuration, such that together these give a smooth 
 supersymmetric solution of minimal gauged supergravity. Moreover, assuming this metric is asymptotically locally (Euclidean) AdS, 
 we will show that on the conformal boundary the four-dimensional solution reduces to a three-dimensional geometry solving the rigid Killing spinor equations of
 \cite{Klare:2012gn,Closset:2012ru}, in the form presented in \cite{Alday:2013lba}. We will illustrate this construction through 
 several examples, including previously known as well as new solutions. We will also discuss how all the examples that we will present can be understood as arising from 
 an infinite-dimensional  family of explicit  ``$m$-pole'' metrics \cite{CP}.
 
We will be able to answer the second question, regarding the computation of the holographic  free energy,  \emph{independently} of the details
of a specific solution. Namely, assuming only  that a smooth solution with given boundary conditions exists, we will show that the holographically renormalized on-shell action 
takes the form 
\bea
\label{freetoric}
I & = &  \frac{\pi}{2G_4}\cdot \frac{(|b_1|+|b_2|)^2}{4|b_1b_2|}~, 
\eea
precisely matching the large $N$ field theory prediction from localization  (\ref{prediction})! We emphasize that (\ref{freetoric}) will be derived without reference to
a specific solution, and that it receives non-zero contributions from the boundary (as expected), as well as from the bulk, specifically from the ``centre'' of the ball. 
This latter contribution may be understood as arising from the fixed point of the torus action, and can be determined from a fixed point theorem, using the 
Berline-Vergne formula. We will also present formulas relating the renormalized on-shell action to topological invariants of the bulk and conformal invariants of the boundary, by using the Atiyah-Patodi-Singer index theorem, which may be of independent interest.

The rest of this paper is organized as follows. In section \ref{sec:local} we discuss the local geometry of 
Euclidean supersymmetric solutions of minimal four-dimensional gauged supergravity.
In section \ref{sec:global} we turn to global and smooth 
asymptotically locally Euclidean AdS solutions, with the topology of the four-ball. 
Section \ref{sec:free} contains the derivation of the general formula (\ref{freetoric}) 
for the holographic free energy. In section \ref{sec:explicit}
we present examples.
 In section \ref{sec:conclusions}
we conclude by discussing possible extensions of this work. Appendices \ref{spinconnec} and  \ref{sec:Weyl}
contain details about the  geometry, while  in appendix  \ref{sec:construct} we present a unified view of all the examples, 
arising as particular cases of  the $m$-pole metrics \cite{CP}.

\section{Local geometry of self-dual solutions}
\label{sec:local}

The action for the bosonic sector of four-dimensional $\mathcal{N}=2$ gauged supergravity \cite{Freedman:1976aw} is
\bea\label{4dSUGRA}
I^{\mathrm{SUGRA}} &=& -\frac{1}{16\pi G_4}\int \left(R + 6 - F^2\right)\sqrt{\det g}\, \diff^4x~,
\eea 
where $R$ denotes the Ricci scalar of the four-dimensional metric $g_{\mu\nu}$, we have 
defined $F^2\equiv F_{\mu\nu}F^{\mu\nu}$,
and the cosmological constant 
has been normalized to $\Lambda=-3$. The graviphoton is an Abelian gauge field $A$ with field strength 
$F=\diff A$.  The equations of motion derived from (\ref{4dSUGRA}) are
\bea\label{EOM}
R_{\mu\nu} + 3g_{\mu\nu}  &=& 2\left(F_\mu^{\ \rho}F_{\nu\rho}-\tfrac{1}{4}F^2 g_{\mu\nu}\right)~,\nonumber\\
\diff *_4F &=&0~. 
\eea
This is simply Einstein-Maxwell theory with a cosmological constant $\Lambda=-3$. 
Notice that when $F$ is anti-self-dual the right hand side of the Einstein equation in (\ref{EOM}) is zero, 
so that the metric $g_{\mu\nu}$ is  necessarily Einstein.

A solution is supersymmetric provided it admits a (not identically zero) Dirac 
spinor $\epsilon$ satisfying the Killing spinor equation
\bea\label{KSE}
\left( \nabla_\mu - \ii A_\mu + \frac{1}{2} \Gamma_\mu  + \frac{\ii}{4} F_{\nu\rho} \Gamma^{\nu\rho} \Gamma_\mu \right) \epsilon &=& 0~.
\eea
This takes the same form as in Lorentzian signature, except that here 
the gamma matrices generate the Clifford algebra $\mathrm{Cliff}(4,0)$ in an orthonormal frame, so 
$\{\Gamma_\mu,\Gamma_\nu\}=2g_{\mu\nu}$. Notice that we may define the charge conjugate 
of the spinor $\epsilon$ as $\epsilon^c\equiv B\epsilon^*$, where $B$ is the charge conjugation 
matrix satisfying $B^{-1}\Gamma_\mu B=\Gamma_\mu^*$, $BB^*=-1$ and may be chosen to be antisymmetric 
$B^T=-B$ \cite{Martelli:2011fu}. Then provided the 
gauge field $A$ is real (as it will be in the present paper) $\epsilon^c$ satisfies (\ref{KSE})
with $A\rightarrow -A$. 

In \cite{Dunajski:2010uv, Dunajski:2010zp} the authors studied the 
local geometry of Euclidean supersymmetric solutions to the above theory 
for which $F$ is anti-self-dual, $*_4F=-F$. It follows that the metric $g_{\mu\nu}$ 
then has anti-self-dual Weyl tensor, and adopting a standard abuse of 
terminology we shall refer to such solutions as ``self-dual''.\footnote{Einstein 
four-manifolds with anti-self-dual Weyl tensor and non-zero scalar curvature are also sometimes called 
\emph{quaternionic K\"ahler four-manifolds}, with the condition on the Weyl tensor being referred to as \emph{half-conformally flat}. See, for example, \cite{lebrun}.} Supersymmetry 
also equips this background geometry with a Killing vector field $K$. 
Self-dual Einstein metrics with a Killing vector have a rich geometric structure that 
has been well-studied (see for example \cite{Tod}), and are well-known to be related 
by a Weyl rescaling to a (local) K\"ahler metric with zero Ricci scalar. Such 
metrics are described by a solution to a single PDE, known as the Toda
equation, and this solution also  specifies  uniquely the background 
gauge field $A$. In fact we will show that $F=\diff A$ is $\tfrac{1}{2}$ 
the Ricci-form of the conformally related K\"ahler metric, so that 
$A$ is the natural connection on $\mathcal{K}^{-1/2}$, where $\mathcal{K}$ denotes 
the canonical bundle of the K\"ahler manifold.  Moreover, 
we will reverse the direction of implication in \cite{Dunajski:2010uv, Dunajski:2010zp} 
and show that any self-dual Einstein metric with a choice of Killing vector 
field admits (locally) a solution to the Killing spinor equation (\ref{KSE}). 
This may be constructed from the canonically defined spin$^c$ spinor 
that exists on any K\"ahler manifold. 

\subsection{Local form of the solution}\label{sec:review}

In this section we briefly review the local geometry determined in  \cite{Dunajski:2010uv, Dunajski:2010zp}. 
The existence of a non-trivial solution to the Killing spinor equation (\ref{KSE}), together 
with the ansatz that $F$ is anti-self-dual and real, implies that the metric $g_{\mu\nu}$ is 
Einstein with anti-self-dual Weyl tensor. There is then a canonically defined local 
coordinate system in which the metric takes the form
\bea\label{SDE}
\diff s^2_{\mathrm{SDE}} &=& \frac{1}{y^2}\left[V^{-1}(\diff\psi+\phi)^2 + V\left(\diff y^2 + 4\ex^w \diff z\diff\bar{z}\right)\right]~,
\eea
where 
\bea
V &=& 1 - \frac{1}{2}y\partial_y w~,\label{V} \\
\diff\phi &=& \ii \partial_zV\diff y \wedge \diff z - \ii \partial_{\bar{z}}V\diff y \wedge \diff\bar{z} + 2\ii \partial_y(V\ex^w)\diff z\wedge \diff\bar{z}~, \label{dphi}
\eea
and $w=w(y,z,\bar{z})$ satisfies  the Toda equation
\bea\label{Toda}
\partial_z\partial_{\bar{z}} w + \partial^2_y\ex^w &=& 0~.
\eea
Notice that the function $w$ determines entirely the metric. 
The two-form $\diff\phi$ is easily verified to be closed provided the 
Toda equation (\ref{Toda}) is satisfied, implying the existence of a local 
one-form $\phi$.

The vector $K=\partial_\psi$ is a Killing vector field, and arises canonically 
from supersymmetry as a bilinear $K^\mu \equiv \ii \epsilon^\dagger \Gamma^\mu\Gamma_5\epsilon$, where $\epsilon$ is the Killing spinor solving (\ref{KSE}) and $\Gamma_5\equiv\Gamma_{0123}$. 
Notice that the corresponding bilinear in the charge conjugate spinor $\epsilon^c$ 
is $ \ii (\epsilon^c)^\dagger \Gamma^\mu\Gamma_5\epsilon^c=-K^\mu$. Thus 
as in the discussion after equation (\ref{KSE}) we may change 
variables to $\tilde{\epsilon}=\epsilon^c$, $\tilde{A}=-A$. In the tilded 
variables the equations of motion (\ref{EOM}) and Killing spinor equation (\ref{KSE})
are identical to the untilded equations, but now $\tilde{A}=-A$ and $\tilde{K}=-K$.
Thus the sign of the instanton is correlated with a choice of sign for the supersymmetric 
Killing vector, with charge conjugation of the spinor changing the signs of both $A$ and $K$.

As we shall see in the next section, the coordinate $y$ 
determines the conformal factor for the conformally related K\"ahler metric, and is also the Hamiltonian function
for the vector field $K=\partial_\psi$ with respect to the associated symplectic form. The graviphoton field is given (in our conventions) by
\bea\label{A}
 A &=& -\frac{1}{4}V^{-1}\partial_y w(\diff\psi+\phi) +\frac{\ii}{4}\partial_zw\diff z - \frac{\ii}{4}\partial_{\bar{z}}w\diff \bar{z}~.
\eea
We are of course free to make gauge transformations of $A$, and we stress that (\ref{A}) 
is in general valid only locally.

Having summarized the results of  \cite{Dunajski:2010uv, Dunajski:2010zp}, in the next 
two sections we study this local geometry further. In particular we show 
that any self-dual Einstein metric with Killing vector $K\equiv \partial_\psi$, which then 
takes the form (\ref{SDE}), admits a Killing spinor $\epsilon$ solving (\ref{KSE}), 
where $A$ is given by (\ref{A}).

\subsection{Conformal K\"ahler metric}\label{sec:Kahler}

As already mentioned, every self-dual Einstein four-metric with a Killing vector is conformally related 
to a scalar-flat K\"ahler metric. This is given by
\bea\label{Kahler}
\diff s^2_{\mathrm{Kahler}} &\equiv & \diff \hat{s}^2 \ = \ y^2 \diff s^2_{\mathrm{SDE}} \nonumber \\ 
&=& V^{-1}(\diff \psi+\phi)^2 + V\left(\diff y^2 + 4\ex^w\diff z\diff\bar{z}\right)~.
\eea\label{ONkahler}
Introducing an associated local orthonormal frame of one-forms
\bea\label{Kahlerframe}
\hat{e}^0 &=& V^{1/2}\diff y~, \quad \hat{e}^1 \ = \ V^{-1/2}(\diff\psi+\phi)~, \quad \hat{e}^2 + \ii \hat{e}^3 \ = \ 2(V\ex^w)^{1/2}\diff z~,
\eea
the K\"ahler form is
\bea\label{omega}
\omega &=& \hat{e}^{01}+ \hat{e}^{23}~,
\eea
where we have denoted $\hat{e}^0\wedge\hat{e}^1=\hat{e}^{01}$, {\it etc}.
That (\ref{omega}) is indeed closed follows immediately from the expression for $\diff\phi$ in (\ref{dphi}). The K\"ahler form 
is self-dual with respect to the natural orientation on a K\"ahler manifold, namely $\hat{e}^{0123}$ above, and it is with 
respect to this orientation that the curvature $F$ and Weyl tensor are anti-self-dual. 
We denote the corresponding orthonormal frame for the self-dual Einstein metric (\ref{SDE}) 
as $e^a=y^{-1}\hat{e}^a$, $a=0,1,2,3$. 

Next we introduce the Hodge type $(2,0)$-form
\bea
\Omega &\equiv & (\hat{e}^0+\ii \hat{e}^1)\wedge (\hat{e}^2+\ii \hat{e}^3)~,
\eea
and recall that the metric (\ref{Kahler}) is K\"ahler if and only if 
\bea
\diff\Omega &=& \ii \mathcal{P} \wedge \Omega~,
\eea 
where $\mathcal{P}$ is then the Ricci one-form, with Ricci two-form $\mathcal{R}=\diff \mathcal{P}$. 
It is straightforward to compute $\diff\Omega$ for the metric (\ref{Kahler}), and one finds that
\bea
\mathcal{P} &=& 2 A~,
\eea
where $A$ is given by (\ref{A}).
Thus the gauge field is the natural connection on $\mathcal{K}^{-1/2}$, where 
$\mathcal{K}$ denotes the canonical line bundle for the K\"ahler metric. The curvature 
is correspondingly $F=\diff A  = \tfrac{1}{2}\mathcal{R}$, 
where recall that $\mathcal{R}_{\mu\nu} = \tfrac{1}{2}\hat{R}_{\mu\nu\rho\sigma}\omega^{\rho\sigma}$ where
$\hat{R}_{\mu\nu\rho\sigma}$ denotes the Riemann tensor for the K\"ahler metric. A computation gives
\bea\label{Rprimitive}
-2\mathcal{R}\wedge \omega &=& \frac{1}{V\ex^w}\left[\partial_z\partial_{\bar{z}}w + \partial^2_y\ex^w\right]\hat{e}^{0123}~,
\eea
so that the K\"ahler metric is indeed scalar flat if the Toda equation holds.
Since the Ricci two-form has Hodge type $(1,1)$ and the metric is scalar 
flat, it follows immediately that $F= \tfrac{1}{2}\mathcal{R}$ 
is anti-self-dual. This is because the anti-self-dual two-forms 
on a K\"ahler four-manifold are precisely 
the primitive $(1,1)$-forms ({\it i.e.} having zero wedge product with $\omega$, as in (\ref{Rprimitive})), so $\Lambda^2_-\cong \Lambda^{(1,1)}_0$. 
An explicit computation shows that with respect to the frame (\ref{Kahlerframe}) 
\bea
 F &=& -\frac{1}{4}\de_y\left[V^{-1}\de_y w\right]\left(\hat{e}^{01}-\hat{e}^{23}\right) 
+ \frac{1}{8\ex^{w/2}}\Big[\ii(\partial_z-\partial_{\bar{z}})[V^{-1}\partial_yw] \left(\hat{e}^{02}+\hat{e}^{13}\right) 
\nonumber\\ 
&& - (\partial_z+\partial_{\bar{z}})[V^{-1}\partial_yw]\left(\hat{e}^{03}-\hat{e}^{12}\right)\Big]~,
\eea
which is then manifestly anti-self-dual.
One can also derive the  formula
\be
F\ =\ -\left(\tfrac12 y \dd K^\flat + y^2 K^\flat \wedge J K^\flat\right)^-~,
\ee
where $K^\flat$ denotes the one-form dual to the Killing vector $K$ (in the self-dual Einstein metric), and $J$ is the complex structure tensor for the K\"ahler metric (\ref{Kahler}), 
and a further short computation leads to
\bea\label{Fddbar}
F &=& \left(\frac{1}{y}\ii \partial\bar\partial y\right)^- \ = \  \frac{1}{y}\ii\partial\bar\partial y + \frac{1}{4y}\big(\hat\Delta y\big)\omega~,
\eea
where $\bar\partial$ denotes the standard operator on a K\"ahler manifold, 
the superscript ``$-$'' in (\ref{Fddbar}) denotes anti-self-dual part, 
and $\hat\Delta$ denotes the scalar Laplacian for the K\"ahler metric.

Let us note that the K\"ahler form is explicitly
\bea
\omega &=& \diff y\wedge (\diff\psi+\phi) + 2\ii V\ex^w \diff z\wedge \diff\bar{z}~.
\eea
Thus $\diff y = -\partial_\psi\lrcorner\omega$, which identifies the coordinate $y$ as the Hamiltonian function
for the Killing vector $K=\partial_\psi$. Of course, $y^2$ is also the conformal factor 
relating the self-dual Einstein metric to the K\"ahler metric in (\ref{Kahler}).

\subsection{Killing spinor: sufficiency}\label{sec:sufficiency}

In this section we show that a self-dual Einstein metric with Killing vector $K=\partial_\psi$, 
which necessarily takes the form (\ref{SDE}), admits a solution to the Killing spinor 
equation (\ref{KSE}) with gauge field given by (\ref{A}). The key to this construction 
is to begin with the canonically defined spin$^c$ spinor that exists on any K\"ahler manifold. 

The positive chirality spin bundle on a K\"ahler four-manifold takes the form 
$\mathcal{S}_+\cong \mathcal{K}^{1/2}\oplus \mathcal{K}^{-1/2}$, where $\mathcal{K}$ denotes 
the canonical bundle. The spin bundle then exists globally only if the latter admits a square root, 
but the spin$^c$ bundle 
$\mathcal{S}_+\otimes \mathcal{K}^{-1/2} \cong 1 \oplus \mathcal{K}^{-1}$ always exists 
globally.
 In particular the first 
factor in $\mathcal{S}_+\otimes \mathcal{K}^{-1/2} \cong 1 \oplus \mathcal{K}^{-1}$ is 
a trivial complex line bundle, whose sections may be identified with complex-valued functions, 
and there is always a section $\zeta$ satisfying the spin$^c$ Killing spinor equation
\bea\label{spinc}
\left(\hat{\nabla}_\mu - \tfrac{\ii}{2}\mathcal{P}_\mu\right)\zeta &=& 0~.
\eea
Here the hat denotes that we will apply this to the conformal K\"ahler 
metric (\ref{Kahler}) in the case at hand, 
and $\mathcal{P}$ is the Ricci one-form potential we encountered above. 
The connection term in (\ref{spinc}) precisely corresponds 
to twisting the spin bundle $\mathcal{S}_+$ by $\mathcal{K}^{-1/2}$. Using the result earlier 
that $\mathcal{P}=2A$ the spin$^c$ equation (\ref{spinc}) may be rewritten as
\bea\label{spincA}
\left(\hat{\nabla}_\mu -{\ii} A_\mu\right)\zeta &=& 0~,
\eea
which may already be compared with the Killing spinor equation (\ref{KSE}).

More concretely, the solution to (\ref{spinc}), or equivalently (\ref{spincA}), is 
simply given by a constant spinor $\zeta$, so that $\partial_\mu \zeta=0$. This equation makes 
sense globally  as $\zeta$ may be identified with a complex-valued function.
To see this it is useful to take the following projection conditions
\bea
\hat{\Gamma}_1 \zeta &=& \ii \hat{\Gamma}_0 \zeta~, \qquad \hat{\Gamma}_3 \zeta \ = \ \ii \hat{\Gamma}_2 \zeta~,\label{projections}
\eea
following {\it e.g.} reference \cite{Gauntlett:2003cy}. Here $\hat{\Gamma}_a$, $a=0,1,2,3$, denote 
the gamma matrices in the orthonormal frame (\ref{Kahlerframe}).\footnote{Strictly speaking the 
hats are redundant, but we keep them as a reminder that in this section the orthonormal frame is for the K\"ahler metric.}
 The covariant derivative of $\zeta$ is then computed to be
\bea
\hat\nabla_\mu \zeta & =& \left(\de_\mu+\frac{1}{4}\hat\omega_\mu^{\:\,\nu\rho}\hat\Gamma_{\nu\rho}\right) \zeta \ = \ \de_\mu\zeta + \tfrac{\ii}{2}\left(\hat\omega_\mu^{\:01}+\hat\omega_\mu^{\:23}\right) \zeta \ = \ \de_\mu\zeta+\ii A_\mu \zeta~,
\eea
where $\hat\omega_\mu^{\: \,\nu\rho}$ is the spin connection of the conformal K\"ahler metric, 
and we have used the explicit form of this in  appendix \ref{spinconnec} together with the formula (\ref{A})
for $A$.  It follows that simply taking $\zeta$ to be constant, $\partial_\mu\zeta=0$, solves 
(\ref{spinc}). This is a general phenomenon on any K\"ahler manifold.

Using  the canonical spinor $\zeta$ we may construct a spinor $\epsilon$ that is a solution to the Killing spinor equation \eqref{KSE}. Specifically, we find
\bea\label{KS4d}
\epsilon & =& \frac{1}{\sqrt{2y}}\left(1+V^{-1/2}\hat\Gamma_0\right)\zeta~.
\eea
To verify this one first notes that
the spin connections of the K\"ahler metric and the self-dual Einstein metric are related by 
\bea
\hat{\nabla}_\mu \zeta &=& \nabla_\mu \zeta + \frac{1}{2}\hat\Gamma_\mu^{\ \, \nu}(\partial_\nu \log y)\zeta~,
\eea
where $\hat\Gamma_\mu = y\Gamma_\mu$ in a coordinate basis. The Killing spinor equation then takes the form
\bea\label{KSEre}
\left[ \de_\mu+\frac{1}{4}{\hat\omega_\mu}^{\:\:\nu\rho}\hat\Gamma_{\nu\rho}- \frac{1}{2}\hat\Gamma_\mu^{\ \, \nu}(\partial_\nu \log y)\ - \ii  A_\mu + \frac{1}{2y} \hat\Gamma_\mu  + \frac{\ii}{4} y F_{\nu\rho} \hat\Gamma^{\nu\rho} \hat\Gamma_\mu \right]\epsilon \  = \  0~.
\eea
To verify this is solved by (\ref{KS4d})
one simply substitutes (\ref{KS4d}) directly into the left-hand-side of \eqref{KSEre}. Using 
the explicit expressions for the spin connection, the gauge field, the field strength, as well as the projection 
conditions on the canonical spinor $\zeta$ and (\ref{spinc}), one sees that (\ref{KSEre}) indeed holds.

From this analysis we can conclude that the self-dual Einstein metric \eqref{SDE} and the gauge field \eqref{A}, which are solutions to Einstein-Maxwell theory in four dimensions, yield a Dirac spinor $\epsilon$ that is solution to the Killing spinor equation \eqref{KSE}. This implies that these self-dual Einstein backgrounds are always locally supersymmetric solutions of Euclidean $\mathcal{N}=2$ gauged supergravity. We turn to global issues in the next section.


\section{Asymptotically locally AdS solutions}\label{sec:global}

In this section  and the next we will assume that we are given 
a complete (non-singular) self-dual Einstein metric with a Killing vector, which then 
necessarily takes the local form (\ref{SDE}). Moreover, we shall assume this 
metric is asymptotically locally Euclidean AdS,\footnote{Since the 
metric has Euclidean signature one might more accurately describe this boundary condition 
as \emph{asymptotically locally hyperbolic}, which is often used in the mathematics literature.} and in later subsections also that the 
four-manifold $M_4$ on which the metric is defined is topologically a ball. 
A two-parameter family of such self-dual solutions on the four-ball, generalizing all previously known 
solutions of this type, was constructed in \cite{Martelli:2013aqa}. In section 
\ref{sec:explicit} we shall review these solutions, and also introduce 
a number of further generalizations. In particular, the results of the current 
section allow us to deform the choice of Killing vector (which was essentially fixed 
in previous results), and we will also explain  how to generalize 
to an \emph{infinite-dimensional} family of solutions satisfying the above properties, 
starting with the local metrics in \cite{CP}.

With the above assumptions in place, we begin in this section by showing that 
if the Killing vector $K=\partial_\psi$ is nowhere zero in a neighbourhood 
of the conformal boundary three-manifold $M_3$ then it
is a Reeb vector field for an almost contact 
structure on $M_3$. We then reproduce the same geometric structure on $M_3$ studied 
from a purely three-dimensional viewpoint in \cite{Closset:2012ru}. 
In particular   the asymptotic
expansion of the Killing spinor $\epsilon$ leads to the same Killing spinor equation as \cite{Closset:2012ru}. This is important, 
as it shows that the dual field theory is defined on a supersymmetric background 
of the form studied in  \cite{Closset:2012ru}, for which the exact
partition function of a general $\mathcal{N}=2$ supersymmetric gauge theory was computed in \cite{Alday:2013lba} using localization. Having studied 
the conformal boundary geometry, we then turn to the bulk in section 
\ref{sec:globalKahler}.  In particular we show that, with an appropriate 
restriction on the Killing vector $K$, the conformal K\"ahler structure 
of section \ref{sec:Kahler} is everywhere non-singular. 
This allows us to prove in turn that the instanton and Killing spinor
defined by the K\"ahler structure are everywhere non-singular. 

In particular 
this means that each of the self-dual Einstein metrics  in 
section \ref{sec:explicit} leads to a one-parameter family (depending on 
the choice of Killing vector $K$) of smooth supersymmetric solutions.
 In other words, if the self-dual Einstein metric depends on $n$ parameters, the complete 
solution will depend on $n+1$ parameters. We emphasize that in the previously known solutions 
the only example of this phenomenon is the solution of \cite{Martelli:2011fu}. 
There the Einstein metric was simply AdS$_4$,  which doesn't have any parameters. 

\subsection{Conformal boundary at $y=0$}\label{sec:y0}

We are interested in self-dual Einstein metrics of the form (\ref{SDE}) which are asymptotically 
locally Euclidean AdS (hyperbolic), in order to apply to the gauge/gravity correspondence. From the assumptions 
described above there is a single asymptotic region where the metric approaches 
$\frac{\diff r^2}{r^2}+r^2\diff s^2_{M_3}$ as $r\rightarrow\infty$, where $M_3$ is a smooth compact 
three-manifold. In fact the metrics (\ref{SDE}) naturally have such a conformal boundary at $y=0$. 
More precisely, we impose boundary conditions such that $w(y,z,\bar{z})$ is analytic 
around $y=0$, so
\bea\label{wanalytic}
w(y,z,\bar{z}) &=& w_{(0)}(z,\bar{z})+y w_{(1)}(z,\bar{z})+\tfrac{1}{2}y^2w_{(2)}(z,\bar{z})+\mathcal{O}(y^3)~.
\eea
It follows that 
\bea
V(y,z,\bar{z}) &=& 1 - \tfrac{1}{2}y w_{(1)}(z,\bar{z})-\tfrac{1}{2}y^2 w_{(2)}(z,\bar{z})+\mathcal{O}(y^3)~,
\eea
and that the metric (\ref{SDE}) is 
\bea
\dd s^2_{\mathrm{SDE}} & = & \left[1+\mathcal O(y)\right] \frac{\dd y^2}{y^2} + \frac1{y^2}\left[(\diff \psi+\phi_0)^2 + 4\ex^{w_{(0)}}\diff z\diff\bar{z}+\mathcal O(y)\right]~.
\eea
Setting $r=1/y$ this is to leading order
\bea
\diff s^2_{\mathrm{SDE}} & \simeq & \frac{\diff r^2}{r^2} + r^2\left[(\diff \psi+\phi_0)^2 + 4\ex^{w_{(0)}}\diff z\diff\bar{z}\right]~,
\eea
as $r\rightarrow\infty$, so that the metric is indeed asymptotically locally Euclidean AdS around $y=0$.  
Here we have also expanded the one-form tangent to $M_3$
\bea
\phi (y,z,\bar z)\mid_{M_3}  & = & \phi_{(0)}(z,\bar z) + y \phi_{(1)}(z,\bar z) + \mathcal O(y^2)\label{phiexp}.
\eea
In fact by expanding (\ref{dphi}) one can show that $\phi_{(1)}=0$.
Of course, as usual one is free to redefine $r\rightarrow r \Omega(\psi,z,\bar{z})$, 
where $\Omega$ is any smooth, nowhere zero function on $M_3$, resulting in a conformal transformation of the 
boundary metric $\diff s^2_{M_3}\rightarrow \Omega^2\diff s^2_{M_3}$. However, in the present
context notice that $r=1/y$ is a \emph{natural} choice of radial coordinate.

With the analytic boundary condition (\ref{wanalytic}) for $w$ it follows automatically 
that $K=\partial_\psi$ is nowhere zero in a neighbourhood of the conformal boundary 
$y=0$. As we shall see, this will reproduce the same structure on $M_3$ as \cite{Closset:2012ru}, but we should stress that this is not the general situation. For example, 
one could take the standard hyperbolic metric for Euclidean AdS, conformally 
embedded as a unit ball in $\R^4$, and take $K$ to be the Killing vector that 
rotates the first factor in $\R^2\oplus \R^2\cong \R^4$. In fact this will be the \emph{natural}
choice of $K$ that arises in the two-monopole solution described in appendix \ref{2pole}.
The ansatz (\ref{wanalytic}) is thus certainly a restriction on the class of possible 
globally regular solutions, although all examples in section \ref{sec:explicit} have 
choices of Killing vector for which this expansion holds.

Returning to the case at hand,
the conformal boundary is a compact three-manifold $M_3$ (by assumption), 
and from the above discussion a natural choice of representative for the metric is 
\bea\label{3metric}
\diff s^2_{M_3} &=& (\diff \psi+\phi_0)^2 + 4\ex^{w_{(0)}}\diff z\diff\bar{z}~.
\eea
Notice that the form of the metric
(\ref{3metric}) is precisely of the form studied in \cite{Alday:2013lba}. 
In that reference an important role is played by the one-form
\bea\label{eta}
\eta\  \equiv \ \diff\psi+\phi_0~,
\eea
which has exterior derivative
\bea\label{phi0}
\diff\eta \ = \ \diff\phi_0 &=& 2\ii \partial_y(V\ex^w)\mid_{y=0}\diff z\wedge \diff \bar{z} \ = \  \ii w_{(1)}\ex^{w_{(0)}}\diff z\wedge \diff\bar{z}~.
\eea
The form $\eta$ is a global \emph{almost contact one-form} on $M_3$. The most straightforward 
way to derive this in the case at hand is to 
note the form of the boundary Killing spinor equation in section \ref{sec:bKSE} and 
appeal to the results of \cite{Closset:2012ru}.

The Killing vector $K=\partial_\psi$ is the \emph{Reeb vector} for the almost contact form $\eta$, as follows from the 
equations
\bea
K\lrcorner \eta &=& 1~, \qquad K\lrcorner \diff\eta \ = \ 0~.
\eea
The orbits of $K$ thus foliate $M_3$, and moreover this foliation is transversely holomorphic 
with local complex coordinate $z$. When the orbits of $K$ all close it generates a 
$U(1)$ symmetry of the boundary structure, and the orbit space $M_3/U(1)$ is in general 
a compact orbifold surface, on which $z$ may be regarded as a local complex coordinate. 
These are generally called Seifert fibred three-manifolds in the literature.
On the other hand, if $K$ has at least one non-closed orbit then since the isometry group 
of a compact manifold is compact, we deduce that $M_3$ admits at least a $U(1)\times U(1)$ 
symmetry, and the structure defined by $\eta$ is a \emph{toric} almost contact structure.  
 In this case
 we may introduce 
standard $2\pi$-period coordinates $\varphi_1$, $\varphi_2$ on the torus $U(1)\times U(1)$ and write 
\bea\label{Ktoric}
K &=& \partial_\psi \ = \  b_1\partial_{\varphi_1}+b_2\partial_{\varphi_2}~.
\eea

From (\ref{phi0}) we deduce that the Taylor coefficient $w_{(1)}$ is a globally defined \emph{basic} function on 
$M_3$ -- that is, it is invariant under $K=\partial_\psi$. Moreover, the almost contact
form $\eta$ is a \emph{contact form} precisely when the function $w_{(1)}$ is everywhere positive. 
We shall see later that there are examples for which $\eta$ is contact and not contact.
 On the other hand, the coefficient $w_{(0)}$ is in general only a locally 
defined function of $z,\bar{z}$, as one sees by noting that the 
transverse metric $g_T = \ex^{w_{(0)}}\diff z\diff\bar{z}$ is a global two-tensor, 
but in general the complex coordinate $z$ is defined only locally.\footnote{For example, for Euclidean
AdS$_4$ realized as a hyperbolic ball and with  $K=\partial_\psi$ generating the Hopf 
fibration of the boundary $S^3$ then $g_T$ is the standard metric on the round two-sphere,
implying that $w_{(0)}(z,\bar{z})=-2\log (1+|z|^2)$ which blows up at $z=\infty$ (which is 
a smooth copy of $S^1\subset M_3\cong S^3$).} It will be useful in what follows to define 
a corresponding transverse volume form
\bea
\vol_T & \equiv & 2\ii \ex^{w_{(0)}}\diff z\wedge \diff \bar{z}~.
\eea
Again, this is a global tensor on $M_3$, with
\bea
\diff\eta &=& \diff\phi_0 \ = \ \frac{w_{(1)}}{2}\vol_T~.
\eea

\subsection{Boundary Killing spinor}\label{sec:bKSE}

In this section we show that the Killing spinor $\epsilon$ induces 
a Killing spinor $\chi$ on the conformal boundary $M_3$ that solves 
the Killing spinor equation in \cite{Closset:2012ru}.

We begin by recalling the orthonormal frame of one-forms
\bea\label{SDEframe}
{e}^0 &=& \frac{1}{y}V^{1/2}\diff y~, \quad {e}^1 \ = \ \frac{1}{y}V^{-1/2}(\diff\psi+\phi)~, \quad {e}^2 + \ii {e}^3 \ = \ \frac{2}{y}(V\ex^w)^{1/2}\diff z~,
\eea
for the self-dual Einstein metric (\ref{SDE}).  We introduce a corresponding frame for the three-metric $\diff s^2_{M_3}$ on the conformal boundary:
\be\label{fra1}
e^1_{(3)} \ = \ \dd \psi + \phi_{(0)}~, \hspace{12 mm} e_{(3)}^2+\ii e_{(3)}^3 \ = \ 2 \ex^{w_{(0)}/2} \dd z~,
\ee
and will use indices $i,j,k=1,2,3$ for this orthonormal frame. 

We next expand the four-dimensional Killing spinor equation \eqref{KSE} as a Taylor series in $y$.
One starts by noting that $\Gamma^\mu = {e^\mu}_a \Gamma^a=\mathcal O(y)$. But as $\Gamma_\mu = {e^a}_\mu \Gamma_a=\mathcal O(1/y) $ and the field strength expands as $F=F_{(0)}+yF_{(1)}+\mathcal O(y^2)$ we see that 
\bea
\frac{\ii}{4} F_{\nu\rho} \Gamma^{\nu\rho} \Gamma_\mu & =& \mathcal O(y)~.
\eea
After a computation we then obtain
\bea\label{kseexpanint}
\Bigg[ \nabla^{(3)}_\mu - \ii A_{(0)\mu}  +  \frac1{2y} \left(1+\frac14 y w_{(1)}\right) e^i_{(3)\mu} (\Gamma_i-\Gamma_{i0}) +\mathcal O(y) \Bigg] \epsilon \ = \ 0~,
\eea
where $\mu=\psi,z,\bar z$, and where 
\bea\label{A0frame}
A_{(0)} \ = \ - \frac{1}{4}w_{(1)} e_{(3)}^1+ \frac{\ii}{8} \ex^{-w_{(0)}/2} (\de_z - \de_{\bar z}) w_{(0)} e_{(3)}^2-\frac{1}{8}  \ex^{-w_{(0)}/2} (\de_z +\de_{\bar z}) w_{(0)}    e_{(3)}^3~,
\eea
is the lowest order term in the expansion of $A$ given by (\ref{A}). 
The Killing spinor $\epsilon$ then  expands as
\bea\label{KSEexpa}
\epsilon & =& \frac{1}{\sqrt{2 y}} \Big[1+\Gamma_0+\frac14 y w_{(1)} \Gamma_0 +\mathcal O(y^{2}) \Big] \zeta_0~,
\eea
where $\zeta_0$ is the lowest order ($y$-independent) part of the K\"ahler spinor $\zeta$. 
Substituting this into \eqref{kseexpanint} gives a leading order term that is identically zero. The subleading term then reads
\bea
\left[ \left(\nabla^{(3)}_i - \ii  A_{(0)i} \right) (1+\Gamma_0) +  \frac1{8}  w_{(1)} (\Gamma_{i0}-\Gamma_{i}) \right] \zeta_0 & =& 0~.
\eea

The projections (\ref{projections}), in the current context, read
\bea
\Gamma_1 \zeta_0 & =&  \ii \Gamma_0 \zeta_0~, \hspace{20 mm} \Gamma_3 \zeta_0 \ = \ \ii \Gamma_2 \zeta_0~.
\eea
We may choose the following representation of the gamma matrices:
\bea
\Gamma_i & =& \left(\begin{array}{c c} 0 & \sigma_{i} \\  \sigma_{i} & 0 \end{array} \right)~, \hspace{15 mm} \Gamma_0 \ =\ \left(\begin{array}{c c} 0 & \ii \mathbb I_2 \\ -\ii \mathbb I_2 & 0 \end{array} \right)~,
\eea
with $\sigma_i$ the Pauli matrices.\footnote{In this basis the charge conjugation matrix $B$, appearing in $\epsilon^c\equiv B\epsilon^*$, is $B=\left(\begin{array}{cc}\varepsilon & 0 \\ 0 & -\varepsilon\end{array}\right)$ where $\varepsilon=\left(\begin{array}{cc}0 & -1 \\ 1 & 0\end{array}\right)$. } The projection conditions then force $\zeta_0$ 
to take the form\footnote{Notice 
that although our frame coincides with that of \cite{Closset:2012ru}, our three-dimensional 
gamma matrices are a permutation of those in the latter reference, which is why 
the spinor solution takes a slightly different form.}
\bea\label{zeta0}
\zeta_0 & =& \begin{pmatrix}\chi\\ 0\end{pmatrix}\quad\text{where}\quad \chi \ = \ \begin{pmatrix}\chi_0\\ \chi_0\end{pmatrix}~.
\eea
Here $\chi$ is a two-component spinor and $\chi_0$ is simply a constant. The three-dimensional Killing spinor equation then becomes
\bea\label{3dKSE}
\left(\nabla^{(3)}_i - \ii  A_{(0)i} -\frac{\ii}{8} w_{(1)}   \sigma_i \right) \chi & =& 0~. 
\eea
This three-dimensional Killing spinor equation is precisely of the form found in 
\cite{Closset:2012ru}, and studied  in \cite{Alday:2013lba}. More 
precisely, this is the form of the Killing spinor equation in the case where the background 
geometry has real-valued fields, with the metric given by (\ref{3metric}), and the Killing spinor $\chi$ and its charge conjugate $\chi^c$ 
give rise to a
supersymmetric background admitting two supercharges of opposite R-charge. In the notation 
of these references we have that the three-dimensional gauge field $V=0$ (or rather 
there exists a gauge in which this is true -- see appendix \ref{sec:Weyl}), 
while $A=A_{(0)}$ and the function $H=-\frac{\ii}{4}w_{(1)}$. This result shows that there indeed exists a spinor 
$\chi$ with the required properties to construct supersymmetric field theories on $M_3$.

We close this subsection by remarking that supersymmetry singled out a natural representative 
(\ref{3metric}) of the conformal class of the boundary metric. However, one is free 
to make the change in radial coordinate $r\rightarrow r\Omega$, with $\Omega$ any smooth, nowhere zero function on $M_3$, 
resulting in a conformal transformation of (\ref{3metric}) by $\diff s^2_{M_3}\rightarrow 
\Omega^2\diff s^2_{M_3}$. In particular, in the metric (\ref{3metric}) the Killing vector 
$K=\partial_\psi$ has length 1, while the latter conformal rescaling gives $\|K\|_{M_3}=\Omega$. 
In this case one instead finds that the vector $V$ in \cite{Closset:2012ru, Alday:2013lba} is non-zero, with gauge-invariant 
and generically non-zero components $V_2=\de_3\log\Omega$ and $V_3=-\de_2\log\Omega$. 
This is then in agreement with the three-dimensional results of \cite{Alday:2013lba}. For 
further details of this conformal rescaling we refer the reader to appendix \ref{sec:Weyl}.

\subsection{Non-singular gauge}\label{sec:A0}

In a neighbourhood of the conformal boundary the K\"ahler metric 
is defined on $[0,\epsilon)\times M_3$, for some $\epsilon>0$. This follows 
since via the conformal rescaling (\ref{Kahler}) the K\"ahler metric
asymptotes to
\bea\label{Kahlercollar}
\diff s^2_{\mathrm{Kahler}} & \simeq & \diff y^2 + \diff s^2_{M_3}~,
\eea
near to the conformal boundary $y=0$. In particular the K\"ahler structure is 
smooth and globally defined in a neighbourhood of this boundary.
Recall also that the gauge field $A$ is a connection on $\mathcal{K}^{-1/2}$. 
Since every orientable three-manifold is spin the canonical bundle $\mathcal{K}$ 
admits a square root in this neighbourhood, and so $A$ restricts 
to a \emph{bona fide} connection one-form on $M_3$. 
The corresponding $U(1)$ principal bundle can certainly be 
non-trivial for generic topology of $M_3$. In this 
section we analyse the simpler case where $M_3\cong S^3$. 
Here $A$ necessarily restricts to a \emph{global} one-form 
$A_{(0)}$ on the conformal boundary, but as we shall see the 
explicit representative (\ref{A0frame}) is in a singular gauge. 
Correspondingly, since the boundary Killing spinor $\chi$ 
is a spin$^c$ spinor, the solution (\ref{zeta0}) to (\ref{3dKSE}) is similarly in a singular gauge. 
In this section we correct this by writing $A_{(0)}$ 
as a global one-form on $M_3\cong S^3$.

The expression (\ref{A0frame}) for the restriction of $A$ to the conformal boundary is 
of course only well-defined up to gauge transformations. We may rewrite the expression in 
(\ref{A0frame}) as 
\bea\label{Alocal}
A_{(0)}^{\mathrm{local}} &=&  -\frac{1}{4}w_{(1)}(\diff\psi+\phi_0) +\frac{\ii}{4}\partial_zw_{(0)}\diff z - \frac{\ii}{4}\partial_{\bar{z}}w_{(0)}\diff \bar{z}~,
\eea
adding the superscript label ``local'' to emphasize that in general this is only a local one-form.
The first term is $-\frac{1}{4}w_{(1)}\eta$, which is always a global one-form on $M_3$, 
independently of the topology of $M_3$. 
However, the last two terms are not globally defined in general.
 We may remedy this in the case where $M_3\cong S^3$
 by making a gauge transformation, adding an appropriate multiple of $\diff\psi$:
\bea\label{Aglobal}
A_{(0)} &=&  -\frac{1}{4}w_{(1)}\eta +\gamma \left[\diff\psi + \frac{\ii}{4\gamma}\partial_zw_{(0)}\diff z - \frac{\ii}{4\gamma}\partial_{\bar{z}}w_{(0)}\diff\bar{z}\right]~.
\eea
This is then a global one-form on $M_3\cong S^3$ if and only if the curvature two-form of the connection in square 
brackets lies in the 
same basic cohomology class as $\diff\eta=\diff\phi_0$. Concretely, 
we write
\bea\label{B}
\gamma\diff\psi + \frac{\ii}{4}\partial_zw_{(0)}\diff z - \frac{\ii}{4}\partial_{\bar{z}}w_{(0)}\diff\bar{z}
& \equiv & \gamma\diff \psi + B \ \equiv \ \gamma \eta+\alpha~,
\eea
and compute
\bea\label{dB}
\diff B & =&   -\frac{\ii}{2}\partial_{z}\partial_{\bar{z}}w_{(0)}\diff z\wedge \diff \bar{z} \ =\   \left(w_{(1)}^2+w_{(2)}\right) \ex^{w_{(0)}}\frac{\ii}{2}\diff z\wedge \diff\bar{z} \nn\\
&=& \frac{1}{4}\left(w_{(1)}^2+w_{(2)}\right) \vol_T~,
\eea
where we used the Toda equation (\ref{Toda}) and Taylor expanded. 
Since $\eta$ is a global one-form on $M_3\cong S^3$, it follows that 
(\ref{Aglobal}) is a global one-form precisely if $\alpha$ defined via (\ref{B})
is a global \emph{basic} one-form, {\it i.e.}  $\alpha$ is invariant 
under $\mathcal{L}_{\partial_\psi}$ and satisfies $\partial_\psi\lrcorner\alpha=0$. 
In this case we have
\bea\label{basicclass}
\int_{M_3}\eta\wedge \frac{1}{\gamma}\diff B &=& \int_{M_3} \eta \wedge \diff\eta~,
\eea
which may be interpreted as saying that $[\frac{1}{\gamma}\diff B]=[\diff\eta]\in H^2_{\mathrm{basic}}(M_3)\cong \R$ 
lie in the same basic cohomology class.  Indeed, this is the case if and only if 
$\frac{1}{\gamma}\diff B$ and $\diff\eta$ differ by the exterior derivative of a global basic one-form. 

The integral on the right hand side of (\ref{basicclass}) is the \emph{almost contact volume}
of $M_3$:
\bea\label{contactvol}
\Vol_{\eta} &\equiv & \int_{M_3}\eta \wedge \diff\eta \ = \ \int_{M_3} \frac{w_{(1)}}{2}\eta \wedge \vol_T 
\ = \ \int_{M_3}\frac{w_{(1)}}{2}\sqrt{\det g_{M_3}}\, \dd^3x~.
\eea
This played an important role in computing the classical localized Chern-Simons action 
in \cite{Alday:2013lba}, which contributes to the field theory partition function on $M_3$. 
Using (\ref{dB}), (\ref{basicclass}) and (\ref{contactvol}) we see that $A_{(0)}$ in (\ref{Aglobal}) is a global one-form 
if we choose the constant $\gamma$ via
\bea\label{gammaintegral}
\frac{1}{4\gamma}\int_{M_3}\left(w_{(1)}^2+w_{(2)}\right)\, \sqrt{\det g_{M_3}}\, \dd^3x &=& \Vol_{\eta}~.
\eea
We shall return to this formula in section \ref{sec:equivariant}

\subsection{Global conformal K\"ahler structure}\label{sec:globalKahler}

Recall that at the beginning of this section we assumed we were given 
a complete self-dual Einstein metric with Killing vector $K=\partial_\psi$, of the local form (\ref{SDE}). We would like to understand when the  conformal K\"ahler structure, studied locally 
in section \ref{sec:Kahler}, is then globally non-singular. As we shall see, this is not 
automatically the case. Focusing on the case of
toric metrics on a four-ball (all examples in section \ref{sec:explicit} are of this type),  with an appropriate restriction on $K$ 
we will see that the conformal K\"ahler structure is indeed everywhere regular. It follows in this case that the K\"ahler spin$^c$ spinor and instanton $F=\frac{1}{2}\mathcal{R}$ are globally
non-singular, and thus that the Killing spinor $\epsilon$ given by (\ref{KS4d}) is also globally defined and non-singular. Before embarking on this section, 
we warn the reader that the discussion is a little involved, and this section is 
probably better read in conjuction with the explicit examples in section \ref{sec:explicit}. 
In fact the Euclidean AdS$_4$ metric in section \ref{AdS4} displays almost all of the generic 
features we shall encounter.

The self-dual Einstein metrics of section \ref{sec:explicit} are all toric, and we may thus parameterize a choice of toric 
Killing vector $K$ as
\bea\label{Ktoricagain}
K &=& b_1\partial_{\varphi_1}+b_2\partial_{\varphi_2}~,
\eea
where we have introduced 
standard $2\pi$-period coordinates $\varphi_1$, $\varphi_2$ on the torus $U(1)\times U(1)$. 
It will be important to fix carefully the orientations here.
Since the metrics are defined on a ball, diffeomorphic to $\R^4\cong \R^2\oplus \R^2$ 
with $U(1)\times U(1)$ acting in the obvious way, we choose $\partial_{\varphi_i}$ 
so that the orientations on $\R^2$ induce the given orientation on $\R^4$ 
(with respect to which the metric has anti-self-dual Weyl tensor). This fixes 
the relative sign of $b_1$ and $b_2$. Given that we have also assumed that 
$K$ has no fixed points near the conformal boundary, we must also have $b_1$ and 
$b_2$ non-zero. Thus $b_1/b_2\in \R\setminus\{0\}$, and its sign will be important 
in what follows. 

Since the self-dual Einstein metric is assumed regular, the one-form 
$K^\flat$ and its exterior derivative $\diff K^\flat$ are both globally defined and regular. 
The self-dual two-form
\bea\label{twistor}
\Psi & \equiv & \left(\diff K^\flat\right)^+ \ \equiv \ \frac{1}{2}(\diff K^\flat + *\diff K^\flat)~,
\eea
is a \emph{twistor} \cite{CP}, and the invariant definition of the function/coordinate 
$y$ in section \ref{sec:local} is given in terms of its norm  by
\bea\label{ydefinition}
\frac{2}{y^2} & =  & \|\Psi\|^2 \ \equiv \ \frac{1}{2!}\Psi_{\mu\nu}\Psi^{\mu\nu} ~.
\eea
The complex structure tensor for the conformal K\"ahler structure is correspondingly
\bea\label{Jkahleromega}
J^{\mu}_{\ \, \nu} &=& -y\Psi^{\mu}_{\ \, \nu}~,
\eea
where indices are raised and lowered using the self-dual Einstein metric. It is then 
an algebraic fact that $J^2=-1$. 
The conformal K\"ahler structure will thus be everywhere regular, provided 
the functions $y$ and $1/y$ are not zero. Of course $y=0$ is the conformal boundary 
(which is at infinity, and is not part of the self-dual Einstein space). 
We are  free to choose the sign when taking a square root of 
(\ref{ydefinition}), and without loss of generality we take $y>0$ 
in a neighbourhood of the conformal boundary at $y=0$. Since everything is regular, 
in particular the norm of the twistor $\Psi$ cannot diverge anywhere (except 
at infinity), and thus $y\neq 0$ in the interior of the bulk $M_4$. It follows 
that $y$ is everywhere positive on $M_4$.

The Killing vector $K$ is zero only at the ``NUT'', namely the fixed origin of 
$\R^4\cong \R^2\oplus \R^2$. At this point the two-form $\diff K^\flat$, in an orthonormal 
frame, is a skew-symmetric $4\times 4$ matrix whose weights are precisely 
the coefficients $b_1$, $b_2$ in (\ref{Ktoricagain}).\footnote{This is perhaps easiest to see 
by noting that to leading order the metric is flat at the NUT, so one can compute 
$\diff K^\flat$ in an orthonormal frame at the NUT using the flat Euclidean metric on 
$\R^2\oplus\R^2$.} It follows from the definitions 
(\ref{twistor}) and (\ref{ydefinition}), together with a little linear algebra 
in such an orthonormal frame, that
\bea\label{yNUT}
y_{\mathrm{NUT}}&=& \frac{1}{|b_1+b_2|}~.
\eea

The conformal K\"ahler structure will thus be regular everywhere, except 
potentially where $1/y=0$. Suppose 
that $1/y=0$ at a point $p\in M_4\setminus\{\mathrm{NUT}\}$. 
Then $K=\partial_\psi\mid_p\, \neq 0$, and thus from the metric (\ref{SDE}) we see that $1/(Vy^2)\mid_p\, \neq 0$. It follows that the function $V$ must tend to zero 
as $1/y^2$ as one approaches $p$. We may thus write $V=\frac{c}{y^2}+o(1/y^2)$, 
where $c=c(z,\bar{z})$ is non-zero at $p$. Using the definition of $V$ in terms of 
$w$ in (\ref{V}) we thus see that $\partial_y w=\frac{2}{y}-\frac{2c}{y^3}+o(1/y^3)$. 
There are then various ways to see that the corresponding supersymmetric 
supergravity solution is \emph{singular}. Perhaps the easiest is to note from the 
Killing spinor formula (\ref{KS4d}), together with the fact that we may normalise $\zeta^\dagger\zeta=1$, we have
\bea
\epsilon^\dagger\epsilon &=& \frac{1}{2y}\left(1+{V}^{-1}\right)~,
\eea
which from the above behaviour of $V$ then diverges as we approach the point $p$.
It follows that the Killing spinor $\epsilon$ is divergent at $p$, and the solution is singular.

The solutions are thus singular on $M_4\setminus \{\mathrm{NUT}\}$ if and only if  $\{1/y=0\}\setminus \{\mathrm{NUT}\}$ is non-empty. 
Since $y_{\mathrm{NUT}}=1/|b_1+b_2|$, the analysis will be a little different 
for the cases $b_1/b_2=-1$ and $b_1/b_2\neq -1$. We thus assume the latter (generic)
case for the time being. As in the last paragraph, let us suppose $1/y\mid_p=0$.
Due to the behaviour of $V$ and $w$ near $p$, it follows from the form of the metric 
(\ref{SDE}) that $p$ must lie on one of the axes, {\it i.e.}  at $\rho_1=0$ or
at $\rho_2=0$, where $(\rho_i, \varphi_i)$ are standard polar coordinates on 
each copy of $\R^2\oplus \R^2\cong \R^4\cong M_4$, $i=1,2$.\footnote{Notice that 
when $b_1/b_2=-1$ in fact $1/y=0$ at the NUT itself, $\rho_1=\rho_2=0$.} 
In either case there is then an $S^1\ni p$ locus of points where $1/y=0$, 
as follows by following the orbits of the Killing vector $\partial_{\varphi_2}$ or 
$\partial_{\varphi_1}$, respectively. 

To see when this happens, 
our analysis will be based on the fact that, since the Killing vector has finite norm in the interior of $M_4$, one can straightforwardly show that $y$ diverges if and only if $||\dd y||=0$. 
It is then convenient to consider the function $y$ restricted to the relevant axis, {\it i.e.} 
$y\mid_{\{\rho_1=0\}}\equiv y_2(\rho_2)$ or $y\mid_{\{\rho_2=0\}}\equiv y_1(\rho_1)$. 
We have $y_1(0)=y_2(0)=y_{\mathrm{NUT}}>0$. Suppose that $y_i(\rho)$ (for either $i=1,2$) 
starts out decreasing along the axis as we move away from the NUT. Then 
in fact it must remain monotonic decreasing along the whole axis, until 
it reaches $y=0$ at conformal infinity where $\rho=\infty$. The reason for this is simply that if 
$y_i(\rho)$ has a turning point then\footnote{Notice that $\diff y$ necessarily points 
along the axis, given the form of the metric (\ref{SDE}).} $\dd y=0$, which we have already seen can happen 
only where $y$ diverges: but this contradicts the fact that $y_i(\rho)$ is decreasing 
from a positive value at $\rho=0$ (and is bounded below by 0). On the other hand, 
suppose that $y_i(\rho)$ starts out increasing at the NUT. Then since 
at conformal infinity $y_i(\infty)=0$, it follows that $y_i(\rho)$ must have a turning 
point at some finite $\rho>0$. At such a point $y$ will diverge, and from our above 
discussion the solution is singular.

This shows that the key is to examine $\dd y$ at the NUT itself. Recall that the coordinate $y$ is a Hamiltonian function for the Killing vector $K$, {\it i.e.} $\diff y = -K\lrcorner\omega$. From \eqref{Jkahleromega}, we also know that $\omega$ is related to the two-form $\Psi = \left(\diff K^\flat\right)^+$ by $\omega=-y^3\Psi$, yielding $\dd y= y^3 K\lrcorner\left(\diff K^\flat\right)^+$. At the NUT we may again use the polar coordinates $(\rho_i,\varphi_i)$ for the two copies of $\mathbb{R}^2$, where the metric is to leading order the metric 
on flat space.  In the usual orthonormal frame for these polar coordinates, using the above formulae we then compute 
to leading order
\bea\label{signs}
(\dd y)|_{\text{NUT}} & \simeq & \begin{pmatrix}- \frac{b_1}{(b_1+b_2)^2}\ \sign(b_1+b_2) \rho_1\\ 0\\ - \frac{b_2}{(b_1+b_2)^2}\ \sign(b_1+b_2) \rho_2\\ 0\end{pmatrix}\ .
\eea
Thus when $b_1/b_2>0$ we see that $y_i(\rho)$ starts out decreasing at the NUT, for 
\emph{both} $i=1,2$, and from the previous paragraph it follows that the solution 
is then globally non-singular! On the other hand, the case $b_1/b_2<0$ 
splits further into two subcases. For simplicity let us describe the case where $b_2>0$ (with the case $b_2<0$ being similar). Then when $b_1/b_2<-1$ we have $y_2(\rho)$ starts 
out increasing at the NUT, which then leads to a singularity along the axis 
$\rho_1=0$ at some finite value of $\rho_2$; on the other hand, when $-1<b_1/b_2<0$ 
we have that $y_1(\rho)$ starts 
out increasing at the NUT, which then leads to a singularity along the axis $\rho_2=0$ at some finite value of $\rho_1$. Notice these two subcases meet where $b_1/b_2=-1$, when 
we know that $1/y=0$ at the NUT itself, $\rho_1=\rho_2=0$. 

This leads to the simple picture that all solutions with $b_1/b_2>0$ are globally 
regular, while all solutions with $b_1/b_2<0$ are singular, \emph{except} 
when $b_1/b_2=-1$. In this latter case $y$ is infinity at the NUT. As one moves out along either axis $y$
is then necessarily monotonically decreasing to zero, by similar 
arguments to those above. Thus the $b_1/b_2=-1$ solution is in fact also non-singular, although 
qualitatively different from the solutions with $b_1/b_2>0$. 
One can show that, regardless of the values of $b_1$ and $b_2$, the complex structure (\ref{Jkahleromega}) is always the standard complex structure on flat space at the NUT, 
meaning that when $b_1/b_2>0$ the induced complex structure at the NUT is 
$\C^2$, while when $b_1/b_2=-1$ the NUT becomes a point at infinity in the 
conformal K\"ahler metric, with the K\"ahler metric being asymptotically Euclidean. 
In particular the instanton is zero at the NUT in this case, and so is regular there.

Notice that, for the regular solutions, since $K$ is nowhere zero away from the NUT
we may deduce that also $\diff y=-K\lrcorner\omega$ is nowhere zero (as 
$\omega$ is a global symplectic form on $M_4\setminus\{\mathrm{NUT}\}$). In particular $y$ is a global 
Hamiltonian function for $K$, and in particular it is a Morse-Bott function on $M_4$. 
This implies that $y$ has no critical points on $M_4\setminus\{\mathrm{NUT}\}$,
and thus that $y_{\mathrm{NUT}}$ is the \emph{maximum} value of $y$ on $M_4$. 
Moreover, the Morse-Bott theory tells us that constant $y$ surfaces on $M_4\setminus\{\mathrm{NUT}\}$ are all diffeomorphic to $M_3\cong S^3$.

We shall see all of the above behaviour very explicitly in section \ref{sec:explicit} for the case when the self-dual Einstein metric is simply Euclidean AdS$_4$. The more complicated 
Einstein metrics in that section of course also display these features, although the corresponding formulae 
become more difficult to make completely explicit as the examples become more complicated.

\subsection{Toric formulae}\label{sec:equivariant}

In this section we shall obtain some further formulae, valid for any toric self-dual Einstein 
metric on the four-ball. These will be useful for computing the holographic 
free energy in the next section.

We first note that for $M_3\cong S^3$ with Reeb vector (\ref{Ktoric}) the 
almost contact volume in (\ref{contactvol}) may be computed using 
equivaraint localization to give
\bea\label{contacttoric}
\Vol_{\eta} \ = \  \int_{M_3}\eta\wedge \diff \eta &=& -\frac{(2\pi)^2}{b_1b_2}~.
\eea
This formula also appeared in \cite{Alday:2013lba}, although in the present 
paper we have been more careful with sign conventions. 
One proves (\ref{contacttoric}) by an analogous computation 
to the Duistermaat-Heckman formula in \cite{Martelli:2006yb}. 
Specifically, we define a two-form 
\bea
\tilde\omega &\equiv & \tfrac{1}{2}\diff (\varrho^2\eta)~,
\eea
on $M_4$, where $\varrho$ is a choice of radial coordinate with the NUT at 
$\varrho=0$ and the conformal boundary at $\varrho=\infty$, and notice that
\bea\label{DH}
\Vol_{\eta} &=& -\int_{M_4} \ex^{-\varrho^2/2}\frac{1}{2!}\tilde\omega\wedge\tilde\omega~.
\eea 
The minus sign arises here because the natural orientation on $M_3$ 
defined in our set-up is opposite to that on the right hand side of (\ref{DH}). 
Specifically, $y$ is decreasing towards the boundary of $M_4$, so that 
$\diff y$ points inwards from $M_3=\partial M_4$, while $\varrho$ is increasing towards the boundary, 
with $\diff\varrho$ pointing outwards.\footnote{Notice that we could have 
avoided this by choosing $y$ to be strictly negative on the interior of $M_4$, rather than 
strictly positive.} One then evaluates the right hand side of (\ref{DH}) using 
equivariant localization. Specifically, the integrand is 
\bea
\exp\left[-\frac{\varrho^2}{2}+\tilde\omega\right]~,
\eea
which since $K\lrcorner \tilde\omega=-\diff(\tfrac{\varrho^2}{2})$ is an equivariantly 
closed form for $K$, {\it i.e.} is closed under $\diff + K\lrcorner$. The Berline-Vergne 
equivariant integration theorem then localizes the integral to the fixed point 
set of $K$, and one obtains precisely (\ref{contacttoric}), with the $b_i$ appearing 
as the weights of the action of $K$ at the NUT.\footnote{This is then the Duistermaat-Heckman 
formula when $\tilde\omega$ is a symplectic form, {\it i.e.} when $\eta$ is a contact form.}

Finally, let us return to the equation (\ref{gammaintegral}). 
In fact there is another interpretation of the constant $\gamma$,  in terms of the charge of the Killing spinor under $K$. To see this, recall that the solution (\ref{zeta0}) 
to the three-dimensional Killing spinor equation (\ref{3dKSE}) is simply constant 
in our frame, but that was for the case 
where the gauge field $A_{(0)}$ is given by (\ref{Alocal}), which as we saw in section \ref{sec:A0} 
is always in a singular gauge on $M_3\cong S^3$. The gauge transformation 
$A_{(0)}\rightarrow A_{(0)}+\gamma\diff\psi$ that we made in (\ref{Aglobal}) to 
obtain a non-singular gauge implies that the correct global spinor $\chi$ has a phase dependence
\bea\label{spinorpsi}
\chi^{\mathrm{global}} &=& \ex^{\ii\gamma\psi} \begin{pmatrix}\chi_0\\ \chi_0\end{pmatrix}~,
\eea
where $\chi_0$ is a constant complex number. Since the frame is invariant 
under $K=\partial_{\psi}$, we thus deduce that $\gamma$ is precisely 
the charge of the Killing spinor under $K$. 

On the other hand, the total four-dimensional spinor is constructed 
from the canonical spinor $\zeta$ on the conformal K\"ahler manifold, 
via (\ref{KS4d}). Thus $\gamma$ is also the charge of $\zeta$ under $K$. 
This immediately allows us to write down that
\bea\label{gamma}
|\gamma| &=& \frac{|b_1|+|b_2|}{2}~.
\eea
This formula may be fixed by 
looking at the behaviour at the NUT, where recall that the complex structure 
is that of $\C^2$. In terms of complex coordinates $z_1=|z_1|\ex^{\ii \psi_1}$, 
$z_2=|z_2|\ex^{\ii \psi_2}$, the K\"ahler spinor $\zeta$, and hence also our Killing spinor, has charges $\tfrac{1}{2}$ under 
each of $\partial_{\psi_i}$, $i=1,2$. 
However, one must be careful to correctly fix the orientations, which leads to the modulus 
signs in (\ref{gamma}). More precisely, for $b_1/b_2>0$ the conformal K\"ahler metric 
fills the interior of a ball in $\C^2$, while for $b_1/b_2=-1$ instead it is the exterior -- see,
 for example, the discussion at the end of section \ref{AdS4}.


\section{Holographic free energy}
\label{sec:free}

In this section we compute the regularized holographic free energy for 
a supersymmetric self-dual asymptotically locally Euclidean AdS solution defined on the four-ball, 
deriving  the remarkably simple formula (\ref{freetoric}) quoted in the introduction. 

\subsection{General formulae}

The computation of the holographic free energy follows  standard holographic renormalization methods \cite{Emparan:1999pm, Skenderis:2002wp}. 
The total on-shell action is
\bea
I &=&  I^{\text{grav}}_{\text{bulk}} + I^F + I^{\text{grav}}_{\text{bdry}}+ I^{\text{grav}}_{\text{ct}} ~.
\label{alli}
\eea
Here the first two terms are the bulk (Euclidean) supergravity action (\ref{4dSUGRA}) 
\bea\label{incrediblebulk}
I^{\mathrm{SUGRA}} \ = \ I^{\text{grav}}_{\text{bulk}} + I^F &\equiv & -\frac{1}{16\pi G_4}\int_{M_4} \left(R + 6 - F^2 \right)\sqrt{\det g}\, \diff^4x~,
\eea
evaluated on a particular solution with topology $M_4$. The boundary term $ I^{\text{grav}}_{\text{bdry}}$ 
in (\ref{alli})
is the Gibbons-Hawking-York term, required so that the equations of motion (\ref{EOM}) 
follow from the bulk action (\ref{incrediblebulk}) for a manifold $M_4$ with boundary. 
This action is divergent, but we may regularize it using holographic renormalization. Introducing 
a cut-off at a sufficiently small value of $y=\delta>0$, with corresponding hypersurface $\mathcal{S}_\delta
=\{y=\delta\}\cong M_3$, we  have the following total boundary terms
\bea
 I^{\text{grav}}_{\text{bdry}} + I^{\text{grav}}_{\text{ct}} &=&  \frac{1}{8\pi G_4} \int_{\mathcal{S}_\delta}  \left( - K+ 2 + \tfrac{1}{2} R(\induced) \right)\sqrt{\det \induced}\, \diff^3x ~.
\eea
Here $R(\induced)$ is the Ricci scalar of the induced metric $\induced_{ij}$ on $\mathcal{S}_\delta$, and $K$ is the trace of
 the second fundamental form of $\mathcal{S}_\delta$, the latter being the Gibbons-Hawking-York boundary term. 
 It is convenient to rewrite the latter using
 \bea
\int_{\mathcal{S}_\delta} K\sqrt{\det \induced}\, \diff^3x \ = \ \mathcal{L}_n\int_{\mathcal{S}_\delta} \sqrt{\det \induced}\, \diff^3x~,
 \eea
 where $n$ is the outward pointing normal vector to the boundary $\mathcal{S}_\delta$.

\subsection{The four-ball}

In this section we evaluate the total free energy (\ref{alli}) in the case of a supersymmetric 
self-dual solution on the four-ball $M_4\cong B^4\cong \R^4$. 
 
 We deal with each term in (\ref{alli}) in turn, beginning with the gauge field contribution
\bea\label{IF}
I^F &=& \frac{1}{16\pi G_4}\int_{M_4} F^2\sqrt{\det g}\, \diff^4 x \ = \ -\frac{1}{8\pi G_4}\int_{M_4}F \wedge F  \ = \ \int_{M_3} 
A_{(0)}\wedge F_{(0)}~. 
\eea
Here in second equality we have used the fact that $*_4 F = -F$ is anti-self-dual, 
while 
in the last equality we used the fact that on the four-ball $M_4=B^4\cong \R^4$ the curvature $F=\diff A$ is 
globally exact. Thus we may apply Stokes' theorem
with $M_3=\partial M_4$, recalling that the natural orientation on $M_3$ is induced 
from an inward-pointing normal vector, as in the discussion of (\ref{DH}).\footnote{Concretely, the integral over $y$ is $\int_{y_{\mathrm{NUT}}}^0\diff y$, where we chose 
the convention that $y_{\mathrm{NUT}}>0$.}
Notice also that here the gauge field action is already finite, so there is 
no need to realize the conformal boundary $M_3$ as the limit $\lim_{\delta\rightarrow 0}\mathcal{S}_\delta$. 
Next we compute the integrand in (\ref{IF}) using the global form of $A_{(0)}$  (\ref{Aglobal})
in section \ref{sec:A0}. Recall that this reads
\bea
A_{(0)} &=& -\frac{1}{4}w_{(1)}\eta + \gamma\diff\psi + B \ = \  -\frac{1}{4}w_{(1)}\eta+\gamma\eta +\alpha~,
\eea
where in particular $\alpha$ is a global basic one-form. We then compute
\bea\label{AF}
A_{(0)}\wedge F_{(0)} & =&  \frac{w_{(1)}^3}{32} \eta\wedge \vol_T-\frac{1}{4}w_{(1)}\eta
\wedge \diff B - \frac{\gamma}{8}w_{(1)}^2\eta\wedge \vol_T \nn\\
&&+ \gamma \eta\wedge \diff B -\frac{1}{4}\alpha\wedge \diff w_{(1)}\wedge \eta~.
\eea
When we integrate this over $M_3$, the last term may be integrated by parts,
giving an integral that is equal to the integral of $-\frac{1}{4}w_{(1)}\eta\wedge \diff \alpha$, which then 
combines with the first line of (\ref{AF}). 
On the other hand, the first term on the second line of (\ref{AF}) may be evaluted 
in the $U(1)\times U(1)$ toric case using (\ref{dB}), the integral (\ref{gammaintegral}) and 
the formula (\ref{gamma}) for $|\gamma|$. This leads to
 \bea\label{IA}
I^F & = &   - \frac{\pi}{2G_4}\cdot \frac{(|b_1|+|b_2|)^2}{4b_1b_2}+\frac{1}{8\pi G_4}\int_{M_3}\frac{w_{(1)}^3}{32}\sqrt{\det g_{M_3}}\, \dd^3 x\nn\\
&&-\frac{1}{8\pi G_4}\int_{M_3}\frac{1}{8}(w_{(1)}^3+w_{(1)}w_{(2)})\sqrt{\det g_{M_3}}\, \dd^3 x~.
\eea
Notice that the first term closely resembles the free energy appearing in (\ref{freetoric}) --
we shall see momentarily that this combines with a term coming from the gravitational
contribution.

We turn next to the bulk gravity part of the action, which when evaluated on-shell is
\bea
 I^{\text{grav}}_{\text{bulk}} &=& \frac{1}{16\pi G_4}\int_{M_4^\delta} 6 \vol_4~.
\eea
Here $M_4^\delta$ is cut off along the boundary $\mathcal{S}_\delta=\{y=\delta\}\cong M_3$, 
which is necessary as the volume is of course divergent. 
The volume form of interest is 
\bea
\vol_4 &=& \frac{1}{y^4}\diff y \wedge (\diff\psi+\phi)\wedge V\ex^w2\ii \diff z\wedge \diff\bar{z}~.
\eea
A computation reveals that this may be written as the exact form
\bea
-3\vol_4 &=& \diff\Gamma~,
\eea
where we have defined the three-form
\bea\label{Gamma}
\Gamma & \equiv & \frac{1}{2y^2}(\diff\psi+\phi)\wedge \dd\phi + \frac{1}{y^3}(\diff\psi+\phi)\wedge V\ex^w2\ii \diff z\wedge \diff\bar{z}~.
\eea
We may then integrate over $M_4^\delta$ using Stokes' theorem. 
To do this let us define $\varrho$ to be geodesic distance from the NUT -- the origin 
of $M_4\cong B^4\cong \R^4$ that is fixed by the Killing vector $K=\partial_\psi$. 
We then more precisely cut off the space also at small $\varrho>0$ and 
let $\varrho\rightarrow 0$, so that we are integrating over $M_4^{\delta,\varrho}$. 
The form $\Gamma$ may be written
\bea\label{Gammaagain}
\Gamma &=&  \frac{1}{2y^2}(\diff\psi+\phi)\wedge \dd\phi + \frac{1}{y^3}(\diff\psi+\phi)\wedge \omega~,
\eea
where $\omega$ is the conformal K\"ahler form. As argued in section \ref{sec:globalKahler}, 
when $y_{\mathrm{NUT}}$ is finite $\omega$ is everywhere a smooth two-form, and thus in particular in polar coordinates 
near the NUT at $\varrho=0$ it takes the form $\omega\simeq\varrho\diff\varrho\wedge \beta_1 + 
\varrho^2\beta_2$ to leading order, where $\beta_1$ and $\beta_2$ are pull-backs of smooth forms 
on the $S^3= S^3_{\mathrm{NUT}}$ at constant $\varrho>0$. Because of this, the second term in (\ref{Gammaagain}) 
does not contribute to the integral around the NUT. However, notice that
\bea
\int_{S^3_{\mathrm{NUT}}} (\diff\psi +\phi)\wedge \diff\phi &=& \int_{M_3^{y=0}}  (\diff\psi +\phi)\wedge \diff\phi \ = \ -\frac{(2\pi)^2}{b_1b_2}~,
\eea
follows from a simple application of Stokes' theorem, where we have used the almost contact volume 
(\ref{contacttoric}). Using the fact (\ref{yNUT}) that $y_{\mathrm{NUT}}=1/|b_1+b_2|$ 
one thus obtains
\bea
\int_{M_4^{\delta}}\vol_4 &=& \frac{(2\pi)^2|b_1+b_2|^2}{6b_1b_2}+\int_{M_3^{y=0}}\Big[\frac{1}{3\delta^3}+\frac{w_{(1)}}{4\delta^2}\Big]\sqrt{\det g_{M_3}}\, \dd^3x 
~,
\eea
so that
\bea\label{Igrav}
 I^{\text{grav}}_{\text{bulk}} &=& \frac{\pi}{2G_4}\cdot \frac{|b_1+b_2|^2}{2b_1b_2} + \frac{1}{8\pi G_4}\cdot \frac{1}{\delta^3} \int_{M_3^{y=0}} \sqrt{\det g_{M_3}}\, \dd^3x \nn\\
 &&+\frac{3}{32\pi G_4}\cdot \frac{1}{\delta^2} \int_{M_3^{y=0}} w_{(1)}\sqrt{\det g_{M_3}}\, \dd^3x~.
\eea
In particular notice that 
 the $\mathcal{O}(0)$ term at the conformal boundary is zero. This follows from the identity
 \bea\label{Todaintegral}
\int_{M_3}\left(w_{(1)}^3+ 3w_{(1)}w_{(2)}+w_{(3)}\right)\sqrt{\det g_{M_3}}\, \dd^3x &=& 0~,
\eea
which arises from Taylor expanding the Toda equation (\ref{Toda}) as
\bea
0&=& \partial_z\partial_{\bar{z}} w_{(0)} + \ex^{w_{(0)}}\left(w_{(1)}^2+ w_{(2)}\right)\nn\\
&&+ y\left[ \partial_z\partial_{\bar{z}} w_{(1)} + \ex^{w_{(0)}}\left(w_{(1)}^3+ 3w_{(1)}w_{(2)}+w_{(3)}\right)\right] + \mathcal{O}(y^2)~.
\eea
In particular, because $w_{(1)}$ is a smooth global function on $M_3$, the second line 
implies (\ref{Todaintegral}).

It remains to evaluate the boundary terms $ I^{\text{grav}}_{\text{bdry}}+ I^{\text{grav}}_{\text{ct}}$. 
After a computation, and again using (\ref{Todaintegral}), one obtains
\bea\label{Iboundary}
I^{\text{grav}}_{\text{bdry}}+ I^{\text{grav}}_{\text{ct}} &=& -\frac{1}{8\pi G_4\delta^3}\int_{M_3^{y=0}}
\sqrt{\det g_{M_3}}\, \dd^3x - \frac{3}{32\pi G_4\delta^2}\int_{M_3^{y=0}}w_{(1)}\sqrt{\det g_{M_3}}\, \dd^3 x\nn\\
&&+\frac{1}{256\pi G_4}\int_{M_3}\left(3w_{(1)}^3+4w_{(1)}w_{(2)}\right)\sqrt{\det g_{M_3}}\, \diff^3 x~.
\eea
Adding (\ref{Iboundary}) to the bulk gravity term (\ref{Igrav}) we see that the divergent terms 
do indeed precisely cancel, and further combining with (\ref{IA}) we see that the terms involving 
the integrals of $w_{(i)}$ also all cancel.

The computations we have done are valid only for globally regular solutions, and recall 
these divide into the two cases $b_1/b_2>0$, and $b_1/b_2=-1$. In the first case 
the first term in  (\ref{IA}) combines with the first term in (\ref{Igrav}) 
to give
\bea\label{Ifinal}
I &=&  \frac{\pi}{2G_4}\cdot \frac{(|b_1|+|b_2|)^2}{4|b_1b_2|}~,
\eea
where notice $|b_1+b_2|=|b_1|+|b_2|$. On the other hand the isolated case 
with $b_1/b_2=-1$ has $b_1+b_2=0$, so that the free energy comes entirely 
from the first term in (\ref{IA}), which remarkably is then also given by 
the formula (\ref{Ifinal}). Thus for all regular supersymmetric solutions we have shown that 
(\ref{Ifinal}) holds.

\subsection{Index theory formulae}

Although our main result (\ref{Ifinal}) is extremely simple, 
it is also possible to derive another interesting formula for the holographic free energy (that  however  seems less practically useful). 
We begin by following  \cite{anderson}, which rewrites the gravitational contribution 
\bea
I^{\text{grav}} &=&  I^{\text{grav}}_{\text{bulk}}+ I^{\text{grav}}_{\text{bdry}}+ I^{\text{grav}}_{\text{ct}}~,
\eea
to the total holographic free energy  $I=I^{\text{grav}}+I^F$. Specifically, 
we may use  the Gauss-Bonnet formula to rewrite $I^{\text{grav}}$ as \cite{anderson}
\bea\label{and}
I^{\text{grav}} &=& \frac{\pi}{2G_4}\chi(M_4) -\frac{1}{16\pi G_4}\int_{M_4}|W|^2\sqrt{\det g}\, \diff^4 x~,
\eea
where $W$ denotes the Weyl tensor and $\chi(M_4)$ is the Euler number of $M_4$.
For example, for Euclidean AdS, which is conformally flat and has the topology of a four-ball, 
(\ref{and}) immediately gives $I=\frac{\pi}{2G_4}$. 

When the metric on $M_4$ 
is also anti-self-dual one can go further, using the Atiyah-Patodi-Singer index theorem \cite{APS}. This was first 
applied, in the current context, in \cite{hitchin}. The index theorem for the signature operator in general reads \cite{APS}
\bea\label{APSsignature}
\sigma(M_4) &=& -\frac{1}{24\pi^2}\int_{M_4}\mathrm{Tr} \left(R\wedge R\right)+\frac{1}{24\pi^2}\int_{\partial M_4}
\mathrm{Tr}\left(\Pi\wedge R\right) - \eta(\partial M_4)~.
\eea
Here $\sigma(M_4)$ is the signature of $M_4$, $R$ is the curvature tensor of $M_4$, $\Pi$ is the second fundamental form of the boundary, 
and  $\eta(\partial M_4)$ denotes the eta invariant\footnote{We hope that 
no confusion arises between this and the almost contact form on $M_3$, which we have also called $\eta$.} of the boundary conformal 
structure on $\partial M_4 $. Recall that the latter  is defined in terms of the analytic continuation of the series 
\bea\label{etas}
\eta(s) &=& \sum_{\lambda\neq 0}\frac{\mathrm{sign}\, \lambda}{|\lambda|^s}~,
\eea
where the summation is over non-zero eigenvalues $\lambda$ of the first order differential operator 
$B=(-1)^p(*\diff - \diff  *)$ acting on even forms $\Omega^{2p}(\partial M_4)$. Specifically, one 
defines $\eta(\partial M_4)=\eta(0)$, which may thus be thought of as a regularization of the number 
of positive eigenvalues of $B$ minus the number of negative eigenvalues. This is a conformal invariant of the boundary, but 
in general depends on the conformal class. 

We may apply (\ref{APSsignature}) in the case at hand \cite{hitchin} by noting that on a four-manifold
\bea
\mathrm{Tr} \left(R\wedge R\right) &=& - 2\left(|W^+|^2-|W^-|^2\right)\sqrt{\det g}\, \diff^4x~,
\eea
where $W^\pm$ denotes the self-dual/anti-self-dual parts of $W$. Moreover, the boundary term in (\ref{APSsignature}) 
involving the second fundamental form $\Pi$ is zero; this follows because $\Pi$ is proportional to the boundary metric\footnote{That is, the 
boundary is \emph{totally umbilical}.} for the  asymptotically locally Euclidean AdS boundary condition, 
and the trace is then zero on using the Bianchi identity for the curvature $R$. We thus conclude that
\bea\label{APS}
\frac{1}{12\pi^2}\int_{M_4} \left(|W^+|^2-|W^-|^2\right)\sqrt{\det g}\, \diff^4 x &=& \sigma(M_4) +
\eta(\partial M_4)~,
\eea
When $(M_4,\diff s^2_{\mathrm{SDE}})$ is anti-self-dual one can combine (\ref{APS}) with 
(\ref{and}) to obtain (we have corrected a sign in \cite{Anderson:2004yi})
\bea\label{Igraveta}
I^{\text{grav}} &=& \frac{3\pi}{4G_4}\eta(\partial M_4)+\frac{\pi}{4G_4}\left(2\chi(M_4)+3\sigma(M_4)\right)~.
\eea
This expresses the gravitational contribution to the free energy as a conformally 
invariant local contribution from the boundary $\partial M_4$, plus a purely topological part depending 
on the filling $M_4$. 

In the supergravity setting there is also the gauge field contribution $I^F$ to the action. Given that the 
Killing spinor is charged under the graviphoton field $A$, the natural operator on $M_4$ to consider 
is the index of the associated twisted Dirac operator $\mathcal{D}_A$. The index theorem in this case reads
\bea\label{APSdirac}
\mathrm{Ind}_{\mathcal{D_A}}&=&\frac{1}{24\cdot 8\pi^2}\int_{M_4}\Tr\left(R\wedge R\right)-\frac{1}{24\cdot 8\pi^2}\int_{\partial M_4} \Tr\left(\Pi\wedge R\right)-\frac{1}{8\pi^2}\int_{M_4} F\wedge F\nn\\
&-&\frac{1}{2}(\eta_{\mathcal{D}_A}(\partial M_4)+h_{\mathcal{D}_A}(\partial M_4))~.
\eea
Here $\mathrm{Ind}_{\mathcal{D_A}}$ is the index of the Dirac operator on $M_4$, twisted by the 
graviphoton $A$, with APS boundary conditions. The eta invariant is defined analogously to 
(\ref{etas}), replacing the operator $B$ by the restriction of the Dirac operator to the boundary, 
while $h_{\mathcal{D}_A}(\partial M_4)$ is the number of zero modes for that operator. 
As for the signature operator, the boundary term in (\ref{APSdirac}) involving the second fundamental form is 
zero, and we thus find the total holographic free energy $I=  I^{\text{grav}}+I^F$ may be written
\bea
I \, =\, \frac{\pi}{2G_4}\left\{\big[\eta_{\mathcal{D}_A}(\partial M_4)+h_{\mathcal{D}_A}(\partial M_4)+\tfrac{7}{4}\eta(\partial M_4)\big]+\big[\chi(M_4)+2\mathrm{Ind}_{\mathcal{D}_A}+\tfrac{7}{4}\sigma(M_4)\big]\right\}~.
\label{freeindexform}
\eea
Here the terms in the first square bracket depend only on the conformal boundary, via 
eta invariants of the boundary twisted Dirac operator and signature operator, while 
the terms in the second square bracket are topological invariants of $M_4$ 
(each of $\chi(M_4)$, $\sigma(M_4)$ and $\mathrm{Ind}_{\mathcal{D}_{A}}$ is 
of course an integer).

Finally, as a simple corollary of our results notice that we obtain a formula 
for the eta invariant of $M_3\cong S^3$, arising 
as the conformal boundary of a toric self-dual Einstein metric on the ball:
\bea\label{etaformula}
\eta(M_3) &=& \frac{|b_1+b_2|^2}{3b_1b_2}-\frac{2}{3} + \frac{1}{192\pi^2}\int_{M_3}\left(3w_{(1)}^3+4w_{(1)}w_{(2)}\right)\sqrt{\det g_{M_3}}\diff^3 x~.
\eea
For example, in section \ref{TNAdS} below we will see how the general formulae derived thus far 
apply when one takes the self-dual Einstein metric on the four-ball to be the Euclidean Taub-NUT-AdS
metric. In this case the conformal boundary is a biaxially squashed three-sphere. One can then 
use (\ref{etaformula}) to compute the $\eta$ invariant of this conformal geometry to obtain 
$\eta=-\frac{2}{3}(1-4s^2)^2$, where $s$ is the squashing parameter of section \ref{TNAdS}.
This agrees with a direct computation of the eta invariant in \cite{hitchinetaold}.


\section{Examples}
\label{sec:explicit}

In this section we illustrate our general results by discussing three explicit families of solutions. 
These consist of three sets of self-dual Einstein metrics on the four-ball, 
studied previously by some of the authors in \cite{Martelli:2011fu,Martelli:2011fw,Martelli:2012sz,Martelli:2013aqa}. 
We begin with  AdS$_4$ in section \ref{AdS4}. Although the metric 
is trivial, the one-parameter family of instantons given by our general results 
is non-trivial, and it turns out that this family is identical to that in \cite{Martelli:2011fu}. 
The solutions in sections \ref{TNAdS} and \ref{sec:PD} each add a deformation parameter, 
meaning that the metrics in each subsequent section generalize that in the previous section. 
\emph{Particular} supersymmetric instantons on these backgrounds were found in 
\cite{Martelli:2011fw,Martelli:2012sz,Martelli:2013aqa}, but our general results allow us 
to study the most general choice of instanton, leading to new solutions.
Furthermore, in section \ref{infinite} we indicate how to generalize these
metrics further by adding an \emph{arbitrary} number of parameters. 
This is discussed in more detail in appendix \ref{sec:construct}.

\subsection{AdS$_4$}\label{AdS4}

The metric on Euclidean AdS$_4$ can be written as
\bea\label{AdS4metric}
\diff s^2_{\mathrm{EAdS}_4} &=& \frac{\diff q^2}{q^2+1}+q^2\left(\diff\ppsi^2 + 
\cos^2\ppsi \diff\varphi_1^2+\sin^2\ppsi\diff\varphi_2^2\right)~.
\eea
Here $q$ is a radial variable with  $q\in[0,\infty)$, so that the NUT 
is at $q=0$ while the conformal boundary is at $q=\infty$. 
The coordinate $\ppsi\in[0,\tfrac{\pi}{2}]$, with the endpoints being the two 
axes of $\R^2\oplus\R^2\cong \R^4$.  The AdS$_4$ metric is of course both self-dual and anti-self-dual. 

Writing a general choice of Reeb vector field as $K=b_1\partial_{\varphi_1}+b_2\partial_{\varphi_2}$, as in 
our general discussion (\ref{Ktoricagain}), the function $y$ is then 
defined in terms of $K$ via (\ref{twistor}) and (\ref{ydefinition}). Using 
these formulae one easily computes
\bea\label{yAdS}
y(q,\ppsi) \ = \ \frac{1}{\sqrt{(b_2+b_1\sqrt{q^2+1})^2\cos^2\ppsi +(b_1+b_2\sqrt{q^2+1})^2\sin^2\ppsi }}~.
\eea
Notice that indeed $y_{\mathrm{NUT}}=1/|b_1+b_2|$, in agreement 
with (\ref{yNUT}). Using (\ref{yAdS}) one can also verify the general behaviour 
in section \ref{sec:globalKahler} very explicitly. In particular we see 
the very different global behaviour, depending on the sign 
of $b_1/b_2$.  If $b_1/b_2>0$ then $1/y$ is nowhere zero, while 
if $b_1/b_2<0$ instead $1/y$ has a zero on $M_4$. More precisely, 
if $-1<b_1/b_2<0$ then $1/y=0$ at $\{\ppsi=0,q=\sqrt{b_2^2-b_1^2}/|b_1|\}$, 
while if $b_1/b_2<-1$ then $1/y=0$ at $\{\ppsi=\tfrac{\pi}{2},q=\sqrt{b_1^2-b_2^2}/|b_2|\}$. 
These are each a copy of $S^1$ at one or other of the ``axes'' of $\R^2\oplus \R^2$, at the corresponding 
radius given by $q$.
In the special case that $b_1=-b_2$ we have $1/y=0$ at the NUT itself, where the axes meet.
These comments of course all agree with the general analysis in section \ref{sec:globalKahler}, except here all formulae can be made completely explicit.

We thus indeed obtain smooth solutions for all $b_1/b_2>0$, as well as the isolated 
non-singular solution with $b_1/b_2=-1$. 
In fact it is not difficult to check that the former are precisely 
the solutions first found in \cite{Martelli:2011fu}, where the parameter $b^2=b_2/b_1$ 
(compare to the formulae at the beginning of section 2.5 of  \cite{Martelli:2011fu}).
To see this we may compute the instanton using
 the formulae
in section \ref{sec:local}, finding
\bea
A &=& \frac{\left(b_1+b_2\sqrt{q^2+1}\right)\diff\varphi_1+\left(b_2+b_1\sqrt{q^2+1}\right)
\diff\varphi_2}{2\sqrt{(b_2+b_1\sqrt{q^2+1})^2\cos^2\ppsi +(b_1+b_2\sqrt{q^2+1})^2\sin^2\ppsi }}~,
\eea
which agrees with the corresponding formula in \cite{Martelli:2011fu}. In particular 
one can check that this gives a regular instanton when $b_1/b_2>0$, with the 
particular cases that $b_1/b_2=\pm 1$ giving a \emph{trivial} instanton, and 
correspondingly the conformal K\"ahler structure is flat. We shall comment 
further on this below. Moreover, one can also check that the singular instantons with $b_1/b_2<0$
are singular at precisely the locus that $1/y=0$, again in agreement with our general discussion. 

In this case we may also compute all other functions appearing in sections \ref{sec:local}, \ref{sec:global} and \ref{sec:free} explicitly. 
For example, we find
\bea
V(q,\ppsi) \ = \ \frac{(b_2+b_1\sqrt{q^2+1})^2\cos^2\ppsi +(b_1+b_2\sqrt{q^2+1})^2\sin^2\ppsi}{q^2(b_1^2\cos^2\ppsi + b_2^2\sin^2\ppsi)}~,
\eea
while the functions $w_{(1)}$ and $w_{(2)}$ on $\partial M_4=M_3\cong S^3$
appearing in the free energy computations are given by
\be
w_{(1)} \ =\ \frac{-4b_1b_2}{\sqrt{b_1^2\cos^2\ppsi + b_2^2\sin^2\ppsi}}~, 
~~~~ w_{(2)} \ = \ \frac{-2\left(3b_1^2b_2^2+b_1^4\cos^2\ppsi + b_2^4\sin^2\ppsi\right)}{b_1^2\cos^2\ppsi + b_2^2\sin^2\ppsi}~.
\ee
Using these expressions one can verify all of the key formulae in our general 
analysis. For example, the integrals in (\ref{contacttoric}), (\ref{IA}), (\ref{Igrav}) and 
(\ref{Iboundary}) are all easily computed in closed form.

Finally, let us return to discuss the special cases $b_1/b_2=\pm 1$, where recall 
that the instanton is trivial and the conformal K\"ahler structure is flat. 
The latter is thus locally the flat K\"ahler metric on $\C^2$, but in fact 
in the two cases $b_1/b_2=\pm 1$ the Euclidean AdS$_4$ metric is conformally
embedded into \emph{different} regions of $\C^2$. Notice this has to be the case, 
because the conformal factor $y$ of the $b_1/b_2=+1$ solution has $y_{\mathrm{NUT}}=1/(2|b_1|)$, while for the $b_1/b_2=-1$ solution instead $y_{\mathrm{NUT}}=\infty$. 
We may see this concretely by writing the flat K\"ahler metric on $\C^2$ as
\be
\diff s^2_{\mathrm{flat}} = \diff R^2 + R^2\left(\diff\ppsi^2 + 
\cos^2\ppsi \diff\varphi_1^2+\sin^2\ppsi\diff\varphi_2^2\right)~.
\ee
In both cases the change of radial coordinate to (\ref{AdS4metric}) is 
\be 
q(R)\ = \ \frac{2R}{|R^2-1|}~.
\ee
However, for the $b_1/b_2=+1$ case the range of $R$ is $0\leq R<1$, with the NUT being at 
$R=0$ and the conformal boundary being at $R=1$; while for the  $b_1/b_2=-1$ 
case the range of $R$ is instead $1<R\leq \infty$, with the NUT being at $R=\infty$ (and 
the conformal boundary again being at $R=1$). In particular the two conformal factors 
are 
\be 
y(R)  =  \tfrac{1}{2|b_1|}|R^2-1|~.
\ee
The two solutions $b_1/b_2=\pm 1$ thus effectively fill opposite sides of the unit sphere 
in $\C^2$, and because of this they induce opposite orientations on $S^3$. Again, this may 
be seen rather explicitly in various formulae. For example, $w_{(1)}=\mp 4|b_1|$ 
in the two cases, so that the boundary Killing spinor equation (\ref{3dKSE}) on the round $S^3$ becomes 
respectively $\nabla^{(3)}_i\chi = \mp \frac{\ii}{2}|b_1|\gamma_i\chi$, where one can take the gamma matrices to be the Pauli matrices $\gamma_i=\sigma_i$ in an orthonormal frame.

\subsection{Taub-NUT-AdS$_4$}\label{TNAdS}

The Taub-NUT-AdS$_4$ metrics are a one-parameter family of self-dual Einstein metrics on
the four-ball, and have been studied in detail in \cite{Martelli:2011fw, Martelli:2012sz}.
The metric may be written as
\bea\label{TNAdSmetric}
\dd s^2_4 & =& \frac{r^2-s^2}{\Omega(r)}\diff r^2 + (r^2-s^2)(\ttau^2_1+\ttau_2^2) + \frac{4s^2\Omega(r)}{r^2-s^2}\ttau_3^2~,
\eea
where 
\bea
\Omega(r)  &= & (r\mp s)^2[1+(r\mp s)(r\pm3s)]~,
\eea
and $\ttau_1,\ttau_2,\ttau_3$  are left-invariant one-forms on $SU(2) \simeq S^3$. The latter may 
 be written in terms of Euler angular variables as
\bea
\ttau_1+\ii\ttau_2 & =& \ex^{-\ii\anglepsi}(\dd\theta+\ii\sin{\theta}\dd\varphi)\ ,\qquad \ttau_3 \ = \ \dd\anglepsi+\cos{\theta}\dd\varphi\ .
\eea
Here $\anglepsi$ has period $4\pi$, while $\theta\in[0,\pi]$ with $\varphi$ having period $2\pi$. The radial coordinate $r$ 
lies in the range $r\in [s,\infty)$, with the NUT (origin of the ball $\cong\R^4$) being at $r=s$. 
 The parameter $s>0$ is referred to as the \emph{squashing parameter}, 
with $s=\frac{1}{2}$ being the Euclidean AdS$_4$ metric studied in the previous section. 
Indeed, the metric is asymptotically locally Euclidean AdS as $r\to\infty$, with
\bea
\dd s^2_4 & \approx & \frac{\dd r^2}{r^2}+r^2(\ttau_1^2+\ttau_2^2+4 s^2\ttau_3^2)\ ,
\eea
so that the conformal boundary at $r=\infty$ is a biaxially squashed $S^3$.

Using the results of this paper we may write 
 a general choice of Reeb vector field as $K=(b_1+b_2)\partial_\varphi+(b_1-b_2)\partial_\anglepsi$, as in 
our general discussion (\ref{Ktoricagain}), and the function $y$ is then 
defined in terms of $K$ via (\ref{twistor}) and (\ref{ydefinition}). Using 
these one computes
\bea\nn
\frac{1}{y(r,\theta)^2} &=&\left[2(b_1-b_2)(r-s)s +(b_1+b_2)(1+2(r-s)s)\cos{\theta}\right]^2\\
&&+(b_1+b_2)^2 \left[1+(r-s) (r+3 s)\right]\sin^2{\theta}\ .\label{yTNUTAdS}
\eea
Notice that indeed $y_{\text{NUT}}=\lim_{r\to s}y(r,\theta)=1/|b_1+b_2|$. We see that if $b_1/b_2>0$ or $b_1/b_2=-1$ then $1/y$ is indeed never zero (except at the NUT in the latter case), as expected. In this way we obtain a \emph{two-parameter} family of regular supersymmetric 
solutions, parametrized by the squashing parameter $s$ and $b_1/b_2$. 
One can also compute explicitly the corresponding instanton  $F$ for a general choice of $s$ and $b_1/b_2$, 
although in practice it turns out to be more convenient to derive this as a special limit of the Plebanski-Demianski solutions, discussed in section \ref{sec:PD}. 
We do this in appendix \ref{TNlimit}, where the resulting expression for $F$ is given in (\ref{longtn}). In the remainder of this subsection
we shall instead discuss further some special cases, making contact with the previous results \cite{Martelli:2011fw,Martelli:2012sz}.

While the Taub-NUT-AdS metric (\ref{TNAdSmetric}) has $SU(2)\times U(1)$ isometry, a 
 generic choice of the Killing vector $K=(b_1+b_2)\partial_\varphi+(b_1-b_2)\partial_\anglepsi$ breaks the symmetry of the full solution to $U(1)\times U(1)$. In particular, this symmetry is also broken by the corresponding instanton $A$.
On the other hand, in \cite{Martelli:2011fw, Martelli:2012sz} the $SU(2)\times U(1)$ symmetry of the metric was \emph{also} imposed 
on the gauge field, which results in two one-parameter subfamilies of the above two-parameter family of solutions, which are $1/4$ BPS and $1/2$ BPS, respectively. In each case this effectively fixes the  Killing 
vector $K$ (or rather the parameter $b_1/b_2$) as a function of the squashing parameter $s$.

\paragraph{1/4 BPS solution:}This solution is simple enough that it can be presented in complete detail. The coordinate transformation to the \eqref{SDE} form for the $1/4$ BPS solution reads
\bea\label{herman}
r-s \ = \ 1/y~,~~~~ -2s \ttau_3 \ = \ \diff \psi + \phi~,
\eea
and
\bea
y^2(r^2-s^2) \ = \  \ex^w  V (1+|z|^2)^2~,~~~~\frac{r^2-s^2}{\Omega(r)} \ = \ y^2 V~.
\eea
Notice immediately that at the NUT $r=s$ we have $1/y=0$, so that this solution must have $b_1=-b_2$ -- 
we shall find this explicitly below. The metric $(\ttau^2_1+\ttau^2_2)$  is  diffeomorphic to the Fubini-Study metric  on $\mathbb{CP}^1 \cong S^2$:
\bea
\ttau^2_1+\ttau^2_2&=& \frac{4\diff z\diff \bar z}{(1+|z|^2)^2}~.
\eea 
The metric functions then simplify to
\bea
V(y) &=& \frac{1+2sy}{1+4sy+y^2}~, \qquad w(y,z,\bar{z}) \ = \ \log\frac{1+4sy+y^2}{(1+|z|^2)^2}~,
\eea
and it is straightforward to check these satisfy the defining equation (\ref{V}) and Toda equation (\ref{Toda}). The conformally related scalar-flat K\"ahler metric is
\bea
\diff s^2_{\mathrm{Kahler}} &=& \frac{1+2sy}{1+4sy+y^2}\diff y^2 + (1+2sy)(\ttau_1^2+\ttau_2^2) + \frac{4s^2(1+4sy+y^2)}{1+2sy}\ttau_3^2~,
\eea
with K\"ahler form
\bea
\omega &=& -\diff y \wedge 2s\ttau_3 + (1+2sy)\ttau_1\wedge \ttau_2 \ = \ -\diff\left[(1+2sy)\ttau_3\right]~.
\eea
Using the formula (\ref{A}) for the gauge field $A$, we compute
\bea
A &=& \frac{1}{2}(4s^2-1)\frac{r-s}{r+s}\ttau_3 + \mbox{pure gauge}~,
\label{hello}
\eea
which we see reproduces the 1/4 BPS choice of instanton in section 3.3 of \cite{Martelli:2012sz}.\footnote{Notice that in  \cite{Martelli:2012sz} the opposite orientation convention was 
chosen, so that that instanton in  \cite{Martelli:2012sz} is self-dual, rather than anti-self-dual. 
Recall also from the discussion above equation (\ref{A}) that the overall sign of the instanton 
is correlated with the sign of the supersymmetric Killing vector $K$. Here $K=-\frac{1}{2s}\partial_\TNnu$, 
which is minus the expression in \cite{Martelli:2012sz}, hence leading to the opposite sign for the 
instanton gauge field $A$.}
The supersymmetric Killing vector is $K=\de_\psi=-\frac{1}{2s}\partial_\TNnu$ and so 
 generates the Hopf fibration of $S^3$. Since $\TNnu=\varphi_1-\varphi_2$, $\varphi=\varphi_1+\varphi_2$ we 
 hence find
\bea
b_1 & = & -b_2  \ = \ -\frac{1}{4s}\ ,
\eea
which using \eqref{freetoric} yields
\bea
I_{\mathrm{1/4\, BPS}} &=& \frac{\pi}{2G_4}.
\eea
This formula matches the result of section 5.4 of \cite{Martelli:2012sz}.

\paragraph{1/2 BPS solution:}The Taub-NUT-AdS metric  (\ref{TNAdSmetric}) also admits a 1/2 BPS solution 
\cite{Martelli:2011fw, Martelli:2012sz}. We hence have two linearly independent Killing spinors, 
which may be parametrized by an arbitrary choice of constant two-component spinor 
$\chi_{(0)}=\left(\begin{array}{c}\mathsf{p}\\ \mathsf{q}\end{array}\right)\in \C^2\setminus\{0\}$.\footnote{The full Killing spinor 
is given by substituting this into the right hand side of (2.29) of \cite{Martelli:2012sz}.} The correspondong 
Killing vector is given by the unlikely expression
\bea
K&=& (2s+\sqrt{4s^2-1})\Big[2 \Imag[\ex^{\ii\varphi}\mathsf{p}\bar{\mathsf{q}}]\partial_\theta+\left(|\mathsf{p}|^2-|\mathsf{q}|^2+2 \Real[\ex^{\ii\varphi}\mathsf{p}\bar{\mathsf{q}}]
\cot\theta\right)\partial_\varphi\Big]\label{Knottoric}\\[2mm]
&+ &\left[(|\mathsf{p}|^2+|\mathsf{q}|^2)\Big(\tfrac{1}{2s}-2s-\sqrt{4s^2-1})\Big)-2 \Real[\ex^{\ii\varphi}\mathsf{p}\bar{\mathsf{q}}] (2s+\sqrt{4s^2-1}) \csc\theta\right]\partial_\TNnu~.\nn
\eea
Since multiplying $\chi_{(0)}$ by a non-zero complex number $\lambda\in\C^*$ simply rescales $K$ by $|\lambda|^2$, this leads 
to a $\mathbb{CP}^1$ family of choices of Killing vector $K$ in this case. Of course, the vector 
(\ref{Knottoric}) is not toric for generic choice of $\chi_{(0)}$. Nevertheless, one can still compute 
the various geometric quantities in section \ref{sec:local}. In particular one can check that the formula 
(\ref{Fddbar}) for the instanton gives
\bea
A &=& s\sqrt{4s^2-1}\frac{r-s}{r+s}\ttau_3 + \mbox{pure gauge}~,
\label{hello2}
\eea
for any choice of $K$ in (\ref{Knottoric}), which agrees with the expression in \cite{Martelli:2011fw, Martelli:2012sz}. 
Notice that the instanton is invariant under the $SU(2)\times U(1)$ symmetry of the metric, 
even though a choice of  Killing vector $K$ breaks this symmetry. Indeed, in this case the 
conformal factor $y=y(r,\theta)$ for toric solutions given by (\ref{yTNUTAdS})  depends non-trivially 
on both $r$ and $\theta$, thus also breaking the $SU(2)$ symmetry of the underlying Taub-NUT-AdS metric. This is to be 
contrasted with the 1/4 BPS solution, where instead (\ref{yTNUTAdS}) reduces simply to $y=y(r)=1/(r-s)$ (see (\ref{herman})).

The toric choices of $K$ for these 1/2 BPS solutions correspond to the poles of the $\mathbb{CP}^1$ parameter space. 
For example, choosing $\mathsf{p}=1$, $\mathsf{q}=0$ above gives
\bea
K & =& \left(2 s + \sqrt{4 s^2 - 1}\right)\de_{\varphi}+\left(\tfrac{1}{2 s} - 2 s - \sqrt{4 s^2 - 1}\right)\de_{\TNnu}~,
\eea
so that
\bea
b_1 & =& \frac{1}{4s}\ ,\qquad b_2\ = \ -\frac{1}{4s}+2s+\sqrt{4s^2-1}\ .
\eea
The free energy \eqref{freetoric} is thus
\bea
I & =& \frac{2\pi s^2}{G_4}~,
\eea
which of course matches the result obtained in section 4.4 of \cite{Martelli:2012sz}.

\subsection{Plebanski-Demianski}
\label{sec:PD}

The Taub-NUT-AdS metric has been extended to a two-parameter family of smooth 
self-dual Einstein metrics on the four-ball in \cite{Martelli:2013aqa}, which lie in the Plebanski-Demianski class of local 
solutions \cite{PDpaper} to Einstein-Maxwell theory. We will henceforth refer to the solution of \cite{Martelli:2013aqa} as ``Plebanski-Demianski''.
The metric may be written as
\be\label{PD}
\diff s_\mathrm{PD}^2\ =\ \frac{\mathcal P(q)}{q^2-p^2} (\dd \tau + p^2 \dd \sigma)^2  - 
\frac{\mathcal P(p)}{q^2-p^2} (\dd \tau +q^2 \dd \sigma )^2 + \frac{q^2-p^2}{\mathcal P(q)} \dd q^2 - \frac{q^2-p^2}{\mathcal P(p)}\dd p^2,
\ee
where
\bea
\mathcal{P}(x) & =& (x-p_1)(x-p_2)(x-p_3)(x-p_4)~.
\eea
 The roots of the quartic $\mathcal{P}(x)$ can be expressed in terms of the two parameters of the solution, $\apd$ and $v$, as
\bea\nonumber
p_1&=&-\frac12-\sqrt{1+\apd^2-v^2}\ ,\quad p_3=\frac12-\apd\ ,\\
p_2&=&-\frac12+\sqrt{1+\apd^2-v^2}\ ,  \quad p_4=\frac12+\apd\ .
\eea
The coordinate $p\in [p_3,p_4]$ is essentially a polar angle variable, while $q\in [p_4,\infty)$ plays the role of 
a radial coordinate, with the conformal boundary being at $q=\infty$. The NUT/origin of $\R^4$ is located 
at $p=p_3$, $q=p_4$. 
The Killing vectors $\partial_\tau$, 
$\partial_\sigma$ generate the $U(1)^2$ torus symmetry of the solution, with the coordinates 
related to our standard $2\pi$-period coordinates $\varphi_1$, $\varphi_2$ on $U(1)^2$ via
\bea
\tau &=& \frac{2p_3^2}{\mathcal{P}'(p_3)}\varphi_1-\frac{2p_4^2}{\mathcal{P}'(p_4)}\varphi_2~,\nn\\
\sigma &=& -\frac{2}{\mathcal{P}'(p_3)}\varphi_1+\frac{2}{\mathcal{P}'(p_4)}\varphi_2~.
\eea
In order that the metric is smooth on the four-ball the parameters must obey $v^2>2|\apd|$, with the 
$\apd=0$ limit being the Taub-NUT-AdS metric of the previous section, and further setting $v=1$ one recovers Euclidean AdS$_4$ (we refer the reader to 
\cite{Martelli:2013aqa} for further details). 

It is straightforward, but tedious, to express the metric (\ref{PD}) in the form  \eqref{SDE}, with an 
arbitrary choice of toric Killing vector $K=b_1\partial_{\varphi_1}+b_2\partial_{\varphi_2}$. For the special case
of the Killing vector/instanton in the solution of 
\cite{Martelli:2013aqa}, we work out the change of coordinates 
explicitly towards the end of section \ref{morelabels}, \emph{c.f.} equations (\ref{trans2}), (\ref{upq1}) -- (\ref{upq3}).

In the $(\tau,\sigma)$ coordinates an arbitrary  Killing vector may be written as
\bea
K & = & \btau \partial_\tau+ \bsigma  \partial_\sigma\ ,\eea
where
\bea
 \btau & =& \frac{2 p_3^2}{\mathcal P^{'}(p_3)} b_1- \frac{2 p_4^2}{\mathcal P^{'}(p_4)} b_2\ , \qquad \bsigma \ = \ -\frac{2 }{\mathcal P^{'}(p_3)} b_1+\frac{2 }{\mathcal P^{'}(p_4)} b_2\ .
\eea
Using \eqref{twistor} and \eqref{ydefinition} one can calculate 
\bea\label{conf1y2}
\frac{1}{y(p,q)^2} &=&  \frac{1}{4} \frac1{(q^2-p^2)^2} \Bigg\{\Bigg[ \left( \frac{2 \mathcal P(q) }{q-p}-\mathcal P'(q)\right) (\btau+ \bsigma p^2) \nn\\
&&- \left(\frac{2\mathcal P(p)}{q-p}+\mathcal 
P'(p)\right) (\btau +\bsigma q^2)\Bigg]^2
-4 \bsigma^2 \mathcal P(q) \mathcal P(p) (q+p)^2\Bigg\}~.
\eea
Notice that this is a sum of two non-negative terms. Furthermore, these terms may vanish only when evaluated at the roots $p=p_3$, $p=p_4$ or $q=p_4$, 
which correspond to the axes of $\R^4=\R^2\oplus \R^2$.  Let us calculate these limits:
\bea\nn
\lim_{p\to p_3}\frac{1}{y^2}&=& \left(\frac{(b_1+b_2) v^2+2 \apd b_1 + b_2 (2 q-1)}{v^2+2\apd}\right)^2~,\\
\lim_{p\to p_4}\frac{1}{y^2}&=& \left(\frac{(b_1+b_2) v^2-2\apd b_2  + b_1 (2 q-1)}{v^2-2 \apd}\right)^2~,\\\nn
\lim_{q\to p_4}\frac{1}{y^2}&=& \left(\frac{(b_1+b_2) v^2-2 \apd b_2 + b_1 (2 p-1)}{v^2-2 \apd}\right)^2~.
\eea
A careful analysis of the above limits shows that $1/y$ does not vanish, and hence the metric is regular, whenever $b_1/b_2>0$, while 
$1/y=0$ only at the NUT when $b_1/b_2=-1$. 
 On the other hand, the the solution is indeed singular if $b_1/b_2<0$ and $b_1/b_2\neq -1$.  Notice that we also 
easily recover the formula (\ref{yNUT})
for the conformal factor at the NUT: $\lim_{p\to p_3,\ q\to p_4}y=1/|b_1+b_2|$.

In \cite{Martelli:2013aqa} particular supersymmetric instantons (particular choices of $b_1/b_2$ for fixed $\apd$ and $v$) were studied for this two-parameter family of metrics, 
which by construction lie within the Plebanski-Demianski ansatz. The results of this paper extend these
results to a general choice of instanton on the same background, parametrized by $b_1/b_2$, leading to a \emph{three-parameter} family 
of regular supersymmetric solutions. The general expression for this instanton is lengthy, but computable, 
and the interested reader may find the details in appendix \ref{thegeneral}.

\subsection{Infinite parameter generalization}
\label{infinite}

In each subsection we have generalized the metrics of the previous subsection by adding 
a parameter, and one might wonder whether one can find more general self-dual Einstein metrics 
on the four-ball. In fact from the gauge-gravity point of view it is more natural 
to ask the question of which conformal structures on $S^3$ may be filled by
a self-dual Einstein metric. Of course one expects this problem to be overdetermined, 
and some general results in this direction appear in \cite{biquard}. Roughly 
speaking, as long as the conformal class of the boundary metric $[g_{S^3}]$ 
is sufficiently close to the round metric $[g_{S^3}^0]$, then one can write
$[g_{S^3}]=[g_{S^3}^0]+[g_{S^3}^+]+[g_{S^3}^-]$, where 
$[g_{S^3}^0]+[g_{S^3}^\pm]$ bound self-dual/anti-self-dual Einstein metrics on the 
four-ball $B^4$, respectively. Equivalently, viewed as self-dual fillings these induce opposite orientations on $S^3$. 
This may be regarded as a generalization of the well-known result of Fefferman-Graham \cite{FG} to 
the self-dual case. Another important general result is that these
fillings are \emph{unique}: that is, two self-dual Einstein four-manifolds 
$(M_4^{(1)},g^{(1)})$, $(M_4^{(2)},g^{(2)})$ inducing the same conformal 
structure on $M_3=\partial M_4$ are isometric \cite{Andersonunique}.

However, starting with a particular (conformal) three-metric and trying to construct 
a global filling explicitly is likely to be very difficult. In order to construct further 
explicit examples one might instead attempt to directly generalize the Plebanski-Demianski
metrics of the previous subsection. A natural way to do this is explained in more detail in 
appendix \ref{sec:construct}. Specifically, in \cite{CP} the authors studied the general \emph{local} geometry of toric 
self-dual Einstein metrics, which thus includes all the solutions (locally) above. 
In appropriate coordinates\footnote{Below, and in appendix \ref{sec:construct}, $\eta$ is a coordinate. We hope that 
no confusion arises between this,  the almost contact form on $M_3$, and the $\eta$ invariant. The latter uses will not 
appear in the remainder of the paper.}
the metric takes the form
\bea\label{toricmetricmain}
\diff s^2_\mathrm{toric} &=& \frac{4\rho^2(\func_\rho^2+\func_\eta^2)-\func^2}{4\func^2}\diff s^2_{\mathcal{H}^2}
+ \frac{4}{\func^2(4\rho^2(\func_\rho^2+\func_\eta^2)-\func^2)}\Big[\big(y^{\text{can}}_\rho\diff\spsi \nn \\
&&+ (\eta y^{\text{can}}_\rho - \rho y^{\text{can}}_\eta)\diff\varphi\big)^2
 +\left(y^{\text{can}}_\eta\diff\spsi +(\rho y^{\text{can}}_\rho + \eta y^{\text{can}}_\eta - y^{\text{can}})\diff\varphi\right)^2\Big]~.
\eea
where we have defined
\bea\label{ycanmain}
y^{\text{can}}(\rho,\eta) &\equiv & \sqrt{\rho}\func(\rho,\eta)~,
\eea
and 
\bea\label{H2metricmain}
\diff s^2_{\mathcal{H}^2} &=& \frac{\dd\rho^2+\dd\eta^2}{\rho^2}
\eea
is the metric on hyperbolic two-space $\mathcal{H}^2$, regarded as the upper half plane with boundary at $\rho=0$. 
The metric (\ref{toricmetricmain}) is entirely determined by the choice of function $\func=\func(\rho,\eta)$, and the metric 
is self-dual Einstein if and only if this solves the eigenfunction equation 
\bea\label{Laplacemain}
\Delta_{\mathcal{H}^2} \func &=& \frac{3}{4}\func \qquad \Longleftrightarrow  \qquad \func_{\rho\rho}+\func_{\eta\eta} \ = \ \frac{3}{4\rho^2}\func~,
\eea
where $\func_\rho \equiv \de_\rho \func$, {\it etc}. Unlike the Toda equation (\ref{Toda}) this is linear, and 
one may add solutions.  In particular there is a basic solution
\bea\label{basicpolemain}
\mathcal{F}(\rho,\eta; \lambda) &=& \frac{\sqrt{\rho^2+(\eta-\lambda)^2}}{\sqrt{\rho}}~,
\eea
where $\lambda$ is  any constant. Via linearity 
then
\bea\label{mpolemain}
\mathcal{F}(\rho,\eta) &=& \sum_{i=1}^m \alpha_i \mathcal{F}(\rho,\eta;\lambda_i)~,
\eea
also solves (\ref{Laplacemain}), for arbitrary constants $\alpha_i,\lambda_i$, $i=1,\ldots, m$.
 We refer to (\ref{mpolemain}) as an \emph{$m$-pole solution}.
 Of course, one could also replace the sum in (\ref{mpolemain}) by an integral, 
 smearing the monopoles in some chosen charge distribution. 

Thus the \emph{local} construction of toric self-dual Einstein metrics is very straightforward -- 
the above gives an infinite-dimensional space. However, understanding when 
the above metrics extend to complete asymptotically locally hyperbolic metrics on a ball
(or indeed any other topology for $M_4$) is more involved. In appendix \ref{sec:construct} 
we take some steps in this direction by showing that the general 
2-pole solution is simply (Euclidean) AdS$_4$, while the general 3-pole solution 
is precisely the Plebanski-Demianski solutions of section \ref{sec:PD}. This requires 
taking into account the symmetries of (\ref{toricmetricmain}) (in particular the 
$PSL(2,\R)$ symmetry of $\mathcal{H}^2$), and then making a number of 
rather non-trivial coordinate transformations. We also analyse in detail the global structure 
of Euclidean AdS$_4$ in the $(\rho,\eta)$ coordinates, together with 
some global properties of the Plebanski-Demianski solutions 
in the $(\rho,\eta)$ coordinates. 

Some work has also been done on global properties of the metrics (\ref{toricmetricmain}) in \cite{CS}, 
although the focus in that paper is on constructing complete asymptotically locally Euclidean 
scalar-flat K\"ahler metrics, which are conformal to (\ref{toricmetricmain}). However, these 
have non-trivial Lens space boundaries $S^3/\Gamma$, and correspondingly
the second Betti number $b_2=\mathrm{dim}\, H_2(M_4,\R)$ of the filling $M_4$ is non-zero (they contain ``bolt $S^2$s'').
The corresponding complete self-dual Einstein metrics in Theorem B of that paper 
then also do not have the topology of the ball. Thus it remains an interesting open problem 
to understand when the general $m$-pole metrics extend to complete metrics on the ball.\footnote{At 
the end of reference \cite{CP} it is briefly noted that one can obtain regular $m$-pole metrics 
by deforming, for example, a given 3-pole solution. It would be interesting to examine the details 
of this deformation argument further.} 

Finally, let us remark that in \cite{lebrun} Lebrun has constructed 
infinitely many self-dual Einstein metrics on the four-ball using 
twistor methods. This is essentially a deformation argument, where 
one starts with (the twistor space of) Euclidean AdS$_4$, and perturbs the twistor space.
However, as such this is rather more implicit than the toric metrics above, and in order 
to construct supersymmetric solutions one needs to ensure that the resulting 
self-dual Einstein metric has at least one Killing vector field. Nevertheless, this 
might be an alternative method for analysing regularity of the above $m$-pole solutions, 
at least in a neighbourhood of Euclidean AdS$_4$ in parameter space.


\section{Conclusions}
\label{sec:conclusions}

The main result of this paper is the proof of the formula 
(\ref{freetoric}) for the holographically renormalized on-shell action in minimal four-dimensional  supergravity. 
This result is analogous to the general formula for the volume functional of a toric Sasakian manifold in 
\cite{Martelli:2005tp}. Indeed, the latter was also entirely determined by the Reeb vector field 
of the corresponding Sasakian manifold, and was later shown to agree with the large $N$ limit of the 
trial $a$ function in a dual four-dimensional field theory \cite{Butti:2005vn}.\footnote{A similar general 
result, valid for the trial free energy of a three-dimensional field theory with AdS$_4$ dual, 
was conjectured in \cite{Martelli:2011qj}.}
Moreover, we have provided a general construction of regular
supersymmetric solutions of this theory\footnote{Of course, these uplift to solutions of eleven-dimensional supergravity using the results of \cite{Gauntlett:2007ma}.}, 
based on self-dual Einstein metrics on the four-ball equipped with 
a one-parameter family of instanton fields for the graviphoton. Specifically, if the self-dual Eintein metric
admits $n$ parameters, our constuction produces an $(n+1)$-parameter family of solutions. We have shown that the 
renormalized on-shell action does \emph{not} depend on the $n$ metric parameters, but only on this last 
``instanton parameter''. This matches beautifully the field theory results of \cite{Alday:2013lba}.

We have also shown how all the previous examples in the literature, as well as some new examples that we have presented, can 
be understood as arising from an infinite-dimensional family of local self-dual Einstein metrics with torus symmetry \cite{CP}.
In section \ref{infinite} we have suggested that using this family of local metrics, it should be possible to construct global  
asymptotically locally (Euclidean) AdS self-dual Einstein metrics on the four-ball, thus obtaining an infinite family of 
completely explicit metrics. It will be interesting to analyse these $m$-pole solutions in more detail.
 
In this paper we have achieved a rather general understanding of the gauge/gravity duality for supersymmetric
asymptotically locally Euclidean AdS$_4$ solutions. Nevertheless, there are a number of possible extensions of our work.
First, it is possible to extend the matching of the free energy (\ref{freetoric}) for the class of self-dual 
backgrounds we have considered to other BPS observables. In particular in \cite{wilson} the Wilson loop 
around an orbit of the  Killing vector $K$ is shown to be BPS in the field theory, and may also be computed
via localization. The gravity dual is an M2-brane wrapping a calibrated copy of the M-theory circle
in the internal space \cite{Farquet:2013cwa}, and computing its renormalized action one finds an analogously 
simple formula to (\ref{freetoric}), namely
\bea\label{Wvev}
\lim_{N\rightarrow\infty}\log\, \langle W\rangle &=& \frac{|b_1|+|b_2|}{2}\ell \cdot \log\, \langle W\rangle_1~,
\eea
where $\langle W\rangle_1$ denotes the large $N$ limit of the Wilson loop on the round sphere/AdS$_4$, whose log scales 
as $N^{1/2}$, and $2\pi\ell$ denotes the length of the orbit of $K$ (for example, such orbits always close 
over the poles of the $S^3$, where $\ell=1/|b_1|$ or $\ell=1/|b_2|$, respectively; notice that for these Wilson 
loops (\ref{Wvev}) is again a function only of $|b_1/b_2|$). Details of this computation 
are given in \cite{Farquet:2014bda}.

One might further generalize our results by  relaxing one or more of the assumptions we have made. For example, remaining in the context of minimal gauged supergravity, 
it would be very interesting to investigate the more general class of supersymmetric, but non-(anti-)self-dual 
solutions \cite{Dunajski:2010uv}. Several examples of such solutions were constructed in 
\cite{Martelli:2011fw,Martelli:2012sz}, and  these all turn out to have a bulk topology different from the four-ball. 
This suggests that self-duality  and the topology of supersymmetric asymptotically AdS$_4$ 
solutions are two related issues, and it would be desirable to 
clarify this. On the other hand, at present it is unclear to us what is the precise dual field theory implication of 
non-trivial two-cycles in the geometry, and therefore this direction is both challenging and interesting.
Perhaps related to this, 
one of our main results is that a smooth toric self-dual Einstein metric on the four-ball with supersymmetric Killing 
vector $K=b_1\partial_{\varphi_1}+b_2\partial_{\varphi_2}$  gives rise to a smooth supersymmetric
solution only if $b_1/b_2>0$ or $b_1/b_2=-1$. Specifically, for other choices of $b_1/b_2$ the conformal factor/Killing spinor are singular 
in the interior of the bulk. Nevertheless, the conformal boundary is smooth for all choices of $b_1, b_2$, and 
the question arises as to how to fill those boundaries smoothly within gauged supergravity. A natural 
conjecture is that these are filled with the non-self-dual solutions mentioned above.
 
 Another assumption that should be straightforward to relax is in taking the gauge field $A$ to be real. In general, if $A$ is complex
 the existence of one (Euclidean) Killing spinor does not imply that the metric possesses any isometry \cite{Dunajski:2010uv}. 
 However, we expect that if one requires the existence of \emph{two} spinors of opposite R-charge, then there will be 
 canonically defined Killing vectors,  and therefore 
 it should be possible to analyse the solutions with the techniques of this paper. 

All the above extensions would be important conceptually, in order to address the issue of uniqueness of the
filling of a given conformal boundary geometry. In fact, this could also motivate the study of this problem in 
a more general consistent truncation, or directly in eleven-dimensional supergravity. 

Of course, in any of these more general set-ups a central issue will be to prove a generalized version of the formula
(\ref{freetoric}) for the renormalized on-shell action. In this respect, some of the methods that we employed to derive this
may be more amenable to generalization than others. For example,  the expression (\ref{freeindexform}), given in terms of boundary conformal 
invariants and bulk topological invariants, might extend to the class of non-self-dual metrics and/or non-ball topology.
We also expect that some of the results of the present paper can be adapted to dimensions different from four. In particular, on the one hand
it would be very nice to understand better the structure of the holographically renormalized on-shell action in five dimensions, 
and on the other hand, 
to  enlarge the list of examples, extending the work of \cite{Cassani:2014zwa}.

\subsection*{Acknowledgments}
\noindent  D.~F. is supported by the Berrow Foundation. 
D.~M. and J.~L. are  supported by the ERC Starting Grant N. 304806, ``The Gauge/Gravity Duality and Geometry in String Theory''.
D.~M. also acknowledges partial support from the STFC grant ST/J002798/1. J.~F.~S. is supported by a Royal Society University Research Fellowship. 

\appendix

\section{Spin connection of the K\"ahler metric}
\label{spinconnec}

For the K\"ahler metric \eqref{Kahler} in the frame \eqref{Kahlerframe} the spin connection reads
\bea
\hat\omega^{01}&=&-\frac{ \big( \de_y w + y \de^2_y w\big)}{4 V^{3/2}}\hat e^1+\frac{\ii   y \de_y\left( \de_z -  \de_{\bar z} \right)w}{8 V^{3/2} \ex^{w/2}}\hat e^2-\frac{  y  \de_y \left( \de_z + \de_{\bar z} \right)w}{8 V^{3/2}\ex^{w/2}}\hat e^3~,\nn\\
\hat  \omega^{02}&=&-\frac{ y \de_y(\de_z +\de_{\bar z} )w  }{8 V^{3/2}\ex^{w/2}}\hat e^0+\frac{\ii y \de_y\left( \de_z - \de_{\bar z} \right)w}{8 V^{3/2}\ex^{w/2}}\hat e^1+\frac{ \big( \de_y w + y \de^2_y w\big)-2V \de_ y w}{4 V^{3/2}}\hat e^2~,\nn\\
\hat  	\omega^{03}&=&- \frac{\ii y \de_y(\de_z -\de_{\bar z} )w }{8 V^{3/2}\ex^{w/2}}\hat e^0-\frac{  y \de_y\left( \de_z + \de_{\bar z} \right)w}{8 V^{3/2}\ex^{w/2}}\hat e^1+\frac{  \big( \de_y w + y \de^2_y w\big)-2 V \de_y w}{4 V^{3/2}}\hat e^3~,\nn\\
\hat	\omega^{12}&=&-\hat\omega^{03}~,\nn\\
  \hat \omega^{13}&=& \hat\omega^{02}~, \nn\\
  \hat		\omega^{23}&=& -\frac{  \left(\de_y w - y (\de_y w)^2   -y \de_y^2 w \right)}{4 V^{3/2}}\hat e^1+\frac{\ii  \left[2V (\de_z - \de_{\bar z})w- y \de_y(\de_z -\de_{\bar z} )w\right]}{8 V^{3/2}\ex^{w/2}}\hat e^2\nn\\
  		&&-\frac{ 2V (\de_z + \de_{\bar z}) w- y \de_y(\de_z +\de_{\bar z} )w}{8 V^{3/2}\ex^{w/2}}\hat e^3~.
\eea
Here we have used both \eqref{V} and \eqref{dphi}.

\section{Weyl transformations of the boundary}
\label{sec:Weyl}

 In section \ref{sec:global} of the main text we studied the boundary geometry and Killing spinor equation 
using the radial coordinate $r=1/y$ defined naturally by supersymmetry. This 
gives a preferred representative for the conformal class of the boundary metric on $M_3$. 
In this appendix we study the more general choice $r=1/(\Omega y)$, where 
$\Omega=\Omega(z,\bar{z})$ is an arbitrary smooth, basic, nowhere zero function on $M_3$. This 
results in a Weyl transformation of the boundary geometry and corresponding Killing spinor equation.
We will see that we precisely recover the boundary structure, derived from a purely 
three-dimensional perspective, in \cite{Closset:2012ru, Alday:2013lba}. 

For comparison with \cite{Alday:2013lba}, we begin by rescaling the constant-norm K\"ahler spinor $\zeta $ as
\be\label{RescaledSpinor}
\zeta \ \equiv \ \Omega^{-1/2}(z,\bar z) \ \hat \zeta ~,
\ee
so that the norm of $\hat \zeta$ is $\Omega^{1/2}$ if we normalize $\zeta$ to have unit norm. We then also have a rescaling of the four-dimensional Killing spinor $\epsilon$,
\be
\hat \epsilon \ \equiv \ \Omega^{1/2}  \epsilon  \ =  \ \frac{1}{\sqrt{2y}}\left( 1+ V^{-1/2} \hat \Gamma_0 \right) \hat \zeta~.
\ee
Recall $\epsilon$ solves the Killing spinor equation \eqref{KSE}, with the gauge field $A_\mu$ given by \eqref{A}. Using instead $\hat \epsilon$ this Killing spinor equation reads
\bea
\left( \nabla_\mu - \ii A_\mu -\frac12 \de_\mu \log \Omega + \frac{1}{2} \Gamma_\mu  + \frac{\ii}{4} F_{\nu\rho} \Gamma^{\nu\rho} \Gamma_\mu \right) \hat \epsilon &=& 0~,
\eea
where the third term appears due to the rescaling.\footnote{As this term is a total derivative it can formally be absorbed into a \emph{complex} gauge transformation of $A_\mu$, although 
as we shall see all gauge fields will in the end be real.}

With the new choice of radial coordinate the boundary metric is
\be
\dd s_{M_3}^2 \ =\ \Omega^2(z,\bar z ) \left[(\dd \psi+\phi_0)^2 + 4 \ex^{w_{(0)}} \dd z \dd \bar z \right]~.
\ee
As always, we introduce an orthonormal frame for this metric:
\be\label{fra1again}
e^1_{(3)} \ =\  \Omega (\dd \psi + \phi_0)\ , \hspace{12 mm} e_{(3)}^2+\ii e_{(3)}^3\  =\  2 \Omega  \ex^{w_{(0)}/2} \dd z\ .  
\ee
The four-dimensional geometry is the same as before, namely
\be\label{SDEapp}
\dd s_{\mathrm{SDE}}^2 \ =\  \frac1{y^2} \left[V^{-1} (\dd \psi+\phi )^2 + V(\dd y^2 + 4\ex^w \dd z \dd \bar z )\right]~,
\ee
and we will use the frame
\bea\label{SDEframeapp}
{e}^0 &=& \frac{1}{y}V^{1/2}\diff y~, \quad {e}^1 \ = \ \frac{1}{y}V^{-1/2}(\diff\psi+\phi)~, \quad {e}^2 + \ii {e}^3 \ = \ \frac{2}{y}(V\ex^w)^{1/2}\diff z~.
\eea
Calculating the spin connection of \eqref{SDEframeapp}, expanding in $y$ and comparing to the spin connection of \eqref{fra1again}, we find
\bea 
\omega^{12} &=& \omega_{(3)}^{12}-    \de_2 \log \Omega \, e^1_{(3)}+ \mathcal O(y)~,\nonumber\\
\omega^{13} &=& \omega_{(3)}^{13}-  \de_3 \log \Omega \, e_{(3)}^1+\mathcal O (y)~, \nonumber\\
\omega^{23} &=&\omega^{23}_{(3)} -   \de_3 \log \Omega \, e^2_{(3)}+ \de_2 \log \Omega \, e^3_{(3)} +\mathcal O(y)~,\nonumber\\
\omega^{0i} &=& \frac{1}{y}\Omega^{-1}(1+\tfrac14 y w_{(1)}) \, e^i_{(3)}+\mathcal O(y)~,
\eea
with $i=1,2,3$. 

We next expand the Killing spinor equation with the rescaled spinor, $\hat \epsilon$. As in section \ref{sec:global} the term $\frac{\ii}{4} F_{\nu\rho} \Gamma^{\nu\rho} \Gamma_\mu = \mathcal O(y)$ does not contribute. One gets
\bea\label{KSEconfapp}
&&\left[ \nabla^{(3)}_\mu - \ii A_{(0)\mu} -\frac12 \de_\mu \log \Omega + \frac{1}{2 y} \Omega^{-1} \left(1+\tfrac14 y w_{(1)}\right) e^i_{(3)\mu}(\Gamma_i- \Gamma_{i0}) \right. \\ \nn
&&-\left.\frac{1}{2}     \de_2 \log \Omega e^i_{(3)\mu}\Gamma_{i2}-\frac{1}{2}   \de_3 \log \Omega e_{(3)\mu}^i\Gamma_{i3} +\mathcal O(y)  \right]\hat \epsilon\ =\ 0 \ ,
\eea
where $\mu =\psi,z,\bar z$, and $A_{(0)\mu}$ is the lowest order expansion of the gauge field \eqref{A}, which in the frame \eqref{fra1again} reads
\be\label{A0again}
4 \, A_{(0)} \ = \ - \Omega^{-1}w_{(1)}\, e_{(3)}^1+ \de_3 w_{(0)} \, e_{(3)}^2- \de_2 w_{(0)} \,  e_{(3)}^3~.
\ee
The Killing spinor $\hat \epsilon$ expands as
\bea
\hat \epsilon & =& \frac{1}{\sqrt{2 y}} \Big[1+\Gamma_0+\frac14 y w_{(1)} \Gamma_0 +\mathcal O(y^{2}) \Big] \hat \zeta_0~,
\eea
and when substituted into \eqref{KSEconfapp} gives a vanishing leading order term. The subleading term reads
\bea
&&\left[ \left(\nabla^{(3)}_i - \ii  A_{(0)i}-\frac12 \de_i \log \Omega \right) (1+\Gamma_0) - \frac{1}{8} w_{(1)} \Omega^{-1}( \Gamma_{i}-\Gamma_{i0})\right. \nn\\
&&\left.-\frac{1}{2}\de_2 \log \Omega \Gamma_{i2}(1+\Gamma_0) -\frac{1}{2}  \de_3 \log \Omega \Gamma_{i3} (1+\Gamma_0)  \right] \hat \zeta_0\ =\ 0\ . \label{contracting}
\eea
The projection conditions \eqref{projections} imply the following form for $\hat \zeta_0$,
\bea\label{zeta0again}
\hat \zeta_0 & =& \begin{pmatrix}\hat \chi\\ 0\end{pmatrix}\quad\text{where}\quad \hat \chi \ = \ \begin{pmatrix}\hat\chi_0\\ \hat \chi_0\end{pmatrix}~.
\eea
The three-dimensional Killing spinor equation then becomes
\be\label{3dKSEapp}
\left[ \nabla^{(3)}_i + \ii (V_i- A^{(3)}_i) +\frac{1}2 H\sigma_i +\frac12 \epsilon_{ijk} V_j \sigma_k	\right]\hat \chi\  =\ 0\ ,
\ee
with 
\bea
H&=&-\frac{\ii}4  w_{(1)} \Omega^{-1} +\ii V_1~, \qquad
A^{(3)}_1 \ =\   A_{(0)1} +\frac32 V_1~,\nn\\
A^{(3)}_2&=& A_{(0)2} -\frac{3}2 \ii V_3 -\frac32 \ii \de_2 \log \Omega+ \frac32 \de_3\log\Omega~,\nn\\
A^{(3)}_3&=& A_{(0)3}+\frac32 V_3 ~,\nn\\
V_2+\ii V_3&=&- \ii \de_2 \log \Omega + \de_3 \log \Omega\  . \label{3dfields}
\eea

The Killing spinor equation (\ref{3dKSEapp}) is precisely of the form found in \cite{Closset:2012ru}, which allows for the construction of supersymmetric field theories on $M_3$.
The identifications of $A^{(3)}$, $V$ and $H$ are not unique because equation \eqref{3dKSEapp} has some symmetry properties\footnote{With an abuse of language, in this paper
we refer to this symmetry as a ``gauge'' symmetry. Although $V$ is not a gauge field, and hence does not transform under gauge transformations. Hopefully 
this will not cause any confusion.}, 
{\it c.f.} (4.2) of \cite{Closset:2012ru}. In particular this gauge freedom allows one to freely choose $V_1$, as shown in (2.10) of \cite{Alday:2013lba}. 
Recall that $A_{(0)}$ is real. If we demand also the boundary gauge field $A^{(3)}$ to be real, one finds from the equations in \eqref{3dfields} that also $V$ is real with
\be\label{BndrV}
V_2 \ =\ \de_3 \log \Omega~, \qquad V_3\ =\ - \de_2 \log \Omega~.
\ee
This is exactly the result obtained for $V$ in \cite{Alday:2013lba} using the purely three-dimensional 
analysis of \cite{Closset:2012ru}. The remaining equations in \eqref{3dfields} then further simplify to
\bea
H&=&-\frac{\ii}4  w_{(1)} \Omega^{-1} +\ii V_1~,\label{H}\\
A^{(3)}_i &=& A_{(0)i} +\frac32 V_i ~. \label{BndrA}
\eea
Again this is consistent with \cite{Alday:2013lba}, where it was found (in our notation) that
\be\label{Acheck}
A^{(3)}_\mu \ = \ -\frac{\ii}2 H e^1_{(3)\mu} + V_\mu + j_\mu ~,
\ee
where
\be\label{J}
j_\mu \ =\  \frac{\ii}{4\Omega^2} \left(s\de_\mu \bar s - \bar s \de_\mu s\right) + \frac12 \omega_{\mu(3)}^{\:23} ~,
\ee
and $|s|=\Omega$ is the square norm of the three-dimensional spinor,
\be
\hat \chi = \sqrt{s(\psi, z, \bar z)} \left( \begin{array}{c} \frac1 {\sqrt{2}} \\ \frac 1 {\sqrt{2}} \end{array}\right)~.
\ee
Hence we have $s=\Omega \ex^{2\ii \newvarphi(\psi, z,  \bar z)}$. Equation \eqref{J} then reads
\bea\label{jOmega}
j_\mu & =& \de_\mu \newvarphi+ \frac12 \omega_{\mu(3)}^{\:23} \\
& = & \de_\mu \newvarphi -\frac{1}{8} \Omega^{-1}w_{(1)}e^1_{(3)}+\frac{1}{4}   \left(\de_3 w_{(0)}+2  \de_3 \log \Omega  \right)e^2_{(3)}-\frac{1}{4}  \left(\de_2 w_{(0)}+ 2\de_2 \log \Omega  \right)e^3_{(3)} ~,\nn
\eea
where we also used equation \eqref{dphi}. Substituting equation \eqref{BndrV}, \eqref{H},  and \eqref{jOmega} into the right hand side  of \eqref{Acheck}, this gives 
\bea
A^{(3)}_\mu\  &=& \  - \frac14\Omega^{-1}w_{(1)}\, e_{(3)\mu}^1+ \frac14 \de_3 w_{(0)} \, e_{(3)\mu}^2- \frac14\de_2 w_{(0)} \,  e_{(3)\mu}^3~  +\frac32 V_\mu + \de_\mu \newvarphi \nn\\
&=& A_{(0)\mu}  +\frac32 V_\mu + \de_\mu \newvarphi~, \label{Acheckagain}
\eea
where in the second line we used equation \eqref{A0again}. As the last term in equation \eqref{Acheckagain} is a total derivative, it can be absorbed into a gauge transformation of $A_{(0)}$. Thus we see that equation \eqref{Acheck} reproduces \eqref{BndrA} up to a gauge transformation. Indeed, such a gauge transformation with $\newvarphi = \gamma \psi$ was shown in section \ref{sec:A0} to be necessary in order for the gauge field to be globally well-defined 
on $M_3\cong S^3$.

\section{Toric self-dual Einstein metrics on the four-ball}
\label{sec:construct}

In this appendix we indicate how the metrics in section \ref{sec:explicit} may be extended 
to include arbitrarily many parameters, leaving further details of this construction for another occasion.
The local form of these metrics was 
determined in \cite{CP}, and is given in terms of so-called $m$-pole solutions. We discuss in detail the special cases of $m=2$ and  $m=3$, 
showing that they correspond to Euclidean AdS$_4$ and a particular metric discussed in 
\cite{Martelli:2013aqa}. The latter originates from a class of metrics originally studied by Plebanski-Demianski. 
 Below we will also 
provide more details on the general  instantons associated to a given self-dual Einstein metric, and a choice of Killing vector in the $U(1)^2$ torus of isometries. 

\subsection{Local form of the metrics and instanton}

Following \cite{CP}, the local form of a toric self-dual Einstein metric can be written as\footnote{We have reversed the sign of the metric (1.1) in \cite{CP}, so that
for $4\rho^2(\func_\rho^2+\func_\eta^2)-\func^2>0$ our metric (\ref{toricmetric}) has Euclidean signature $(+,+,+,+)$ and  \emph{negative} scalar curvature.\label{signreverse}}
\bea\label{toricmetric}
\diff s^2_\mathrm{toric} &=& \frac{4\rho^2(\func_\rho^2+\func_\eta^2)-\func^2}{4\func^2}\diff s^2_{\mathcal{H}^2}
+ \frac{4}{\func^2(4\rho^2(\func_\rho^2+\func_\eta^2)-\func^2)}\Big[\big(y^{\text{can}}_\rho\diff\spsi \nn \\
&&+ (\eta y^{\text{can}}_\rho - \rho y^{\text{can}}_\eta)\diff\varphi\big)^2
 +\left(y^{\text{can}}_\eta\diff\spsi +(\rho y^{\text{can}}_\rho + \eta y^{\text{can}}_\eta - y^{\text{can}})\diff\varphi\right)^2\Big]~.
\eea
Here we defined
\bea\label{ycan}
y^{\text{can}}(\rho,\eta) &\equiv & \sqrt{\rho}\func(\rho,\eta)~,
\eea
with $\func=\func(\rho,\eta)$ and the superscript ``can'' indicating that this is a canonical choice 
for the function $y$ (see below). We also have that
\bea\label{H2metric}
\diff s^2_{\mathcal{H}^2} &=& \frac{\dd\rho^2+\dd\eta^2}{\rho^2}
\eea
is the metric on hyperbolic two-space $\mathcal{H}^2$, regarded as the upper half plane with boundary at $\rho=0$. 
Even though the metric (\ref{toricmetric}) is local, this global description of $\mathcal{H}^2$ will be important. 
In particular, in the global construction of \cite{CS} the coordinate singularities 
along which Killing vectors vanish are  mapped onto the boundary $\rho=0$ of $\mathcal{H}^2$, and 
we shall see this for the examples that we study below.
The metric (\ref{toricmetric}) is entirely determined by the choice of function $\func(\rho,\eta)$, and the metric 
is self-dual Einstein if and only if this solves the eigenfunction equation 
\bea\label{Laplace}
\Delta_{\mathcal{H}^2} \func &=& \frac{3}{4}\func \qquad \Longleftrightarrow  \qquad \func_{\rho\rho}+\func_{\eta\eta} \ = \ \frac{3}{4\rho^2}\func~,
\eea
where $\func_\rho \equiv \de_\rho \func$, {\it etc}.
Crucially this is a  linear equation, so we may add solutions as in the more familiar ``multi-centre'' types of solutions in other contexts. 

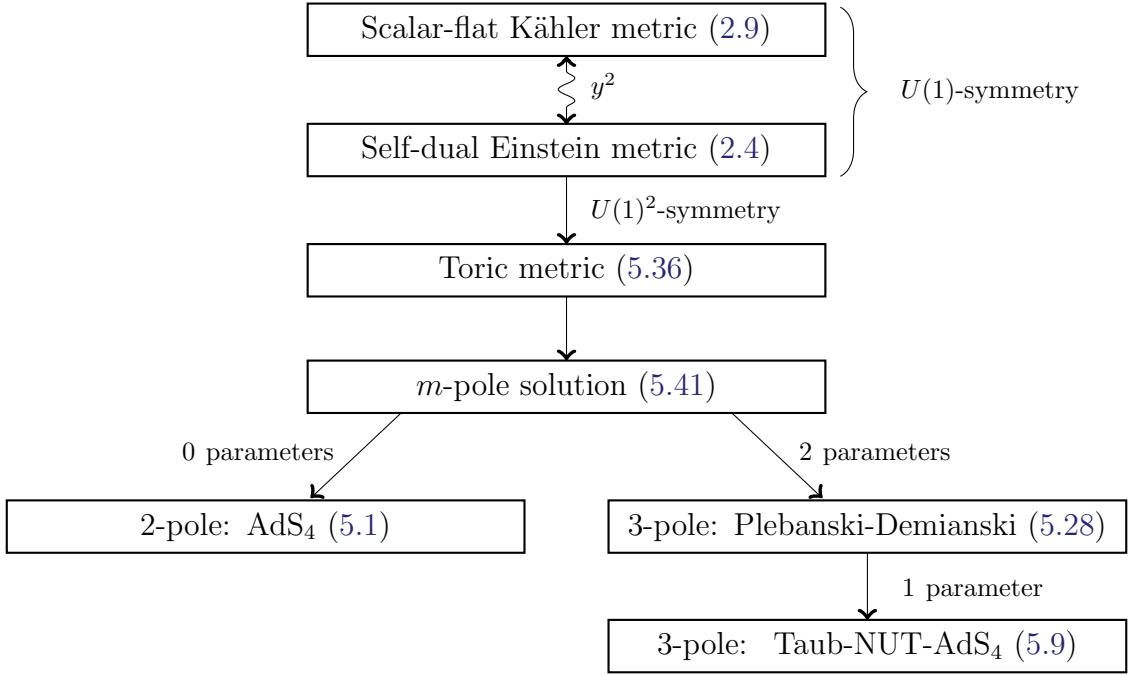
\begin{figure}[t]
\centering
\begin{tikzpicture}
\node[thick,black,draw=black,shape=rectangle,text width=16 em,text centered] at (0,7.3) {Scalar-flat K\"ahler metric \eqref{Kahler}};
\draw[<->,snake=snake,segment amplitude =1.1 mm ,segment length=3 mm,line after snake = 0.9 mm,line before snake=2mm ]  (0,6.95)-- (0,6.05);
\draw[decoration={markings,mark=at position 1 with {\arrow[ultra thick]{>}}},postaction={decorate}]
    (0,6.94) -- (0,6.95);\draw[decoration={markings,mark=at position 1 with {\arrow[ultra thick]{>}}},postaction={decorate}](0,6.05)--(0,6.04);
\node[thick,black,draw=black,shape=rectangle,text width=16 em,text centered] at (0,5.7) {Self-dual Einstein metric \eqref{SDE}};
\draw[decoration={markings,mark=at position 1 with {\arrow[ultra thick]{>}}},postaction={decorate}]
    (0,5.35) -- (0,4.45);
\node[thick,black,draw=black,shape=rectangle,text width=16 em,text centered] at (0,4.1) {Toric metric  \eqref{toricmetricmain}};
\draw[decoration={markings,mark=at position 1 with {\arrow[ultra thick]{>}}},postaction={decorate}]
       (0,3.75)--(0,2.9) ;
\node[thick,black,draw=black,shape=rectangle,text width=16 em,text centered] at (0,2.55) {$m$-pole solution \eqref{mpolemain}};
\node[thick,black,draw=black,shape=rectangle,text width=16 em,text centered] at (4,.69) {3-pole: Plebanski-Demianski \eqref{PD}};
\node[thick,black,draw=black,shape=rectangle,text width=16 em,text centered] at (-4,0.69) {2-pole: AdS$_4$ \eqref{AdS4metric}};
\node[thick,black,draw=black,shape=rectangle,text width=16 em,text centered] at (4,-.9) {3-pole:~ Taub-NUT-AdS$_4$ \eqref{TNAdSmetric}};
\draw[decoration={markings,mark=at position 1 with {\arrow[ultra thick]{>}}},postaction={decorate}]
       (2.2,2.2)--(3.4,1.07) ;
\draw[decoration={markings,mark=at position 1 with {\arrow[ultra thick]{>}}},postaction={decorate}]
       (-2.2,2.2)--(-3.4,1.07) ;
\draw[decoration={markings,mark=at position 1 with {\arrow[ultra thick]{>}}},postaction={decorate}]
       (4,0.34)--(4,-.55) ;
\draw [decorate,decoration={brace,amplitude=10pt,mirror,raise=4pt},yshift=0pt]
(3.5,5.35) -- (3.5,7.6) node [black,midway,right,xshift=0.8cm] {\footnotesize $U(1)$-symmetry};
\node[text width=12 em] at (2.8,4.9) {\footnotesize {$U(1)^2$-symmetry}};
\node[text width=12 em] at (2.8,6.55) {\footnotesize {$y^2$}};
\node[text width=12 em,text centered] at (5.4,-.15) {\footnotesize {$1$ parameter}};
\node[text width=12 em,text centered] at (4.1,1.65) {\footnotesize {$2$ parameters}};
\node[text width=12 em,text centered] at (-4.1,1.65) {\footnotesize {$0$ parameters}};
\end{tikzpicture}
\caption{Overview of the metrics discussed in the main part of the paper and in the present appendix. The arrows point from a metric to a special case of the metric, 
except the wavy arrow which corresponds to a conformal transformation, \emph{i.e.} equation \eqref{Kahler}. }
\label{fig:metrics}
\end{figure}

As discussed in the main part of the paper, any self-dual Einstein metric with a choice of Killing vector gives rise to a conformal
scalar-flat K\"ahler metric, with an associated conformal factor $y$. For the above metric (\ref{toricmetric})
a natural canonical  choice of  Killing vector is $K=\partial_\spsi$, and this  leads to the associated  conformal factor
$y = y^{\text{can}}$ given by (\ref{ycan}).
Depending on global constraints, the Killing vector $\partial_\spsi$ 
may have fixed points, and the associated supersymmetric solution may then be singular.
However, we are also free to pick the supersymmetric Killing vector $K$ to be an arbitrary linear combination 
of Killing vectors 
\be
K \ = \ \aspsi \de_\spsi + \aphi \de_\varphi~,
\label{Kab}
\ee
with real coefficients $\aspsi,\aphi$, giving the following $(\aspsi,\aphi)$-dependent conformal factor 
\be
y (\rho,\eta)\ =\ \frac{\sqrt{\rho}  \func  (\rho,\eta)}{ \sqrt{(\aspsi+\aphi \eta)^2 +\aphi^2 \rho^2}} \ = \ \frac{ y^{\text{can}} (\rho,\eta)}{\sqrt{(\aspsi+\aphi \eta)^2 +\aphi^2 \rho^2}}~.
\label{yab}
\ee
Of course  for $\aspsi=1$ and  $\aphi=0$ one recovers  $y = y^{\text{can}}$. It is simple to check that the conformally rescaled  metric
\be
\diff s^2 \ =\ \frac{\rho \func^2 (\rho,\eta)} {(\aspsi+\aphi \eta)^2 +\aphi^2 \rho^2} \diff s^2_{\text{toric}}
\ee
is K\"ahler and scalar-flat. 

A key feature of this construction will be that the conformal boundary $y^{\text{can}}=0$ will in general only be 
an implicit equation in the $(\rho,\eta)$ coordinates. However, these coordinates  are well-suited for the discussion of regularity of the metric in the interior. The opposite is true in the $y$-coordinates given by supersymmetry in \eqref{SDE}. Before discussing the general family of $m$-pole solutions, and the examples 
$m=2,3$, let us present the general explicit form of the instanton associated to the Killing vector (\ref{Kab}). 
Using the following general formula for the instanton
\be
F\ =\ -\left(\tfrac12 y \dd K^\flat + y^2 K^\flat \wedge J K^\flat\right)^-~,
\ee
where $K^\flat$ denotes the one-form dual to the Killing vector $K$, and $J$ the complex structure (\ref{Jkahleromega}), we  compute
\bea
F &=& \frac{ y \func }{c_{11}\sqrt{\rho } \left(\func ^2-4 \rho ^2 \left(\func_\eta^2+\func_\rho^2\right)\right)^2 }\Bigg[ (g^{13}+g^{24}) \Big[2c_{11}\rho  \Big\{ \func_\rho \left(2 \aphi\rho^2  \func_\eta-\hat \eta\func  \right)\nn\\
&&+2\rho\func_{\rho\rho} \left(2  \aphi \rho^2 \func_\eta  - 2\hat \eta\rho \func_\rho -\hat \eta\func \right)
+2\rho^2\func_{\rho\eta}\left(\aphi\func -2\hat \eta\func_\eta   - 2 \aphi\rho  \func_\rho  \right) \Big\}\nn\\
&&+c_{13}\hat \eta \left(\func^2-4 \rho^2  \func_\eta ^2  \right)-8  c_{12}\rho^2 \hat \eta   \func_\rho^2 \Big]
 +\rho  (g^{12}-g^{34}) \Big[2c_{11}\Big\{ 2\rho^2 \func_{\rho\rho}\big(\aphi\func- 2 \aphi \rho \func_{\rho}\nn\\
 &&-2 \hat\eta \func_\eta  \big)+2\rho \func_{\rho\eta}\left(\hat \eta \func+2 \rho \hat \eta \func_{\rho} -2\aphi\rho^2\func_\eta  \right)  + 3 \aphi\rho \func  \func_\rho + 4 \hat \eta \func  \func_\eta + 2 \rho \hat \eta  \func_\eta \func_\rho \Big\}\nn\\
 &&-8\aphi^3 \rho^4  \func_{\rho}^2 -4  \aphi c_{13} \rho^2 \func_\eta^2-\aphi c_{31}\func^2\Big] \Bigg]~.
\eea
Here $y$ is given by the expression in (\ref{yab}) and we have defined $\hat \eta\equiv \aspsi+\aphi \eta $ 
and $c_{mn}=c_{mn}(\rho,\eta) \equiv  m\, \hat \eta^2 + n\, \aphi^2 \rho^2$. The vielbein are defined as 
\bea
g^1 &=& \sqrt{\frac{4 \rho ^2 \left(\func_\eta^2+\func_\rho^2\right)-\func^2}{4\rho ^2 \func^2}} \  \dd \eta~, \nn\\
g^2 &=& \frac{2}{\sqrt{\func^2 \left(4 \rho ^2 \left(\func_\eta^2+\func_\rho^2\right)-\func^2\right)}} \Big( \left(\eta  y^{\text{can}}_\rho-\rho  y^{\text{can}}_\eta\right)\dd \varphi + y^{\text{can}}_\rho\dd\spsi\Big)~,\nn\\
g^3 &=& \frac{2}{\sqrt{\func^2 \left(4 \rho ^2 \left(\func_\eta^2+\func_\rho^2\right)-\func^2\right)}} \Big( \left(\eta  y^{\text{can}}_\eta +\rho  y^{\text{can}}_\rho -y^{\text{can}}\right)\dd\varphi+y^{\text{can}}_\eta \dd \spsi \Big)~,\nn\\
g^4&=&\sqrt{\frac{4 \rho ^2 \left(\func_\eta^2+\func_\rho^2\right)-\func^2}{4\rho ^2 \func^2}} \ \dd\rho~.
\eea

\subsection{$m$-pole solutions}
\label{mpolesec}

 There is a basic solution to (\ref{Laplace}), namely
\bea\label{basicpole}
\mathcal{F}(\rho,\eta; \lambda) &=& \frac{\sqrt{\rho^2+(\eta-\lambda)^2}}{\sqrt{\rho}}~,
\eea
where $\lambda$ is  any constant. We will refer to this  as a \emph{single monopole} solution, and via linearity 
then
\bea\label{mpole}
\mathcal{F}(\rho,\eta) &=& \sum_{i=1}^m \alpha_i \mathcal{F}(\rho,\eta;\lambda_i)~,
\eea
also solves (\ref{Laplace}), for arbitrary constants $\alpha_i,\lambda_i$, $i=1,\ldots, m$.
 We will refer to (\ref{mpole}) as an $m$-pole solution.
 Of course, one could also replace the sum in (\ref{mpole}) by an integral, 
 smearing the monopoles in some chosen charge distribution. 
 The local construction of infinitely many self-dual Einstein metrics is thus  straightforward
via this construction. 

For the $m$-pole solution (\ref{mpole}) the metric (\ref{toricmetric}) 
depends on only $2m-4$ of the $2m$ constants in 
(\ref{mpole}). This follows from taking into account symmetries. 
Recall that the isometry group of the hyperbolic upper half plane $\mathcal{H}^2$ is 
$PSL(2,\R)$.  In terms of the $(\rho,\eta)$ coordinates in (\ref{H2metric})
this is generated by the three simple transformations
\bea\label{TRI}
\mbox{Translation }: \qquad \eta &\rightarrow & \eta + \mathbf{b}~, \qquad\nn\\
\mbox{Rescaling\  \ \, }: \qquad\eta & \rightarrow &  \mu^2\eta~, \quad \rho\ \rightarrow \ \mu^2\rho~,\nn\\
\mbox{Inversion\   \, \, }: \qquad\eta & \rightarrow & -\frac{\eta}{\rho^2+\eta^2}~, \quad \rho \ \rightarrow \ \frac{\rho}{\rho^2+\eta^2}~.
\eea
We may write these as $SL(2,\R)$ matrices by defining the complex coordinate $Z\equiv\eta+\ii \rho$, so that $PSL(2,\R)$ acts as
\bea
Z & \rightarrow & \frac{\mathbf{a}Z+\mathbf{b}}{\mathbf{c}Z+\mathbf{d}}~, \qquad \left(\begin{array}{cc}\mathbf{a} & \mathbf{b}\\ \mathbf{c} & \mathbf{d}\end{array}\right)\in SL(2,\R)~.
\eea
The above three transformations are then
\bea
T &=& \left(\begin{array}{cc}1 & \mathbf{b} \\ 0 & 1\end{array}\right)~, \qquad R \ = \  \left(\begin{array}{cc}\mu & 0\\ 0 & \frac{1}{\mu}\end{array}\right)~,
\qquad I \ = \  \left(\begin{array}{cc}0 & -1\\ 1 & 0\end{array}\right)~,
\eea
respectively. The symmetries (\ref{TRI}) extend to isometries of the self-dual Einstein metric (\ref{toricmetric})
by also acting on the angular coordinates via
\bea\label{TRIangles}
\mbox{Translation }: \qquad \spsi &\rightarrow & \spsi - \mathbf{b}\varphi~, \qquad\nn\\
\mbox{Rescaling\  \ \, }: \qquad\spsi & \rightarrow &  \mu\spsi~, \quad \varphi\ \rightarrow \ \frac{1}{\mu}\varphi~,\nn\\
\mbox{Inversion\   \, \, }: \qquad\spsi & \rightarrow & -\varphi~, \quad \varphi \ \rightarrow \ \spsi~,
\eea
respectively, and for the $m$-pole solution (\ref{mpole}) one acts on the monopole parameters $\alpha_i$, $\lambda_i$ via
\bea\label{TRImpole}
\mbox{Translation }: \qquad \lambda_i &\rightarrow & \lambda_i + \mathbf{b}~, \qquad\nn\\[2mm]
\mbox{Rescaling\  \ \, }: \qquad\lambda_i & \rightarrow &  \mu^2\lambda_i~, \quad \alpha_i\ \rightarrow \ \frac{1}{\mu}\alpha_i~,\nn\\
\mbox{Inversion\   \, \, }: \qquad\lambda_i & \rightarrow & -\frac{1}{\lambda_i}~, \quad \alpha_i \ \rightarrow \ \alpha_i|\lambda_i|~,
\eea
respectively. Most of these are easily verified, apart from the action of inversion on the angular coordinates. 
Here it is useful to establish a number of transformation properties, such as $\eta\partial_\rho-\rho\partial_\eta \rightarrow -\eta\partial_\rho+\rho\partial_\eta$ under inversion.
In addition to the above $PSL(2,\R)$ transformations, we may also simply rescale
\bea
\mathcal{F}(\rho,\eta) & \rightarrow & \kappa \mathcal{F}(\rho,\eta)~, \quad \spsi \ \rightarrow \kappa\spsi~, \quad \varphi \ \rightarrow \ \kappa\varphi~,
\eea
which for the $m$-pole solution simply scales $\alpha_i\rightarrow \kappa\alpha_i$.

Note  that under  the $PSL(2,\R)$ symmetry action on the coordinates $(\rho,\eta)$, the basic monopole solution transforms as 
\bea
\alpha\frac{\sqrt{\rho^2 + (\eta - \lambda)^2}}{\sqrt{\rho}} ~~\to ~~ \alpha' \frac{\sqrt{\rho^2 + (\eta - \lambda')^2}}{\sqrt{\rho}} ~,
\label{donald}
\eea
where 
\bea
\alpha' \ = \  \alpha |\mathbf{c}\lambda + \mathbf{d} |~, \qquad  \lambda' \ = \ \frac{\mathbf{a}\lambda + \mathbf{b}}{\mathbf{c}\lambda + \mathbf{d}}~,
\label{minnie}
\eea
with $\mathbf{a d} -\mathbf{b c}=1$.  One can then use these symmetries to fix 4 of the $2m$ parameters in (\ref{mpole}). We shall see this explicitly 
for the 2-pole  and 3-pole solutions that we examine in detail below.
 
\subsection{AdS$_4$ from $2$-pole solution}
\label{2pole}

The simplest  example of the construction described above is the 2-pole solution, which turns out to be 
Euclidean AdS$_4$, that is the four-dimensional hyperbolic space. Using the $PSL(2,\R)$ symmetry plus 
the overall scaling symmetry discussed in section \ref{mpolesec}, we can  set 
$\lambda_1=-\lambda_2=1$ and $\alpha_1=-\alpha_2=-\frac{1}{2}$ without loss of generality. Therefore we have
\bea
 \func_{\mathrm{EAdS}} \ = 
 \ \frac{\sqrt{\rho^2+(\eta+1)^2}-\sqrt{\rho^2+(\eta-1)^2}}{2\sqrt{\rho}}~,
\eea
with the conformal factor for a generic choice of Killing vector 
\be
y (\rho,\eta) \,  = \, \frac{\sqrt{\rho ^2+(\eta +1)^2}-\sqrt{\rho ^2+(\eta -1)^2}}{2 \sqrt{(\aspsi+\aphi \eta)^2 +\aphi^2 \rho^2}}~.
\label{yre}
\ee
Identifying $\varphi=\varphi_1$, $\spsi = \varphi_2$ (so that $\aphi=b_1$, $\aspsi=b_2$), and introducing the  change of coordinates
\bea
\rho \ =\  \frac{4 r_1 r_2}{(1+r_1^2 +r^2_2)^2-4r_1^2}~, \qquad \eta\  =\  \frac{(1+r_1^2+r_2^2)(1-r^2_1-r_2^2)}{(1+r_1^2+r_2^2)^2 - 4 r_1^2}~,\label{adscoords}
\eea
the general toric  metric takes the form 
\bea
\diff s^2_{\mathrm{EAdS_4}} &=& \frac{4}{(1-r_1^2-r_2^2)^2}\left(\diff r_1^2 + r_1^2\diff\varphi_1^2 + \diff r_2^2 + r_2^2\diff \varphi_2^2\right)~,\label{EAdS}
\eea 
which is manifestly the metric of Euclidean AdS$_4$, realised as a hyperbolic ball. 
In particular,  $r_1,r_2\geq 0$ are constrained  by $r_1^2+r_2^2<1$, with $\{r_1^2+r_2^2=1\}$ being the conformal boundary $S^3$. 
In these coordinates the conformal factor reads
\be
y (r_1,r_2) \ = \  \frac{1-r_1^2-r_2^2}{\sqrt{2 \left(b_2^2-b_1^2\right) \left(r_2^2-r_1^2\right)+(b_2-b_1)^2  \left(r_1^2+r_2^2\right)^2 +(b_1+b_2)^2}}~.
\label{yads4}
\ee

It is instructive to analyse how the $(\rho,\eta)$ coordinates behave globally, 
in this simple example. 
 First note that the polar axes map precisely to $\rho=0$. 
That is, $\rho=0$ if and only if $r_1=0$ or $r_2=0$. 
Looking more closely at the axes, we  have
\bea
\rho(r_1,0) &=& 0~, \qquad \eta(r_1,0) \ = \ \frac{1+r_1^2}{1-r_1^2}~, \nn\\
\rho(0,r_2) &=& 0~, \qquad \eta(0,r_2) \ = \ \frac{1-r_2^2}{1+r_2^2}~.
\eea
In particular the origin $O=\{r_1=r_2=0\}$, which is the NUT, 
maps to the point $(\rho,\eta)=(0,1)$. Indeed, inserting these values into either (\ref{yads4}) or (\ref{yre}), we recover the general expression (\ref{yNUT}) for 
\be
y_\mathrm{NUT} \ = \ \frac{1}{|b_1+b_2|}~.
\ee
The axis $(r_1,r_2)=(0,r_2)$ for $r_2\in [0,1)$ then maps to $\eta\in (0,1]$, while 
the axis $(r_1,r_2)=(r_1,0)$ for $r_1\in [0,1)$ maps to $\eta\in [1,\infty)$.
Notice
\bea
(1+r_1^2+r_2^2)^2 - 4r_1^2 \ \geq \ (1-r_1^2)^2\  \geq \ 0~,
\eea
with equality holding in the first inequality if and only if $r_2=0$. It follows from this 
that $\rho\geq 0$ and $\eta>0$. 
The coordinate region $\{r_1^2+r_2^2<1\}$ then in fact maps one-to-one to 
the positive quadrant  $\{\rho\geq 0,\eta>0\}$, with the axes mapping to $\rho=0$ in the above way. 
The conformal boundary $\{r_1^2+r_2^2=1\}=\{y^{\text{can}}=0\}$ is mapped to $\{\eta =0 \} \cup  \{ (\rho=\infty,\eta=\infty)\}$, namely the axis $\rho\geq 0$ 
plus the point at infinity in ${\cal H}^2$, the latter corresponding to the point $(r_1, r_2)= (1,0)$. 

Looking at the collapsing Killing vectors for this solution, we see that 
$\partial_{\varphi_2}$ collapses along $\{\rho=0,\eta\geq \lambda_1 = 1\}$, 
while $\partial_{\varphi_1}$ collapses along $\{\rho=0,\eta\in(0,1]\}$, where 
the $(\rho,\eta)-$plane is cut off at $\eta=0$ by the conformal boundary. 
In fact on the whole of $\mathcal{H}^2$ we have that 
$\partial_{\varphi_2}$ collapses on $\{\rho=0,\eta\geq \lambda_1=1\}\cup 
\{\rho=0,\eta\leq \lambda_2 = -1\}$, while 
$\partial_{\varphi_1}$ collapses on the interval $\{\rho=0,\eta\in[\lambda_2,\lambda_1]=[-1,1]\}$. 
We thus see the division of the $\eta$-axis into the three segments $\lambda_3=-\infty<\lambda_2=-1<
\lambda_1=1<\lambda_0=\infty$, with different Killing vectors collapsing in each of the 3 
regions. However, the conformal boundary actually cuts off half this axis. 

\begin{figure}[ht!]
        \centering
        \begin{subfigure}[b]{0.4\textwidth}
                \centering
              \includegraphics[width=\textwidth]{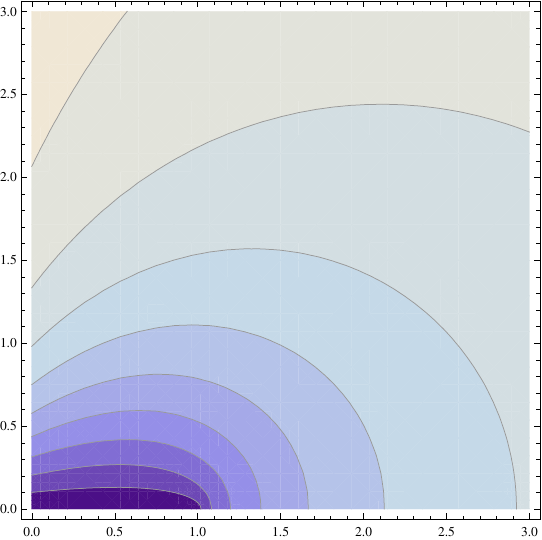}
                \caption{Constant $r_1$ contours in the $(\rho,\eta)$ quadrant. The axis $r_1=0$ maps to 
                $\eta\in (0,1]$ on $\rho=0$.}
                \label{fig:R1}
        \end{subfigure}%
        ~ \qquad
        \begin{subfigure}[b]{0.4\textwidth}
                \centering
                \includegraphics[width=\textwidth]{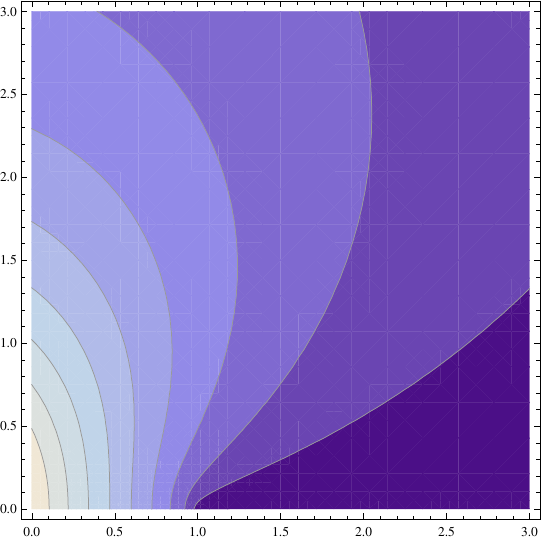}
                \caption{Constant $r_2$ contours in the $(\rho,\eta)$ quadrant. The axis $r_2=0$ maps to 
                $\eta\in [1,\infty)$ on $\rho=0$. }
               \label{fig:R2}
        \end{subfigure}
                	\qquad
        	  \begin{subfigure}[b]{0.04\textwidth}
                \centering
                \includegraphics[width=\textwidth]{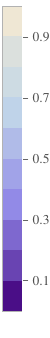}
        \end{subfigure}
                \caption{Contour plots in the $(\rho,\eta)$ quadrant.}\label{fig:contours}
                \end{figure}

It is straightforward to check that the metric in the $(\rho,\eta)$ coordinates is non-singular near the loci where the Killing vectors collapse. 
Using the  expansions, the angular part of the metric reads
\bea
\diff s^2_{\mathrm{angular}} &=& \begin{cases}\ \diff\varphi^2 + \frac{\rho^2}{(\eta^2-1)^2}(\diff\spsi^2-\diff\varphi^2)+\mathcal{O}(\rho^4)~,  & \quad |\eta|>1~, \\ 
\ \diff\spsi^2 + \frac{\rho^2}{(1-\eta^2)^2}(\diff\varphi^2-\diff\spsi^2)+\mathcal{O}(\rho^4)~, & \quad |\eta|<1~,\end{cases}
\eea
and one sees explicitly that $\partial_{\varphi_1}=\partial_\varphi$ collapses 
along $\{\rho=0$, $|\eta|<1\}$, while $\partial_{\varphi_2}=\partial_\spsi$ collapses 
along $\{\rho=0$, $|\eta|>1\}$, which agrees with the statements above. The factor in front of this 
angular part of the metric in (\ref{toricmetric}) is
\bea
\Upsilon(\rho,\eta)^2 \, \equiv \, \frac{4}{ \mathcal{F}^2[4\rho^2(\mathcal{F}_\rho^2+\mathcal{F}_\eta^2)-\mathcal{F}^2]}\, =\,  \frac{4\sqrt{\rho^2+(1-\eta)^2}\sqrt{\rho^2+(1+\eta)^2}}{\left(\sqrt{\rho^2+(1-\eta)^2}-\sqrt{\rho^2+(1+\eta)^2}\right)^2}~.
\eea
Notice that the denominator is non-zero on the quadrant $\{\rho\geq 0,\eta>0\}\subset \mathcal{H}^2$, 
while the numerator is zero precisely at the origin $\rho=0$, $\eta=1$. This is necessary 
in order that the metric is regular at the origin. In particular, we find
\bea
{(1-r_1^2-r_2^2)^2}{\Upsilon^2}\diff s^2_{\mathrm{angular}} &=& 4r_1^2\diff\varphi^2 + 
4r_2^2\diff\spsi^2 + \mathcal{O}(r_1^4, r_2^4)~.
\eea

Notice that the canonically defined Killing vector  $\partial_{\spsi}=\partial_{\varphi_2}$
has a fixed point set along the axis $r_2=0$, which is a copy of $\R^2$. 
Thus the induced Killing vector on the conformal boundary is \emph{not} a Reeb vector 
field.
Indeed, setting $b_1=0, b_2=1$ in (\ref{yads4}) we see that  
\bea
y^{\text{can}}(r_1,r_2) &=& \frac{1-r_1^2-r_2^2}{\sqrt{(1+r_1^2+r_2^2)^2-4r_1^2}}~.
\eea
In particular, $y^{\text{can}}(r_1,0)\equiv 1$ for $r_1\in [0,1)$ is constant along the axis where the associated Killing vector $||\partial_\spsi||=0$. 
But also $y^{\text{can}}=0$ defines the conformal boundary, which contains the point $(r_1,r_2)=(1,0)$. 
Thus actually $y^{\text{can}}$ is not even a continuous 
function on the conformal compactification: it is identically 1 along the axis, which intersects the conformal boundary at infinity,
where it  jumps to 0.  Thus the general expansions we have made 
are not valid and this case is not covered by our analysis.
On the other hand, assuming $b_1\neq 0 $, $b_2\neq 0$ and 
expanding (\ref{yads4})  near the points ${r_1=0,r_2=1}$ and ${r_1=1,r_2=0}$ we find
\bea
y (r_1=0,r_2) & = & \frac{(r_2 - 1)^2}{b_2^2}+ {\cal O }((r_2-1)^3)~,\nonumber\\
y (r_1,r_2=0) & = & \frac{(r_1 - 1)^2}{b_1^2}+ {\cal O }((r_1-1)^3)~,
\eea
respectively, so that $y(r_1,r_2)$ is now a continuous function on the conformal compactification, and we automatically obtain a non-singular instanton.

Doing a further  change of coordinates, setting
\be
r_1 \ = \ \frac{\sqrt{q^2+1}-1}{q} \cos\vartheta ~, \qquad r_2 \  = \ \frac{\sqrt{2+q^2-2\sqrt{1+q^2}}\sin\vartheta}{q}~,
\ee
the metric becomes
\be
\diff s^2_{\mathrm{EAdS}_4}\  = \ \frac{\dd q^2}{q^2+1} +q^2 \left( \dd \vartheta^2 +\cos^2\vartheta \dd \varphi_1^2 + \sin^2 \vartheta \dd \varphi_2\right)~,
\ee
and the conformal factor reads 
\be
y  (q,\vartheta )\ =\ \frac{1}{ \sqrt{\left(b_2+b_1 \sqrt{q^2+1}\right)^2\cos ^2\vartheta+ \left(b_1+b_2 \sqrt{q^2+1}\right)^2\sin ^2 \vartheta } }~,
\ee
in agreement with the formulas in section \ref{AdS4}.

Finally, let us present the instanton in the various coordinate systems introduced. In the original $(\rho,\eta)$ coordinates the general instanton simplifies to 
\bea
F&=&\frac{y (\rho,\eta)(b_1^2-b_2^2)\sqrt{\rho}\func_\mathrm{EAdS} (\rho,\eta)}{2 c_{11} \sqrt{\left(\rho^2+(\eta +1)^2\right)\left(\rho ^2+(\eta -1)^2\right)} } 
\Big[(g^{13}+g^{24}) \left( \hat \eta\left(\eta^2-1\right)  -\rho ^2
    (b_2+3 b_1 \eta )\right)\nn\\
&&+\rho  (g^{12}-g^{34}) \left(2 b_2 \eta +b_1 \left(3 \eta ^2-1\right)-b_1 \rho^2 \right)\Big]~.
\eea
In the $(r_1,r_2)$-coordinates, we have instead
\bea
F&=&\frac{2 (b_2^2-b_1^2) y(r_1,r_2)^3}{\left(1-r_1^2-r_2^2\right)^3}
   \Big[\Big(r_1 \dd r_1 \wedge \dd\varphi_1 -r_2\dd r_2\dd\varphi_2  \Big)\Big(b_2 \left(1 -r_1^2+r_2^2\right) \nn\\
  && + b_1 \left(1 +r_1^2-r_2^2\right) \Big)-2 (b_2-b_1) r_1 r_2 \Big( r_2\dd r_1 \wedge \dd \varphi_2 +r_1 \dd r_2 \wedge \dd\varphi_1  \Big)\Big]~,\label{Fr1r2}
\eea
and the corresponding gauge field reads
\bea
A= \frac{\big[b_1 \left(1-r_1^2-r_2^2\right)+b_2 \left(1+r_1^2+r_2^2\right)\big]\dd\varphi_1+\big[b_1 \left(1+r_1^2+r_2^2\right)+b_2  \left(1-r_1^2-r_2^2\right)\big]\dd \varphi_2}{2 \sqrt{2 (b_2^2-b_1^2) (r_2^2-r_1^2) +(b_1-b_2)^2 (r_1^2+r_2^2)^2 +(b_1+b_2)^2 }} ~. \nonumber
\eea
In the $(\vartheta,q)$ coordinates this becomes 
\be
A\ =\ \frac{\left( b_1+b_2 \sqrt{q^2+1}\right)\dd \varphi_1 + \left( b_2 +b_1 \sqrt{q^2+1} \right) \dd \varphi_2}{2\sqrt{\left( b_2 + b_1 \sqrt{q^2+1}\right)^2 \cos^2 \vartheta + \left(b_1 + b_2 \sqrt{q^2+1}\right)^2 \sin^2 \vartheta }}~,
\ee
which is the expression written in section  \ref{AdS4} and 
originally presented in \cite{Martelli:2012sz}. 

\subsection{Plebanski-Demianski from $3$-pole solution}

\label{morelabels}

\begin{figure}[t]
\centering
\begin{tikzpicture}

\draw[<->,snake=snake,segment amplitude =0.5 mm ,segment length=3 mm,line after snake = 0.9 mm,line before snake=2mm ]  (4,0.36)-- (4,1.25);
\draw[decoration={markings,mark=at position 1 with {\arrow[ultra thick]{>}}},postaction={decorate}]
    (-.75,0) -- (.75,0);
\node[thick,black,draw=black,shape=rectangle,text width=15 em,text centered] at (4,1.6) {\eqref{ortho} ortho-toric $(\xi,\mu)$};
\draw[<->,snake=snake,segment amplitude =0.5 mm ,segment length=3 mm,line after snake = 0.9 mm,line before snake=2mm ]  (4,1.96)-- (4,2.85);
\draw[decoration={markings,mark=at position 1 with {\arrow[ultra thick]{>}}},postaction={decorate}]
    (4,1.96) -- (4,1.95);\draw[decoration={markings,mark=at position 1 with {\arrow[ultra thick]{>}}},postaction={decorate}](4,2.84)--(4,2.85);
\node[thick,black,draw=black,shape=rectangle,text width=15 em,text centered] at (4,3.2) {\eqref{RSmetric} 3-pole $(R,S)$};
\node[thick,black,draw=black,shape=rectangle,text width=15 em,text centered] at (-4,3.2){\eqref{toricmetric} 3-pole $(\rho,\eta)$} ;
\draw[decoration={markings,mark=at position 1 with {\arrow[ultra thick]{>}}},postaction={decorate}]
       (-.75,3.2)--(0.75,3.2) ;\draw[decoration={markings,mark=at position 1 with {\arrow[ultra thick]{>}}},postaction={decorate}](.75,3.2)--(-0.75,3.2);
\node[thick,black,draw=black,shape=rectangle,text width=15 em,text centered] at (-4,0) {\eqref{SDE2} self-dual Einstein $(y,z)$};
\draw[decoration={markings,mark=at position 1 with {\arrow[ultra thick]{>}}},postaction={decorate}]
       (4,0.37)--(4,0.36) ;
       \draw[decoration={markings,mark=at position 1 with {\arrow[ultra thick]{>}}},postaction={decorate}]
       (4,1.24)--(4,1.25) ;
\node[thick,black,draw=black,shape=rectangle,text width=15 em,text centered] at (4,0) {\eqref{PDappendix} Plebanski-Demianski $(p,q)$ };
\node[text width=12 em] at (7,2.4) {\footnotesize {$R_{\text{ortho}}^2$}};
\node[text width=12 em] at (7,.8) {\footnotesize {$R_{\text{ortho}}^2$}};
\end{tikzpicture}
\caption{Overview of the metrics and coordinate transformations in this subsection. Straight arrows denote changes of coordinates, while wavy arrows denote 
conformal transformations (with the indicated conformal factor). The bottom arrow points only one way, to represent the fact that the Plebanski-Demianski metric, with a choice of Killing 
vector $K$,  is a special case of the general self-dual Einstein metric.}
\label{diagram}
\end{figure}
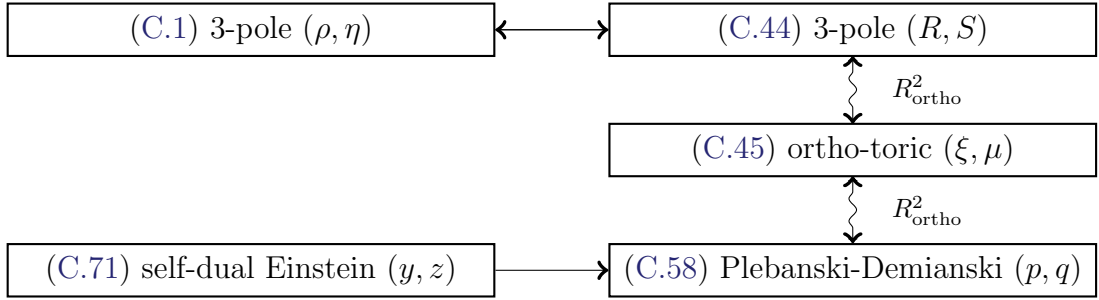

In section \ref{sec:PD} we discussed a two-parameter family of self-dual Einstein metrics on the four-ball.
Although  this was constructed in \cite{Martelli:2013aqa} starting from the local 
Plebanski-Demianski metric, it turns out that it is related  to the 3-pole solutions of  section \ref{mpolesec}. However, the relationship is  complicated
and involves various changes of coordinates and conformal transformations. These are  illustrated in Figure \ref{diagram}, and in the rest of this section will discuss the links in detail. 
Some of these relations were discussed in \cite{CP}, albeit somewhat implicitly. 
In this section we will not be interested in global properties of these metrics, as metrics on the ball; in particular the various angular coordinates that will be introduced do not have  canonical periodicities,
and the action of the associated Killing vectors generically have non-closed orbits. Global properties were discussed in detail in \cite{Martelli:2013aqa}, in the $(p,q)$ coordinate system. 

\subsubsection*{From $(\rho,\eta)$ coordinates  to $(R,S)$ coordinates}

We begin 
noting that applying  the  $PSL(2,\R)$ symmetry transformation in (\ref{donald}), (\ref{minnie})
to the $m=3$ case, we see that we  can choose for example  $\lambda_1'=1$, $\lambda'_2=0$, $\lambda'_3=-1$, so that the general 3-pole solution can be written in the form
\be
\func (\rho,\eta) ~ =~  \frac{b+c}{2}\frac{\sqrt{\rho^2+(\eta+1)^2}}{\sqrt \rho} + \frac{a}{\sqrt{\rho}} +\frac{b-c}{2}\frac{\sqrt{\rho^2+(\eta-1)^2}}{\sqrt \rho}~,\label{3-pole F}
\ee
as presented in \cite{CP}\footnote{In \cite{CP} the 3-pole solution appears written in terms of a parameter $m$, such that $m^2=\mp 1$. 
In this reference these two cases are referred to as Type I and Type II metrics, respectively. Here we are interested only in the case $m^2=1$, corresponding 
to the expression in (\ref{3-pole F}).}.
Using the residual scaling symmetry one of the three parameters $a,b,c$ could be set to an arbitrary non-zero value. However, in the following we will find it convenient 
to keep the 
three parameters in the expressions. The metric \eqref{toricmetric} with $\func(\rho,\eta)$ given by \eqref{3-pole F} corresponds to
the top left corner of Figure \ref{diagram}. Following \cite{CP}, let us define new coordinates\footnote{This change of coordinates may be easily inverted  as
\be
R^2 = \frac{1}{2}\left(1+ \rho^2 + \eta^2 + \sqrt{(1+\rho^2+\eta^2)^2 - 4 \eta^2} \right)\, ,  \quad \sin\hat \theta = \frac{\eta}{R}~. \nn
\ee}
$R,\hat \theta$ as
\be
\rho \ =\  \sqrt{R^2 - 1} \cos \hat \theta ~, \qquad \eta \ = \ R \sin \hat \theta~,
\ee
so that 
\bea
y^\mathrm{can} (R,\theta) \ =  \ a+ b R + c \sin \hat \theta~,
\eea 
and
\be
\rho^{-1} \left( \tfrac14 \func^2 - \rho^2 \left( \func^2_\rho+\func^2_\eta \right)\right) \ =\  \frac{b (a R + b) - c(a \sin \hat \theta + c)}{R^2 - \sin^2 \hat \theta}~.\label{RScoords}
\ee
Then further defining  $S=\sin \hat \theta$, the metric \eqref{toricmetric}
becomes\footnote{We find that the formula corresponding to  \eqref{RScoords} in \cite{CP} (middle of page 18) has a
sign error. Moreover, the  angular variables $\psi$, $\varphi$ in the metric  $g_{RS}$ in \cite{CP} (top of page 21) are inverted with respect to those in our equation 
\eqref{RSmetric}.}
\bea
&& \!\!\!\!\!\! \diff s_{RS}^2 \ =\  \frac{c^2-b^2-a( b R-c S)}{(a+b R+c S)^2} \left(\frac{\dd R^2}{R^2-1} +\frac{\dd S^2}{1-S^2}\right)\nn\\
&& +\frac{1}{(a+b R+cS)^2(c^2-b^2 -a (b R -c S))(R^2-S^2)}\nn\\[2mm]
&&\!\!\! \bigg ( (R^2-1)(1-S^2) \big[ (b R-c S) \dd \spsi + (b S-c R) \dd \varphi\big]^2 +\big[ \big(bS(R^2-1)+c R (1-S^2)\big) \dd \spsi\nn\\
&& -\big (cS(R^2-1)+bR(1-S^2)+a(R^2-S^2)\big) \dd \varphi\big]^2 \bigg) ~.
~\label{RSmetric}
\eea
This form of the 3-pole metric appears at page 21 of \cite{CP}  (with opposite sign --- see footnote \ref{signreverse}) and in Figure \ref{diagram} it corresponds to the  second box in the upper part.
Reference \cite{CP} discusses the moduli space of these metrics, parametrized by $(a,b,c)$, 
  including   different topologies and boundary conditions. 
Here we are only interested in the negative curvature case, and we  note that in general the domain of existence in the $(R,S)$ plane is 
strictly contained in the strip  $R \in [1,+\infty)$, $S\in [-1,1]$, where the conformal boundary is  a segment on the line $a+ b R + c S=0$. 
 \begin{figure}[ht!]
\centering
            \includegraphics[width=0.92\textwidth]{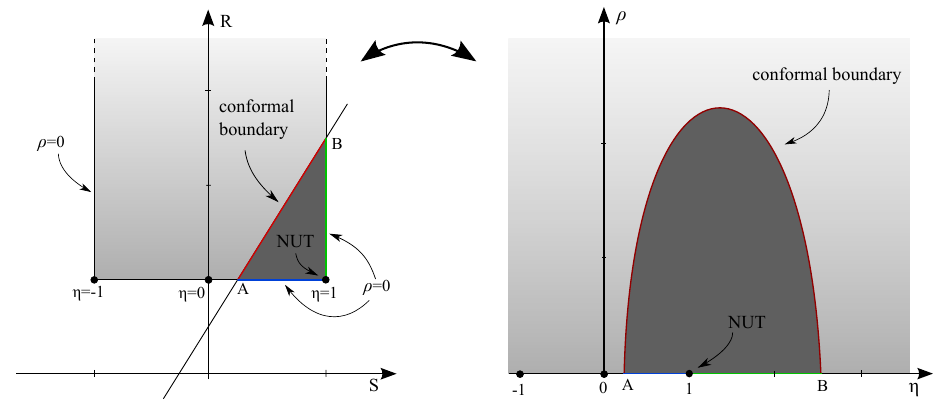}
                \caption{An illustration of the map between $(\rho,\eta)$-coordinates  and $(R,S)$-coordinates:  
                $\rho =   \sqrt{R^2 - 1} \sqrt{1-S^2}$, $\eta  = R S$. The conformal boundary, defined by $y^\mathrm{can}$=0, is simply a segment in the 
                $(R,S)$ plane. This is mapped to an arc intersecting the $\rho=0$ axis at two points ($A$ and $B$). The three marked points on this axis $\eta=-1,0,1$ correspond 
                to the location of the three monopoles in (\ref{3-pole F}), with $\eta=1$ corresponding to the NUT. The parameters  $a,b,c$ are choosen to correspond to region $C$ in 
                Figure 3 of \cite{CP}.}
                \label{nicefigure}
\end{figure}
In Figure \ref{nicefigure} this domain is the triangle on the right-bottom corner of the strip, which maps back to a compact domain in the $(\rho,\eta)$ plane. 
Although this behaviour appears to be qualitatively 
different from that in the AdS$_4$/2-pole case of the previous section, we  notice that 
via $PSL(2,\R)$ transformations we may first shift the point $B$ in Figure \ref{nicefigure} to the origin, 
and then using inversion we may map this to $\infty$. In this way $\ycan=0$ becomes 
a semi-infinite line joining a point $(\rho=0,\eta_0)$ on the $\rho=0$ axis to $\infty$, as in the AdS$_4$/2-pole 
case. However,  we will not further discuss  global issues of the 3-pole metrics in the $(\rho,\eta)$ coordinates, referring 
to \cite{Martelli:2013aqa} for global considerations, in the final $(p,q)$ (Plebanski-Demianski) coordinates.

\subsubsection*{From $(R,S)$ coordinates to  $(\xi,\mu)$ coordinates}

 It was noted in  \cite{CP} that  the 3-pole metric must be  conformally related to the following ortho-toric \cite{ACG}  K\"ahler metric
\bea\label{ortho}
\diff s^2_{\text{ortho}} &= & (\xi-\mu) \left( \frac{\dd \xi^2}{f(\xi)}-\frac{\dd \mu^2}{f(\mu)}\right)+\frac{1}{\xi-\mu} \Big[ f(\xi) \left(\dd t+ \mu \dd v \right)^2 \nn\\
&&-f(\mu) \left(\dd t+ \xi \dd v \right)^2\Big]~,
\eea
with
\be\label{polyf}
f(x) \ =\  (x-x_1)(x-x_2)(x-x_3)(x-x_4)~,
\ee
and $x_1+x_2+x_3+x_4=0$,  so that the quartic polynomial has no cubic term. Here $\xi+\mu$ and $\xi \mu$ are the trace and the Pfaffian of 
the normalized Ricci form of the metric (\ref{RSmetric}), and $(\xi+\mu)^2$ is the 
conformal factor necessary to pass from  the Einstein metric (\ref{RSmetric}) to the K\"ahler metric (\ref{ortho}).
Note that this K\"ahler metric is \emph{not} scalar flat, and therefore it cannot be related by a change of coordinates to the canonical 
  K\"ahler metric of section \ref{sec:Kahler}.  Indeed, it will become clear shortly that one has to make \emph{two} different 
  conformal transformations to relate the K\"ahler metric (\ref{ortho}) to the K\"ahler metric (\ref{Kahler}). 
  
The conformal transformation relating \eqref{RSmetric} and \eqref{ortho} reads
\be
\diff s^2_{\text{ortho}} \ = \ \kappa^2 \left(\frac{a+ b R +c S}{b^2-c^2+a(b R-cS)}\right)^2 \diff s^2_{RS}~,
\label{confRS}
\ee
where $\kappa$ is an arbitrary constant.  In particular, computing the Ricci scalars of the metrics on each side of 
equation \eqref{confRS} and equating these, yields the relation
\bea
\frac{\kappa^3}{a \left(b^2-c^2\right) }(\xi+\mu)   &=&\kappa  \frac{ a+b R+c S}{b^2-c^2 +a (b R-c  S)}~.\label{ricci}
\eea

As noted below \eqref{3-pole F}, using the scaling symmetry the three parameters $a,b,c$ can be multiplied by an overall non-zero constant, while $g_{RS}$
is invariant. For example one could arrange for the conformal factor \eqref{confRS} to be simply $(\xi+\mu)^2$. Instead, 
leaving the arbitrary paramater $\kappa$,  we find that the change of coordinates is given by
\bea
R &=& \frac{1}{2 a b \Delta } \Big[2\kappa ^2\left( \left(b^2-c^2\right)^2-3 a^2  b^2 -a^2c^2 \right)(\xi+\mu) +4 \kappa^4 \left(  a^2+b^2-c^2 \right)\xi \mu\nn\\
&&+\left(b^2-c^2\right)^3  + a^2 \left(4 a^2 b^2 - 3 b^4+  c^4 + 2 b^2 c^2  \right) \Big]~,\\
S&=& \frac{1}{2 a c \Delta}\Big[ 2\kappa ^2\left( \left(b^2-c^2\right)^2-3 a^2  c^2 -a^2b^2 \right)(\xi+\mu) +4 \kappa^4 \left(  a^2-b^2+c^2 \right)\xi \mu\nn\\
&&-\left(b^2-c^2\right)^3  + a^2 \left(4 a^2 c^2 - 3 c^4+  b^4 + 2 b^2 c^2  \right) \Big]~,
\label{eres}
\eea
where
\be
\Delta \ \equiv \ 2 \kappa^2 a^2 (\xi+\mu)-4 \kappa^4 \xi \mu -2 a^2 (b^2+c^2)+(b^2-c^2)^2  ~.
\ee
Note that $R$ and $S$ are rational functions of the trace $\xi+\mu$ and the Pfaffian $\xi \mu$.
The polynomial $f(x)$ takes the form 
\bea
\!f(x)\ =\ \Big(x-\tfrac{-b^2+c^2-2 ab}{2\kappa^2}\Big) \Big(x-\tfrac{-b^2+c^2+2 a b}{2\kappa^2}\Big) \Big( x- \tfrac{b^2-c^2-2 ac}{2\kappa^2}\Big) 
\Big(x-\tfrac{b^2-c^2+2 a c}{2\kappa^2}\Big)\,.
\label{fabc}
\eea
The angular coordinates are linearly related as
\bea
\varphi&=& \varepsilon \frac{abc( c^2-b^2)}{\kappa^6}\left(  2 \kappa^2   \, t + a^2 \,  v\right)~,\nn\\
\spsi &=&  \varepsilon \frac{a (b^2-c^2)}{2 \kappa^6}\left(2 \kappa^2 (a^2 -b^2-c^2) \, t + \left(a^2 \left(b^2+c^2\right)-\left(b^2-c^2\right)^2\right)\, v\right) ~,~\label{angtrans}
\eea
with the choices $\varepsilon = \pm 1$. Setting  $\kappa^2=\tfrac{1}{2}$ by using the scaling symmetry, one recovers the form of $f(x)$ written in \cite{CP}, up to 
a change of sign of the roots $x_i\to -x_i$, due to the different 
overall sign difference in the metrics $\diff s^2_{RS}$ here and in \cite{CP}. 

The inverse change of coordinates is given by
\bea
\xi&=& \frac{1}{2\kappa^2}\, \frac{a  (b^2-c^2)  (a+b R+c S) \pm \sqrt{ W(R,S) }}{b^2-c^2+a( b R-c S)}~,\nn\\
\mu&=& \frac{1}{2\kappa^2}\, \frac{a  (b^2-c^2)  (a+b R+c S) \mp \sqrt{  W (R,S)}}{b^2-c^2+a( b R-c S)}~,
\label{inversecoc}
\eea
where
\bea
 W (R,S) & \equiv & 4a^4b^2 c^2\left( R^2+ S^2\right)+4 a^2bc\left(  \left(b^2-c^2\right)^2 - a^2\left( c^2+ b^2 \right) \right) R \, S\nn\\[2mm]
& & +  (b^2-c^2)^2  (a-b-c)(a+b-c)(a-b+c)(a+b+c)~.
\eea
Let us show that this makes sense, checking that $W (R,S)\geq 0$ everywhere in  the domain of existence of the metric. 
Of course, it is sufficient to show that $W (R,S)\geq 0$  in the strip $ [1,+\infty) \times [-1,1]$.
By explicit computation one finds that for any value of $(R,S)$  the unique solution of $\de_R W = \de_S W =0$ is 
$R=S=0$, so that $W$ does not have an extremal  point in the interior of $[1,+\infty ) \times [-1,1]$. 
On the boundary of this strip we compute
\bea
W (R,S=\pm 1) & = & \left( (b^2-c^2)^2 - a^2 (b^2+ c^2 \mp 2 bc R)\right)^2 ~,\nn\\
W (R=1,S) & = & \left( (b^2-c^2)^2 - a^2 (b^2+ c^2 - 2 bc S)\right)^2~,
\eea
and for $R\to +\infty$ we have  $W\to4 a^4  b^2 c^2  R^2 >0$ for any $S \in [-1,1]$,
therefore for any large value $R_c$ of $R$, $W$ is non-negative.
Since $W (R,S)$ is a continuous bounded function,    by the extreme value theorem it must attain an absolute maximum and an absolute minimum 
on   the boundary of the  domain  $[1,R_c] \times [-1,1]$. 
As there are no extremal points in the interior of the domain, it follows that  the maximum and the minimum must be on the boundary.
Therefore, since at the boundary $W \geq 0$,  the absolute minimum is also non-negative. This is clearly still true when we let the cut-off $R_c\to \infty$.

Notice that the ortho-toric K\"ahler metric \eqref{ortho} strongly resembles the Plebanski-Demianski Einstein  metric \eqref{PD}. In particular, 
both metrics are characterized  by a quartic polynomial without cubic term. Thus it would be tempting to think that, up to a conformal transformation, 
the two sets of coordinates may be simply related.  However, this is \emph{not} the case.

\subsubsection*{From $(\xi,\mu)$ coordinates to $(p,q)$ coordinates}

Here we will show that after undoing the conformal transformation \eqref{confRS},  transforming the ortho-toric K\"ahler metric \eqref{ortho} into an Einstein self-dual metric, this is related
by a non-trivial change of coordinates to the Plebanski-Demianski metric \eqref{PD}. In particular, we will show that 
\be
\diff s^2_{\text{PD}} \ =\  \frac{8 a^2 (b^2 - c^2)^2}{(\xi+\mu)^2} \diff s^2_{\text{ortho}}~,
\ee
displaying the explicit change of coordinates. 
Note that the conformal factor is simply proportional to the square of the Ricci scalar of the ortho-toric K\"ahler metric, namely $R_{\text{ortho}}= -12 (\xi+\mu)$. 

Recall the Plebanski-Demianski metric in the $(p,q)$-coordinates reads
\be
\diff s^2_\text{PD}\ =\ \frac{\mathcal P(q)}{q^2-p^2} (\dd \tau + p^2 \dd \sigma)^2  - 
\frac{\mathcal P(p)}{q^2-p^2} (\dd \tau +q^2 \dd \sigma )^2 + \frac{q^2-p^2}{\mathcal P(q)} \dd q^2 - \frac{q^2-p^2}{\mathcal P(p)}\dd p^2 ~,\label{PDappendix}
\ee
where
\bea
\mathcal{P}(x)&=&(x-p_1)(x-p_2)(x-p_3)(x-p_4)\nn\\[2mm]
& \equiv & x^4 +E x^2-2Mx + \K ~.
\eea
Here we will denote the constant coefficient of the quartic $\mathcal P(x)$ with the symbol $\K$, instead of using the notation ``$-Q^2+\alpha$'' of \cite{Martelli:2013aqa}. 
This is to emphasize the fact that $L$ is a genuine metric parameter, while  $Q$ and $\alpha$ are not, and are meaningful only when discussing the instanton. 
Rewriting the polynamial $f(x)$ in (\ref{fabc}) as
\be
f(x)\ =\  x^4 + H x^2+Tx+U~,
\ee
where $T=8 a^2 (b^2 - c^2)^2$, the non-angular part of the change of coordinates  is then given by
\bea
\xi&=&\left(\frac{T}{4 M^2}\right)^{1/3} \, \frac{- M \pm\tfrac12 \sqrt{ \Xi (p,q)}}{ (p+q)}\, ,\nn\\[2mm]
\mu&=&\left(\frac{T}{4 M^2}\right)^{1/3}\, \frac{- M \mp \tfrac12 \sqrt {\Xi (p,q)} }{ (p+q)}\,,\label{etapq}
\eea
with 
\be
\Xi (p,q) \ =\   4 M^2+ 4 M (p+q) (2 p q-E)+(p+q)^2 \left(E^2- 4L \right)~.
\ee
Let us postpone showing  that $\Xi(p,q)\geq 0$ until the end of this subsection. 
Remarkably, in spite of this complicated relationship, in both sets of coordinates the metrics are characterized by a single quartic polynomial without the cubic term.
The angular coordinates are given by the linear combination,
\be
t\ = \ \left( \frac{|M|}{4 T^2 }\right)^{1/3} \left(E \sigma -2 \tau\right)~,\qquad  v\ = \ \frac{2 |M|}{T} \sigma ~,
\ee
while the remaining parameters are related as
\bea
H&=&-\frac14 \left(\frac{T^{2}}{ 2 M^4} \right)^{1/3} \left(E^2-4L \right)~,\label{S}\\
U&=&\frac1{64} \left(\frac{T^{2}}{2 M^{4}}\right)^{2/3}  \left(32  M^2E +\left(E^2-4 L\right)^2\right)~.\label{U}
\eea
Of course these parameters are  also related to  the original constants $a,b,c$ as
\bea
H & = & -2 ((b^2 - c^2)^2 + 2 a^2 (b^2 + c^2))~,\nn\\
U & = & 16 a^4 b^2 c^2 + (b^2 - c^2)^4 - 4 a^2 (b^2 - c^2)^2 (b^2 + c^2)~.
\eea

Let us now show that $\Xi (p,q) \geq 0$ in the domain of definition of the coordinates $(p,q)$. Adopting the conventions of 
\cite{Martelli:2013aqa}, we have $p \in [p_3, p_4]$ and $q \in [p_4,+\infty)$, with $p_4>0$ 
and $p_4>p_3$, and we must check that $\Xi (p,q) \ge 0$ everywhere in the strip $[p_3,p_4]\times [p_4,+\infty]$. 
On the boundary of the strip, we find\footnote{Here we have used $p_2=-p_1-p_3-p_4$.}
\bea
&&\!\!\!\!\! \!\! \Xi (p_3,q)\, = \, \left(p_3^3+p_1 p_3 p_4+p_1 p_4 (p_1+p_4)-q \left(p_1^2+p_1 p_4+p_4^2-p_3^2+p_1p_3+p_3p_4\right) \right)^2\, , \nn\\
&& \!\!\!\!\! \!\!\Xi (p_4,q)\, = \, \left(p_1^2 (p_3-q)+p_1 (p_3+p_4) (p_3-q)-p_3^2 q+p_4 q(p_4-p_3)+p_4^3\right)^2\, ,\\
&& \!\!\!\!\!\!\! \Xi (p,p_4) \, =\,  \left(-p \left(p_1^2+p_1 (p_3+p_4)+p_3^2+p_3 p_4-p_4^2\right)+p_1 p_3 p_4+p_1 p_3 (p_1+p_3)+p_4^3\right)^2\, ,\nn
\eea
and for $q \rightarrow \infty$ we have
\be
\Xi \ \to\  q^2 (8 M p+E^2-4 L) \ \equiv \ q^2 \Xi_\infty (p) ~.
\ee
One checks that  $\Xi_\infty (p)  \ge 0$ at $p=p_3$ and $p=p_4$, and because $\Xi_\infty (p)$ is linear in $p$ one concludes that $\Xi  (p,q)\ge 0$ for all $p \in [p_3,p_4]$ for $q \rightarrow \infty$.
 Therefore  $\Xi (p,q)\geq 0$ on the boundary of the rectangular domain  $[p_3,p_4]\times [p_4,q_c]$, for any large $q_c$.
 A computation shows that there exist four points  $(p,q)$ where $\de_p \Xi = \de_q \Xi =0$.
Two of these points are
\be
p \ =\ - q \ = \ \pm \sqrt{\frac{-E}{2}}~,
\ee
and exist only if $E\leq 0$. In any case, it's easy to see that the line $p=-q$ intersects the strip  only at $q=p_4$, $p=p_3=-p_4$. 
In \cite{Martelli:2013aqa} it is shown that in order for the instanton to be non-singular, the condition $p_3+p_4>0$ must hold, so these two points are never inside the strip.
Two further points take the form
\be
p \ = \ q \ = \ \alpha + \sqrt{\beta} \qquad \mathrm{and} \qquad  p \ =\  q\  = \ \alpha - \sqrt{\beta}~,
\ee
for some combinations of the parameters denoted $\alpha$ and $\beta$, 
and again it's simple to see that the line $p=q$ intersects the strip only at $p=p_4=q$. 
 Therefore there are no extremal points in the interior of the 
  domain   $[p_3,p_4]\times [p_4,q_c]$ and the argument  to conclude that $\Xi (p,q)\geq 0$ is then exactly the same as that used earlier 
  to show that $W(R,S)\geq 0$.

\subsubsection*{From $(y,u)$ coordinates to $(p,q)$ coordinates}

Finally, let us show that the Plebanski-Demianski metric can be cast in the canonical coordinates characterising the self-dual Einstein metric \eqref{SDE}, which we recall here
\bea\label{SDE2}
\diff s^2_{\mathrm{SDE}} &=& \frac{1}{y^2}\left[V^{-1}(\diff\psi+\phi)^2 + V\left(\diff y^2 + 4\ex^w \diff z\diff\bar{z}\right)\right]~.
\eea
Although in principle this can be done for any choice of Killing vector $K$, the general expressions are  unwieldy.  We will present an expression for the 
instanton constructed from a general Killing vector 
in subsection \ref{thegeneral}.  Here we will consider 
only the  special choices of Killing vector $K$ corresponding to the instantons studied in \cite{Martelli:2013aqa}, for which the  formulas simplify considerably. Moreover, as assumed 
everywhere in this paper, we have to restrict to the \emph{real} solutions in \cite{Martelli:2013aqa}, so that in particular $Q$ and $\sqrt{\alpha}$ are real.

Starting from \cite{Martelli:2013aqa}
\be
K \ =\ \de_\psi \ =  \  2\sqrt \alpha \, \de_\tau + 2\, \de_\sigma ~,
\ee
where 
\be
2 \sqrt{\alpha} \ =\  \frac{M^2}{Q^2} +E~,
\label{seemsusy}
\ee
and using equations \eqref{twistor} and \eqref{ydefinition}, in section \ref{sec:PD} we find that
\be
\frac{1}{y(p,q)^2}\ =\  \frac{4 \big( Q^2 (p+q)+M  p q-M  \sqrt{\alpha}\big)^2}{Q^2}~.
\label{y(p,q)}
\ee
Notice that the relation (\ref{seemsusy}) was derived in \cite{Martelli:2013aqa} by imposing supersymmetry, but in doing so the authors were employing a specific ansatz for $F$. It is the compatibility of that ansatz with 
supersymmetry that yielded (\ref{seemsusy}). However, in our general context we know that supersymmetry is automatic for any choice of Killing vector, and therefore we cannot expect a new relation, such as 
(\ref{seemsusy}) to be found. Thus (\ref{seemsusy}) corresponds  merely to a very special choice of Killing vector. This will become more manifest after writing the general instanton in section \ref{thegeneral}. 

Assuming  the second $U(1)$ isometry of the general metric \eqref{SDE2} may be parametrized by a local angular coordinate $\Theta$, defined through 
$z=u \ex^{\ii \Theta}$,  the self-dual Einstein metric becomes
\bea
\diff s^2_{\mathrm{SDE}} &=& \frac{1}{y^2}\left[V^{-1}(\diff\psi+\phi)^2 + V\left(\diff y^2 + 4\ex^w (\dd u^2+u^2 \dd \Theta^2)\right)\right]~.
\eea
The angular coordinates $(\psi,\Theta)$ must be linearly related to  the angular coordinates $(\sigma,\tau)$ of the Plebanski-Demianski metric \eqref{PDappendix}  as
\be
\left(\begin{array}{c} \tau \\ \sigma \end{array}\right) \ =\ \left(\begin{array}{c c} A & C  \\ B & D \end{array}\right) \left(\begin{array}{c} \psi \\ \Theta \end{array}\right)
\label{trans2} ~,
\ee
where $A= 2\sqrt{\alpha}$ and $B=2$, whereas the entries $C,D$ are arbitrary, provided the transformation  \eqref{trans2} is invertible.
Comparing the relevant terms, we find that the function $V(p,q)$ is  
\be\label{Vpq}
 V (p,q) \ = \ \frac{1}{4 y(p,q)^2} \ \frac{q^2-p^2}{ \mathcal P(q)(\sqrt \alpha  + p^2)^2-\mathcal{P}(p) (\sqrt \alpha + q^2)^2}~,
\ee
and the one-form $\phi$ is
\bea
\label{phipq}
 \phi &  = &   \frac1{2} \ \frac{\mathcal P(q)(\sqrt \alpha +p^2) (C + D\, p^2) -\mathcal{P}(p)(\sqrt \alpha +q^2) (C + D\, q^2  )}{ \mathcal P(q)(\sqrt \alpha  + p^2)^2-\mathcal{P}(p) (\sqrt \alpha + q^2)^2} ~\dd \Theta~.
  \eea 
The coordinate $u$ is found by integrating the following relation
\be\label{dlogu}
\frac{\dd u^2}{u^2}  \   = \  \frac{1}{ Q^2 (C-D \sqrt\alpha)^2 }\left( \frac{    \left(M p+Q^2\right) }{  \mathcal P(p)} \dd p+\frac{  \left(M q+Q^2\right) }{\mathcal P(q)}\dd q \right)^2~.
\ee
If all four roots $p_i$ of the polynomial $\mathcal P (x)$ are distinct this is solved by
\be
u(p,q)\  =\  \tilde C \prod_{i=1}^4  \big[ (p-p_i)(q-p_i) \big]^{\pm  \frac{(M p_i +Q^2 )}{Q(C-D \sqrt\alpha )\mathcal P'(p_i)}}~,
\label{uupq}
\ee
where  $\tilde C$ an integration constant. When $p_1=p_2$, equation \eqref{dlogu} can also be solved by a function $u(p,q)$, but we will not give this expression here. Finally, 
the function $w(p,q)$ is given by 
\be
 \ex^{w (p,q)} = - \frac{y(p,q)^4}{u(p,q)^2} \left( C - D \sqrt \alpha \right)^2 \, \mathcal P(p)\mathcal P(q) ~.\label{PDuexw}
\ee
From these expressions, one can verify that equations \eqref{V}, \eqref{dphi}, and \eqref{Toda} are satisfied.

Notice that choosing   $C=0$ and $D=\frac{1}{2\sqrt{\alpha}}$ so that the angular change of variables \eqref{trans2} 
is an $SL(2,\mathbb{R})$ transformation, the formulas simplify slightly. One can also write more concrete
expressions for $u (p,q)$ in (\ref{uupq}), obtained upon using the various solutions 
for $Q$ \cite{Martelli:2013aqa}:
\bea
\label{onemore}
Q \ =\  \begin{cases}\  \pm \frac{(p_3+p_1)(p_4+p_1)}{2}\\ \  \pm \frac{(p_3+p_4)(p_3+p_1)}{2} \\\  \pm \frac{(p_3+p_4)(p_4+p_1)}{2} \end{cases}~,
\eea
where, as we already noticed, here we  restrict to the region of parameter space where
$Q$ is everywhere real. We refer to  \cite{Martelli:2013aqa} for details.  Setting $\tilde C=1$, and fixing a choice of sign, 
in the first case we have
\bea
u(p,q) & = & \left(\frac{(p-p_2)(q-p_2)}{(p-p_1)(q-p_1)} \right)^{\frac{1}{p_1-p_2}}  \left(\frac{(p-p_3)(q-p_3)}{(p-p_4)(q-p_4)} \right)^{\frac{1}{p_3-p_4}}~,
\label{upq1}
\eea
in the second case we have 
\bea\label{upq2}
u(p,q) & = & \left(\frac{(p-p_2)(q-p_2)}{(p-p_3)(q-p_3)} \right)^{\frac{1}{p_2-p_3}}  \left(\frac{(p-p_4)(q-p_4)}{(p-p_1)(q-p_1)} \right)^{\frac{1}{p_1-p_4}}~,
\eea
and in the third case we have
\bea\label{upq3}
u(p,q) & = & \left(\frac{(p-p_3)(q-p_3)}{(p-p_1)(q-p_1)} \right)^{\frac{1}{p_1-p_3}}  \left(\frac{(p-p_2)(q-p_2)}{(p-p_4)(q-p_4)} \right)^{\frac{1}{p_2-p_4}}~.
\eea
Perhaps not surprisingly these changes of coordinates are very similar to those appearing in equation (25) of \cite{Martelli:2005wy}.

\subsection{General instanton on Plebanski-Demianski}
\label{thegeneral}

Here we illustrate the construction of the general one-parameter instanton, starting directly from the Plebanski-Demianski metric 
\be
\diff s_\mathrm{PD}^2\ =\ \frac{\mathcal P(q)}{q^2-p^2} (\dd \tau + p^2 \dd \sigma)^2  - 
\frac{\mathcal P(p)}{q^2-p^2} (\dd \tau +q^2 \dd \sigma )^2 + \frac{q^2-p^2}{\mathcal P(q)} \dd q^2 - \frac{q^2-p^2}{\mathcal P(p)}\dd p^2,
\ee
with 
\be
\mathcal P (x)\  =\   x^4 +E x^2 -2 M x + \K ~,
\ee
and a Killing vector
\be
K \ = \ \btau  \de_\tau + \bsigma  \de_\sigma ,
\label{genkappa}
\ee
with generic coefficients $\btau,\bsigma$. As before, we will denote the constant coefficient of the quartic $\mathcal P(x)$ with the symbol $\K$, instead of ``$-Q^2+\alpha$''.  This is to emphasize the fact that in our general set up the parameters $\K$, $\btau$, $\bsigma$ are  independent.

Recall that  given the  one-form $K^\flat$, dual to $K$, and the expression for $y$ in (\ref{ydefinition}),
the instanton $F$ can be derived using the following formula
\be
F\ = \ -\left(\tfrac12 y \dd K^\flat + y^2 K^\flat \wedge J K^\flat\right)^-~,
\ee
where the complex structure tensor is 
\be
{J^\mu}_\nu\ =\ -y g^{\mu \rho} \left( \dd K^\flat \right)^+_{\rho \nu} ~.
\ee
Here $g^{\mu\nu}$ is the inverse of the self-dual Einstein metric, and the  contraction with the complex structure is defined as $JK^ \flat = {J^\mu}_\nu K^\flat_\mu \dd x^\nu $.
We then obtain the following general expression
\bea
\frac1{y^2} &=& \frac14 \frac1{(q^2-p^2)^2} \Bigg\{\Bigg[ \left( \frac{2 \mathcal P(q) }{q-p}-\mathcal P'(q)\right) (\btau+ \bsigma p^2) - \left(\frac{2\mathcal P(p)}{q-p}+\mathcal P'(p)\right) 
(\btau + \bsigma q^2)\Bigg]^2 \nn\\
&&-4 \bsigma^2 \mathcal P(q) \mathcal P(p) (q+p)^2\Bigg\}~.
\eea
Inserting the polynomial $\mathcal P(x)$, we see that this is actually  a polynomial of degree two, 
symmetric in $p$ and $q$, namely
\bea
\frac{1}{y(p,q)^2} &=&p^2 q^2 \left(2 \btau  \bsigma -\bsigma^2 E\right)+2 p q \left( \btau  \bsigma E-\btau^2- \bsigma^2 \K\right)+\left(p^2+q^2\right) \left(\btau^2-\bsigma^2 \K\right)\nn\\
&& \!\!\! + \, 2 \bsigma^2 M \left(p^2 q+p q^2\right)-2 \btau \bsigma M (p+q)+2 \btau \bsigma \K+\bsigma^2 (M^2- E \K)~.\label{ypqpd}
\label{newye}
\eea
In the frame
\bea
&& \tilde e^1 \ =\  \sqrt{\frac{q^2-p^2}{-\mathcal P(p)}} \dd p~, \qquad \qquad \qquad \, \tilde e^2 \ =\  \sqrt{\frac{-\mathcal P(p)}{q^2-p^2}} (\dd \tau + q^2 \dd \sigma)~,\nn\\
&& \tilde e^3\ =\ \sqrt{\frac{\mathcal P(q)}{q^2-p^2}} (\dd \tau + p^2 \dd \sigma ) ~, \qquad  \tilde e^4 \ = \ \sqrt{\frac{q^2-p^2}{\mathcal P(q)}} \dd q ~,
\eea
the instanton takes the form 
\bea
F&=&  (\tilde e^{13}+\tilde e^{24}) \,\frac{y(p,q)^3\sqrt{-\mathcal P (p) \mathcal P(q)}}{2 (q+p)}  \left(\bsigma^3 \left(M^2 - E \K \right)-2 \btau^3+\btau^2 \bsigma E+2 \btau 
\bsigma^2 \K \right)\nn\\[2mm]
&+ &  (\tilde e^{12} - \tilde e^{34}) \,\frac{y(p,q)^3}{32 Q^8 (p+q)^2}  \sum_{m,n=0}^3 a_{mn} q^m p^n~,\label{generalpdi}
\eea
with symmetric coefficients, $a_{mn}=a_{nm}$, given by
\bea
a_{00} &=& 2 \btau \bsigma M \left(2 \btau \K+ \bsigma \left(M^2-E \K \right)\right)~,\nn\\
a_{01}&=&-2 \btau^3 \K+\btau^2 \bsigma \left(E \K-6 M^2\right)+2 \btau \bsigma^2 \K^2+\bsigma^3 \K \left(M^2-E\K \right)~,\nn\\
a_{02}&=& M \left(4 \btau^3+\btau^2 \bsigma E-4 \btau \bsigma^2 \K+\bsigma^3 \left(-E \K+M^2\right)\right)~,\nn\\
a_{03}&=&-\btau^3 E+2 \btau^2 \bsigma \K +\btau \bsigma^2 \left(E \K -M^2\right)-2 \bsigma^3 \K^2~,\nn\\
a_{11}&=&6 \btau \bsigma M \left(\btau E-2 \bsigma \K\right)~,\nn\\
a_{12}&=&-\btau^3 E+\btau^2 \bsigma \left(6 \K-E^2\right)+\btau \bsigma^2 \left(E \K +9 M^2\right)+\bsigma^3 \left(E^2 \K-E M^2-6 \K ^2\right)~,\nn\\
a_{13}&=&6 \bsigma M \left(\bsigma^2\K-\btau^2\right)~,\nn\\
a_{22}&=&6 \bsigma^2 M (2 \bsigma\K-\btau E)~,\nn\\
a_{23}&=&-2 \btau^3+\btau^2  \bsigma E+2 \btau  \bsigma^2 \K - \bsigma^3 \left(E \K+5 M^2\right)~,\nn\\
 a_{33}&=&2 \bsigma^2 M (\bsigma E-2 \btau) ~.
\eea 
These  are all homogeneous degree three polynomials in the parameters $\btau$, $\bsigma$, 
but only their ratio is important, so we could set one of them to unity.
We can also express the instanton in terms of $b_1$ and $b_2$, using the relations
\bea
\btau &=& \frac{ 2p_3^2 }{\mathcal P'(p_3)}b_1 - \frac{2p_4^2 }{\mathcal P'(p_4)}b_2~,\\
\bsigma  &=& - \frac{2}{\mathcal P'(p_3)} b_1 + \frac{2}{\mathcal P'(p_4) }b_2~.
\eea
The $(\tilde e^{13}+\tilde e^{24})$ component is rather  simple and reads
\bea
F|_{(\tilde e^{13}+\tilde e^{24})}&=&\frac{y^3 (p_3-p_4) \sqrt{-\mathcal P(p)\mathcal  P(q)}}{(q+p) \mathcal P'(p_3) \mathcal P'(p_4)} \label{b1b2F}\\
&& \!\!\!\! \times  (b_1+b_2) \big(b_1 (p_4-p_1)-b_2 (p_3-p_2)\big) \big(b_1 (p_4-p_2)-b_2 (p_3-p_1)\big)~,\nn
\eea
whereas the  $(\tilde e^{12}-\tilde e^{34})$ component does not simplify and we will not write it here. 

Notice that the second line in (\ref{b1b2F})  vanishes precisely in the three cases corresponding to the solutions in \cite{Martelli:2013aqa}, where this part of the instanton is absent. 
These correspond precisely to the special choice 
\be
\left. \frac{\btau}{\bsigma} \right|_\mathrm{MP} \ =\ \frac12 \left( \frac{M^2}{Q^2} +E \right)~.
\label{speciali}
\ee
Inserting this into (\ref{newye}) one finds that $1/y^2$ factorizes, so that $1/y$ becomes homogeneous of degree one in $p$ and $q$, as in  (\ref{y(p,q)}). Similarly, the symmetric polynomial 
 $ \sum_{m,n=0}^3 a_{mn} q^m p^n $ also becomes the cube of a degree one polynomial, so that the two functions cancel, leaving the 
 enormously simplified instanton
\bea
F & = &  -  \frac{Q }{(q+p)^2}(\tilde e^{12}-\tilde e^{34}) ~,
\eea
in agreement\footnote{Up to an overall sign related to charge conjugation of the spinor  -- see the discussion in the paragraph 
before equation (\ref{A}).}  with (2.28) of \cite{Martelli:2013aqa}.

We also note that  under the exchange of $p$ and $q$ the two terms transform as
\bea
(\tilde e^{12}-e^{34}) &\rightarrow & (\tilde e^{12}-\tilde e^{34})~,\nn\\
(\tilde e^{13}+e^{24}) &\rightarrow & -(\tilde e^{13}+\tilde e^{24})~,
\eea 
respectively, while the functions entering in $F$ are all symmetric.
Therefore, the special instantons in \cite{Martelli:2013aqa} are symmetric under this exchange, while the general instanton is neither symmetric nor antisymmetric, 
thus breaking this symmetry completely.

In conclusion, in this subsection  we have explicitly shown how starting from a metric with two non-trivial parameters ($E,M,L$ mod scaling symmetry) 
we have obtained an instanton, and hence a full  supersymmetric solution, depending
on one further non-trivial parameter ($\btau,\bsigma$ modulo scaling symmetry). By contrast, in the construction of \cite{Martelli:2013aqa}, the full solution depends on only two non-trivial parameters, 
already appearing in the Plebanski-Demianski metric, and the instanton does not introduce a new parameter due to the relation (\ref{speciali}). 

\subsection{Taub-NUT-AdS$_4$ as a limit of Plebanski-Demianski}
\label{TNlimit}

Here we will show how to recover the Taub-NUT-AdS$_4$ metric 
\be
\diff s^2_4 \  = \ \frac{r^2-s^2}{\Omega(r)} \dd r^2 + (r^2-s^2)(\dd \theta^2 + \sin^2 \theta \dd \varphi^2) + \frac{4s^2\Omega(r)}{r^2-s^2}(\dd \TNnu + \cos \theta \dd\varphi)^2 ~,
\label{tnabove}
\ee
with
\be
\Omega(r) \ = \ (r-s)^2 \big(1+(r-s)(r+3s)\big)~,
\ee
from a limit of the Plebanski-Deminaski metric   (\ref{PDappendix}), thus demonstrating that the former is a one-parameter sub-family of the toric 3-pole metric, where the isometry enhances to $SU(2)\times U(1)$.  Applying the same limit to the general instanton on the Plebanski-Demianski metric  
(\ref{generalpdi}) we will also obtain an explicit expression for the general toric instanton on the Taub-NUT-AdS$_4$ metric. 

Following \cite{Martelli:2013aqa} we  parameterise  the four  roots of $\mathcal P(x)$ in terms of two constants $\apd$, $s$ as  
\bea
p_1& =&  - \frac{1}{2} - \sqrt{\apd^2 - 2 M}~, \qquad p_2\ = \ - \frac{1}{2}+ \sqrt{\apd^2 - 2 M }~, \nn\\
 p_3 & = &   \frac{1}{2} -\apd ~,  \qquad \qquad \qquad ~ \, ~p_4 \ = \   \frac{1}{2} +\apd ~,
\eea
with\footnote{In \cite{Martelli:2013aqa} the squashing parameter $s$ was denoted $\tfrac{1}{2v}$.}
\be
2M\ = \ \frac{1}{4s^2}-1 ~.
\ee
Then we   make the following change of coordinates 
\be
p\ =\ \frac12 - \hat a \cos \theta~, \qquad q\ =\  \frac{r}{2s}~, \label{cocpq}
\ee
and 
\be
\tau \ =\  -\left(4 s^2 +\frac{k}{4}\right) \TNnu - \frac{s^2}{\hat a} \varphi~, \qquad \sigma \ = \ k \TNnu+\frac{4s^2}{\hat a} \varphi~,\label{cocts}
\ee
with $k$ an arbitrary real number. Substituting  $p,q,\tau,\sigma$ above into the Plebanski-Demianski metric and taking the limit $\hat a\rightarrow 0$, it is straightforward to verify that  one obtains precisely  the Taub-NUT-AdS$_4$ metric (\ref{tnabove}).

Comparing the expression of the Killing vector in section \ref{thegeneral}, namely 
\be
K\ =\  \de_\psi \ = \ \btau \de_\tau + \bsigma \de_\sigma~,
\ee
with that given in section \ref{TNAdS}, namely 
\be
K\ =\ (b_1+b_2)\partial_\varphi+(b_1-b_2)\partial_\TNnu~, 
\ee
we deduce that the parameters $\btau,\bsigma$ must be related to $b_1,b_2$ as 
\be
\btau \ = \ -\frac{s^2}{\hat a} (b_1+b_2)+\left( \frac{k}{4}   +4s^2 \right)(b_2-b_1)~,\qquad \bsigma \  = \ \frac{4 s^2}{\hat a} (b_1+b_2)+k (b_1-b_2)~.
\ee
Inserting these  into the expression for $y(p,q)$ in  \eqref{ypqpd} along with (\ref{cocpq}) and  (\ref{cocts}), 
 and then taking the limit $\hat a\rightarrow 0$, one finds precisely the $y(r,\theta)$ given in \eqref{yTNUTAdS}. Notice that the final 
 result does not depend on $k$.

Finally, using this change of coordinate/parameters  in the instanton (\ref{generalpdi}),  we find the following  explicit expression for 
  general instanton on the Taub-NUT-AdS$_4$ metric
\bea
F&=&\frac{y^3}{2 } (b_1+b_2)  \left(16 b_1 b_2 s^2-(b_1+b_2)^2\right)(r-s)\sin \theta \Big( \frac{2 s\Omega(r)}{r^2-s^2} \dd\theta \wedge \tau_3+\sin \theta  \dd r \wedge \dd\varphi \Big)\nn\\
&&+\frac{y^3 (r-s)^2}{2  (r+s)}\left(\frac{2 s}{r^2-s^2} \dd r \wedge \tau_3  -\tau_1 \wedge \tau_2\right) \nn\\
&&\Bigg(-(b_1+b_2)^2 \sin ^2\theta   \Big[\left(4 s^2-1\right) (b_1+b_2) \left(4 r^2 s+4 r s^2+r-8 s^3+3 s\right)\cos \theta\nn\\
&&+2 s (b_1-b_2) \left( r^2 \left(8 s^2-1\right)+2 s r \left(4 s^2+1\right)-16 s^4+11 s^2+\frac{2 s}{r-s}\right)\Big]\nn\\
&&+ (2 s (r-s)+1) \left(4s^2-1\right)  \left[s(4s^2-1)-r (4 s^2+1)-2s\right](b_1+b_2)^3\cos^3 \theta  \nn\\
&&-2s(b_1-b_2)(b_1+b_2)^2\cos^2 \theta\nn\\
&&\Big(r^2 \left(48 s^4-4 s^2-1\right)-4 r \left(24 s^5-14 s^3+s\right)+s^2 \left(48 s^4-52 s^2+17\right)+\frac{2 s}{r-s} \Big)\nn\\
&&-8s^3 (b_1-b_2)^2(b_1+b_2) \cos \theta\Big( 1 +2(r-s) \left(6 r s^2-r-6 s^3\right)+\frac{r-s}{2s}(16 s^2-1)\Big)\nn\\
&&-8 s^3 (b_1-b_2)^3  (r-s)^2  \left(4 s^2-1\right) 
\Bigg)~,\label{longtn}
\eea
where  $y(r,\theta)$ is given by \eqref{yTNUTAdS} and $\tau_i$ are the $SU(2)$ left-invariant one-forms
\bea
\tau_1 &=& \cos \TNnu \, \dd \theta +\sin \TNnu \sin \theta \,  \dd \varphi\nn~,\\
\tau_2 &=& -\sin \TNnu \, \dd \theta + \cos \TNnu \sin\theta\,  \dd \varphi\nn~,\\
\tau_3 &=& \dd \TNnu + \cos \theta \, \dd \varphi~.
\eea
Indeed, for $b_1 = -b_2$ this reduces to the 1/4-BPS instanton in (\ref{hello}) 
\be
F_{\tfrac14 \text{BPS}}  \ = \  \frac{1}{2} \left(4 s^2-1\right) \left(\frac{2 s  }{(r+s)^2}\dd r \wedge \tau_3-\frac{ r-s}{r+s}\tau_1 \wedge \tau_2\right)~,
\ee
up to a sign related to charge conjugation of the spinor. While for 
\be
b_1 = \frac{1}{4s}, \qquad b_2 = -\frac{1}{4s} +2s +\sqrt{4s^2-1}~,
\ee
it reduces to the 1/2-BPS instanton in (\ref{hello2}) 
\be
F_{\tfrac12 \text{BPS}} = s\sqrt{4s^2-1}  \left(\frac{ 2s \ }{(r+s)^2}\dd r \wedge \tau_3 -\frac{r-s }{r+s}\tau_1 \wedge \tau_2\right)  ~,
\ee
again up to a sign related to charge conjugation.

Finally, taking the limit $r\to \infty$ of (\ref{longtn}),
it is straightforward to extract the background gauge field induced on the boundary. This has field strength 
\bea
F_{(0)}&=&\frac{\sqrt{2}s}{\mathcal X^{3/2}}\Big[ \tau_1 \wedge \tau_2 \Big(\left(4 s^2-1\right)^2  b_+^3 \cos ^3 \theta +12  s^2 \left(4 s^2-1\right)  b_- b_+^2 \cos ^2\theta \nn\\
&&+2 \left(24 s^4 b_-^2+16s^2 b_1 b_2 -b_+^2\right)b_+\cos \theta + \left(8 s^2-1\right) b_- b_+^2+4 s^2\left(4 s^2-1\right)  b_-^3 \Big)\nn\\
&&+   b_+  \left(16s^2 b_1 b_2 -b_+^2\right)\sin \theta  \ \dd \theta \wedge \tau_3\Big]~,
\eea
where we defined $b_\pm \equiv b_1 \pm b_2$ and
\bea
\mathcal X &=&   b_+^2 \sin^2 \theta +   4  s^2 (b_- + b_+ \cos \theta)^2  ~.
\eea
The corresponding  gauge field takes the form 
\bea
A_{(0)}^\mathrm{local} & = & f_\varphi (\theta) \dd \varphi + f_\anglepsi (\theta) \dd \anglepsi~,
\eea
where 
\bea
f_\varphi (\theta) &=& \frac{s}{\sqrt{\mathcal X} } \left(b_+- ( 4 s^2-1 )(  b_-  + b_+ \cos  \theta )\cos \theta\right)~,\nn\\
f_\anglepsi (\theta) &=& - \frac{s}{\sqrt{\mathcal X} }   \left(4 s^2b_-+(4 s^2-1) b_+ \cos \theta \right)~.
\eea
This provides an explicit one-parameter family of three-dimensional backgrounds interpolating between those of \cite{Hama:2011ea} and 
\cite{Imamura:2011wg}. Of course, in general this preserves only a  $U(1)\times U(1)$ subgroup of the isometry group of the biaxially squashed sphere, which is enhanced to  $SU(2)\times U(1)$  in the two special cases above.

\end{document}